\documentclass[iop,apj,12pt,numberedappendix]{emulateapj}
\usepackage[toc,page]{appendix}
\usepackage[figuresright]{rotating}
\usepackage{natbib}
\usepackage{ulem}
\usepackage{graphicx}
\usepackage{amsfonts}
\usepackage{amsmath}
\usepackage{multirow}
\usepackage{enumerate}
\usepackage[usenames]{xcolor}
\usepackage[plainpages=false, colorlinks=true, anchorcolor=blue, linkcolor=blue, citecolor=blue, bookmarks=false]{hyperref}
\def\sun{_\odot}

\def\Dwa{$\,$\uppercase\expandafter{\romannumeral5}$\,$}

\def\sless{\lower2pt\hbox{$\buildrel {\scriptstyle <}
   \over {\scriptstyle\sim}$}}

\def\sgreat{\lower2pt\hbox{$\buildrel {\scriptstyle >}
   \over {\scriptstyle\sim}$}}
\def\sharpnull#1{}

\begin{document}
\title{Monte Carlo Neutrino Transport through Remnant Disks \\ from Neutron Star Mergers}
\author{Sherwood Richers\altaffilmark{1,+,*}}
\author{Daniel Kasen\altaffilmark{2,3,4}}
\author{Evan O'Connor\altaffilmark{5,++}}
\author{Rodrigo Fern\'andez\altaffilmark{2,3}}
\author{Christian D. Ott\altaffilmark{1}}
\altaffiltext{1}{TAPIR, Mailcode 350-17,
  Walter Burke Institute for Theoretical Physics,
  California Institute of Technology, Pasadena, CA 91125, USA.}
\altaffiltext{2}{Department of Astronomy and Theoretical Astrophysics Center, University of California, Berkeley, CA 94720, USA.}
\altaffiltext{3}{Department of Physics, University of California, Berkeley, 94720, USA.}
\altaffiltext{4}{Nuclear Science Division, Lawrence Berkeley National Laboratory, Berkeley, CA 94720}
\altaffiltext{5}{Department of Physics, Campus Code 8202, North Carolina State University, Raleigh, NC 27695, USA.}
\altaffiltext{*}{srichers@tapir.caltech.edu.}
\altaffiltext{+}{Department of Energy Computational Science Graduate Fellow.}
\altaffiltext{++}{Hubble Fellow}

\begin{abstract}
  We present {\tt Sedonu}, a new open source, steady-state, special relativistic Monte Carlo (MC) neutrino transport code, available at {\tt bitbucket.org/srichers/sedonu}. The code calculates the energy- and angle-dependent neutrino distribution function on fluid backgrounds of any number of spatial dimensions, calculates the rates of change of fluid internal energy and electron fraction, and solves for the equilibrium fluid temperature and electron fraction. We apply this method to snapshots from two-dimensional simulations of accretion disks left behind by binary neutron star mergers, varying the input physics and comparing to the results obtained with a leakage scheme for the case of a central black hole and a central hypermassive neutron star. Neutrinos are guided away from the densest regions of the disk and escape preferentially around 45 degrees from the equatorial plane. Neutrino heating is strengthened by MC transport a few scale heights above the disk midplane near the innermost stable circular orbit, potentially leading to a stronger neutrino-driven wind. Neutrino cooling in the dense midplane of the disk is stronger when using MC transport, leading to a globally higher cooling rate by a factor of a few and a larger leptonization rate by an order of magnitude. We calculate neutrino pair annihilation rates and estimate that an energy of $2.8\times10^{46}\,\mathrm{erg}$ is deposited within $45^\circ$ of the symmetry axis over $300\,\mathrm{ms}$ when a central BH is present. Similarly, $1.9\times10^{48}\,\mathrm{erg}$ is deposited over $3\,\mathrm{s}$ when an HMNS sits at the center, but neither estimate is likely to be sufficient to drive a GRB jet.
\end{abstract}


\maketitle

\section{introduction}
Neutron star-neutron star (NS-NS) and neutron star-black hole (NS-BH) mergers are prime candidates for explaining observed short gamma ray bursts (sGRBs) and their afterglows (see, e.g.,\ \citealt{Berger13} for a recent review). The large amount of extremely neutron-rich matter and available energy make these systems potentially capable of ejecting matter up to $A\sim200$ through the r-process (e.g.,\ \citealt{LS74,Freiburghaus+99,Korobkin+12,Goriely+13}). The thermal and radioactive glow of this ejecta is thought to cause largely isotropic (depeding on the distribution of dynamical ejecta), observable infrared/optical emission lasting on the order of hours to days \citep{LP98,Metzger+10,Berger+13,Barnes+13}. Observation of this so-called kilonova would provide key information about the merger to complement gravitational wave observations (e.g.,\ \citealt{Metzger+12,Nissanke+13,Piran+13}). Some observational evidence of such a kilonova has already been suggested for GRB130603B \citep{Tanvir+13,Berger+13} and GRB060614 \citep{Yang+15}.

Realistic simulations of merging compact objects need to account for general relativity, a hot nuclear equation of state, magnetohydrodynamics (MHD), nuclear reactions, spectral and angle-dependent neutrino transport, and possibly neutrino quantum effects (e.g. flavor oscillations). Neutrinos in particular play an important role in determining the dynamics, brightness, and color of predicted ejecta emission. Neutrino emission and absorption modify the electron fraction and specific entropy of the material, which in turn determine which elements form from the cooling ejecta \citep{Roberts+11,Korobkin+12,Wanajo+14} and the resulting photon opacities \citep{Barnes+13,Kasen13}. Neutrino irradiation can also drive a thermal outflow, generally increasing the amount and electron fraction of ejecta \citep{MS05,SMH06,Wanajo+12,FM13b,Just+15,Martin+15,Goriely+15,Foucart+15}, especially in the presence of a central hypermassive neutron star (HMNS) (\citealt{Dessart+09,Perego+14,Sekiguchi+15,MF14}, hereafter MF14). Neutrino-antineutrino annihilation may generate large amounts of thermal energy in baryon-poor regions and remains a possible engine driving the GRB jet \citep{Eichler+89,Meszaros+92,Popham+99,ZM11,Leng+14}, though many calculations show that the energy production is at best marginally capable of powering GRBs (e.g.\ \citealt{Setiawan+06,Dessart+09})

Due to the great complexity of this problem, all current and past simulation efforts make some level of approximation or evolve only for very short times to make the problem computationally tractable (see, e.g.,\ \citealt{Faber+12,Shibata+11rev} for reviews). Neutrinos are ignored altogether in many studies for simplicity or efficiency (e.g.,\ \citealt{Etienne+12,Kiuchi+14,Bauswein+14,Bernuzzi+15,Takami+15}). \cite{FM13b} approximates self-irradiation from the disk as a gray lightbulb arising from a ring, with optically thin cooling rates corrected for optical depth effects. Various forms of the leakage scheme of \cite{Ruffert+96} can be used to more accurately treat cooling, heating, and electron fraction changes from neutrinos whenever the disk becomes optically thick (\citealt{Janka+99,Ruffert+99,RL03,Sekiguchi+11b,Sekiguchi+11a,Sekiguchi+12,Kiuchi+12,Deaton+13,Galeazzi+13,Neilsen+14,Foucart+14}) although ad-hoc assumptions about the angular distribution of radiation are still needed to compute neutrino absorption (MF14; \citealt{Perego+14,Fernandez+15a,Fernandez+15b}). Moving from simple approximations to actual neutrino transport, \cite{Dessart+09} use Newtonian multi-group flux-limited diffusion (MGFLD) during evolution and multi-group multi-angle transport during post-processing analysis. General relativistic two-moment neutrino transport with an analytic closure (e.g.,\ \citealt{Thorne81,Shibata+11,Cardall+13}) is the state of the art in multi-dimensional simulations and currently gives the most accurate approximation to full Boltzmann transport of all of the methods employed in time-dependent simulations. Initial studies have employed a gray (energy-integrated) transport scheme with general relativity to simulate a merger and remnant \citep{Shibata+12,Sekiguchi+15,Foucart+15}, and \cite{Just+15} recently simulated an axisymmetric remnant disk with Newtonian multi-group two-moment transport.

Unlike the above transport methods, Monte Carlo (MC) transport is continuous in space, direction, and energy, freeing it from many of the grid effects and distribution function approximations used in other methods. Though it is more accurate, it is more computationally expensive and has seen more limited use in the past. MC transport has been used to study neutrino equilibration in a static isotropic background \citep{Tubbs78} and to study transport through static spherically symmetric fluid in the context of core-collapse supernovae (e.g.,\ \citealt{Janka+89,Janka91,Keil+03}. More recently, \cite{Abdikamalov+12} uses a time-dependent MC neutrino transport scheme in static spherically symmetric core-collapse simulation snapshots. MC transport has also been used in the context of photon transport in Ia supernova explosions (e.g.,\ \citealt{KTN06,Wollaeger+13,Roth+15}) and accretion disks (e.g.,\ \citealt{Ryan+15}).

In this paper, we investigate the effect of neutrinos on the rates of change of the composition and thermal energy of the remnant disk and ejecta using time-independent Monte Carlo (MC) neutrino transport calculations. We calculate properties of the neutrino radiation fields to pinpoint the regions of largest error in more approximate schemes and proceed to estimate the effect this would have on dynamical simulations of compact object mergers. We begin by introducing the specifics of our neutrino transport scheme and the background fluid in Section \ref{sec:methods}. We proceed to describe the observed properties and effects of the neutrino radiation field and a comparison to those seen by the leakage scheme in the dynamical calculation with a central black hole in Section \ref{sec:BHresults}. In Section \ref{sec:HMNSresults}, we extend the results to fluid backgrounds with a central hypermassive neutron star (HMNS). We briefly discuss the effects of neutrino pair annihilation for both sets of background data in Section \ref{sec:annihil}. In Section \ref{sec:discussion}, we discuss the potential implications of these results on the dynamical calculation and the implications for nucleosynthesis and kilonovae. In Section \ref{sec:conclusions}, we conclude and list the main points that can be drawn from our results.

\section{methods}
\label{sec:methods}
\subsection{Background Fluid}
\label{sec:background}
\begin{figure*}
  \includegraphics[width=0.525\linewidth]{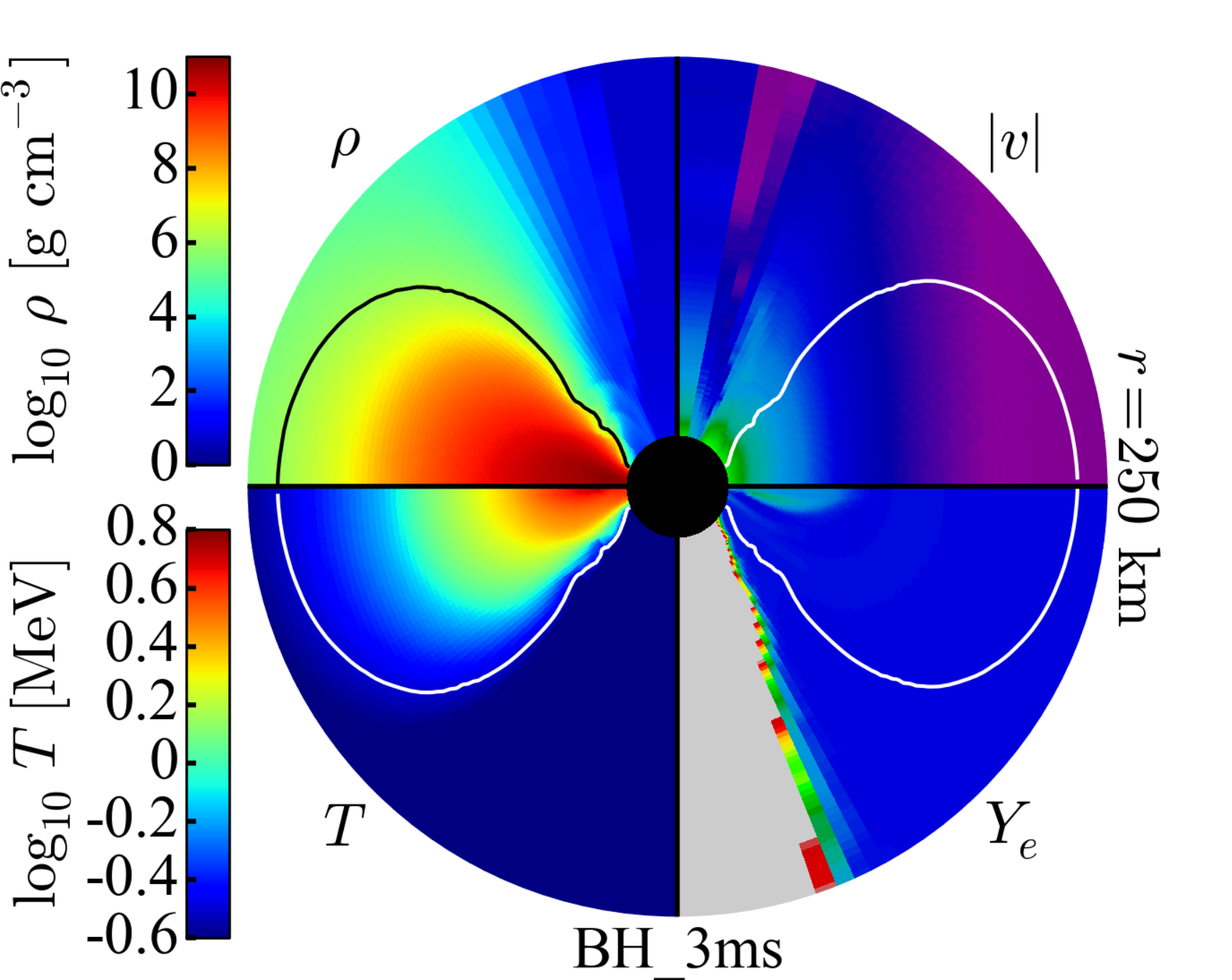}\hspace{-0.2in}
  \includegraphics[width=0.525\linewidth]{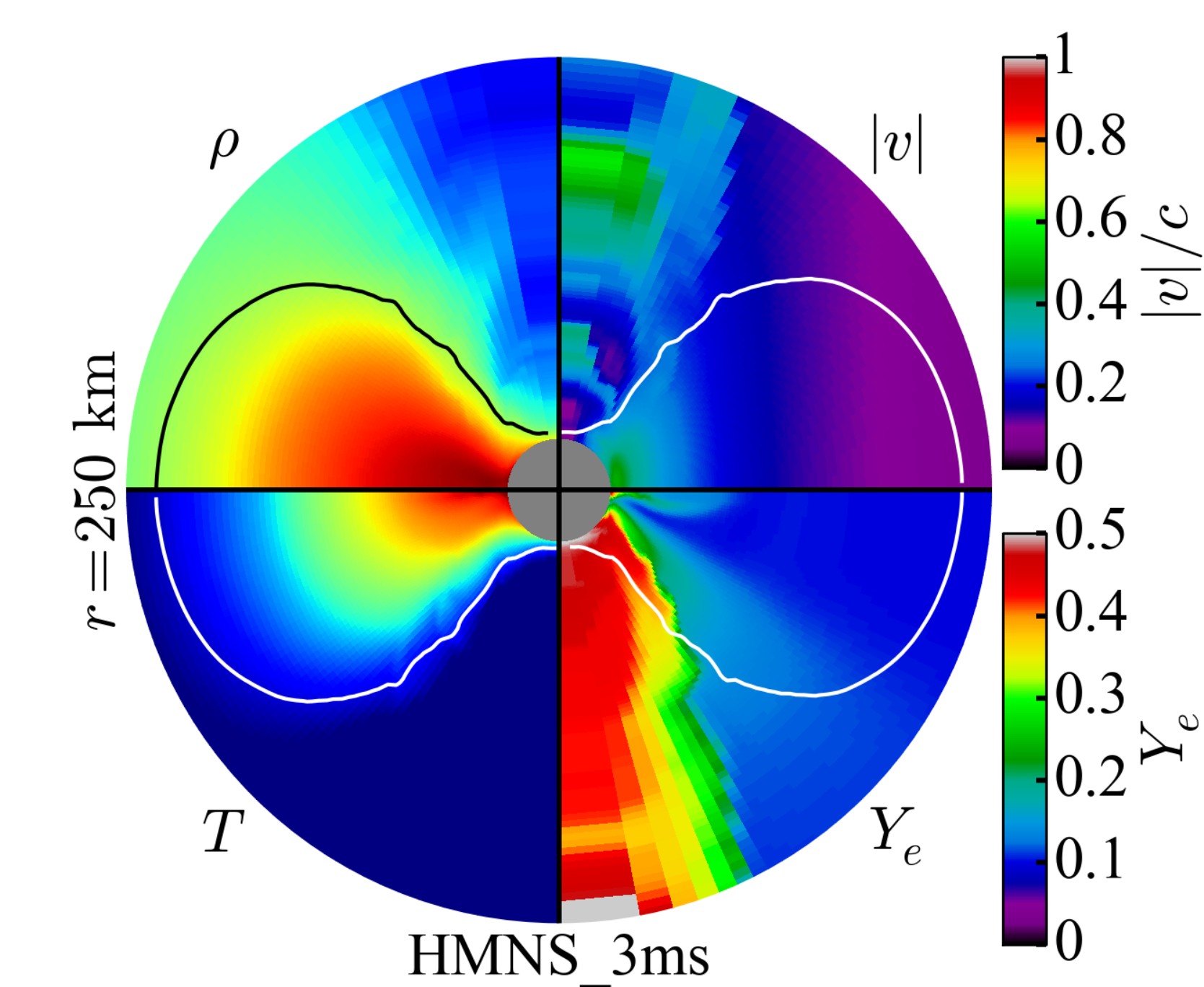}
  \caption{\textbf{Fluid Backgrounds} at $t=3\,\mathrm{ms}$ from the start of the dynamical simulations of MF14. \textit{Left}: central BH \textit{Right}: central HMNS. The outer radius on the plot is at $250\,\mathrm{km}$ and the inner radius is $30\,\mathrm{km}$. Each quadrant covers half of the simulation domain. On the top left of each plot is the density, which peaks at around $5\times10^{10}\,\mathrm{g\ cm}^{-3}$. On the top right is the magnitude of the velocity in units of $c$. On the bottom right is the electron fraction, where grey colors indicate electron fractions larger than 0.5 due to very low-density hydrogen that is present for numerical reasons. On the bottom left is the temperature, which peaks around 5 MeV. The black/white curve is the $\rho=10^6\,\mathrm{g\ cm}^{-3}$ contour, below which {\tt Sedonu} opacities and emissivities are set to zero (see Section \ref{sec:opacities}).\\}
  \label{fig:background}
\end{figure*}
\begin{deluxetable*}{rccccccccccc}
\tablecolumns{13}
\tablecaption{Background Global Properties}
\tablewidth{7in}
\tablehead{\colhead{Time} & \colhead{$M_\mathrm{disk}$}  & \colhead{$\left<T\right>$} & \colhead{$\left<Y_e\right>$} & \colhead{$\mathcal{H}_\mathrm{visc}$} & \colhead{$\mathcal{L}_\mathrm{core}\,(\mathrm{B\ s}^{-1})$} & \colhead{$\mathcal{C}_\nu-\mathcal{H}_\nu$} & \colhead{$\left<\frac{dY_e}{dt}\right>$} & \colhead{$M_\nu$}            & \multicolumn{2}{c}{$\mathcal{L}_\mathrm{SI}\,(\mathrm{B\ s}^{-1})$} & \colhead{$\left<E_{\nu,\mathrm{SI}}\right>$} \\ 
           \colhead{(ms)} & \colhead{$(M_\odot)$} & \colhead{(MeV)}            &                              & \colhead{$(\mathrm{B\ s}^{-1})$}      & \colhead{each $\nu_e,\bar{\nu}_e$}                          & \colhead{$(\mathrm{B\ s}^{-1})$}            & \colhead{$(\mathrm{s}^{-1})$}            & \colhead{$(M_\odot)$} & \colhead{$\nu_e$} & \colhead{$\bar{\nu}_e$}                           & \colhead{(MeV)}                                
}
\startdata
\sidehead{Central BH}
0     & 3.01(-2) & {2.95\phantom{(-2)}}   & 0.10{2} & {3.02(1)\phantom{-}}  & \nodata\phantom{(-)} & \phantom{-}{5.19(1)\phantom{-}} & {2.49(1)\phantom{-}} & {2.24(-6)} & 1.58(1)\phantom{-} & 4.01(1)\phantom{-}    & 17.3\phantom{8} \\
3     & 2.93(-2) & 2.{81\phantom{(-2)}}   & 0.12{1} & {2.84(1)\phantom{-}}  & \nodata\phantom{(-)} & \phantom{-}3.{29(1)\phantom{-}} & 2.47\phantom{(-1)} & 2.27(-7)           & 1.54(1)\phantom{-} & 1.92(1)\phantom{-}    & 15.1\phantom{8} \\
30    & 1.52(-2) & {1.93\phantom{(-2)}}   & 0.11{4} & {7.95\phantom{(-1)}}  & \nodata\phantom{(-)} & \phantom{-}{5.77\phantom{(-2)}} & 4.92(-1)           & 1.35(-8)           & 2.86\phantom{(-2)} & 3.19\phantom{(-2)}    & 10.8\phantom{8} \\
300   & 4.48(-3) & 6.{29(-1)}             & 0.19{0} & {3.63(-1)}            & \nodata\phantom{(-)} & \phantom{-}{1.93(-2)}           & 1.47(-1)           & 0.00\phantom{(-5)} & 4.93(-2)           & 1.44(-2)              & \phantom{1}5.08 \\
\sidehead{Central HMNS}
0     & 3.01(-2) & {2.95\phantom{(-2)}}   & 0.10{2} & {3.05(1)\phantom{-}}  & 1.73(1)              & \phantom{-}{3.58(1)\phantom{-}} & {3.33(1)\phantom{-}}& {2.42(-3)} & 1.60(1)\phantom{-}     & 3.90(1)\phantom{-}    & 15.4\phantom{3} \\
3     & 3.01(-2) & {2.93\phantom{(-2)}}   & 0.1{37} & {2.89(1)\phantom{-}}  & 1.73(1)              & \phantom{-}2.{77(1)\phantom{-}} & 6.51\phantom{(-1)}  & 2.22(-3)  & 2.67(1)\phantom{-}     & 2.64(1)\phantom{-}    & 16.0\phantom{3} \\
30    & 3.01(-2) & 2.{75\phantom{(-2)}}   & 0.1{77} & 2.{11(1)\phantom{-}}  & 1.02(1)              & \phantom{-}{1.87(1)\phantom{-}} & 3.14(-1)            & 6.46(-3)  & 2.59(1)\phantom{-}     & 1.84(1)\phantom{-}    & 16.6\phantom{3} \\
300   & 3.01(-2) & 1.{09\phantom{(-2)}}   & 0.26{4} & {5.85\phantom{(-1)}}  & 3.22\phantom{(1)}    & \phantom{-}{4.75\phantom{(-1)}} & 2.43(-1)            & 5.88(-3)  & 4.99\phantom{(-7)}     & 3.84\phantom{(-6)}    & 13.6\phantom{3} \\
3000  & 1.32(-2) & 4.6{6}(-2)             & 0.300              & 1.1{9(-1)}            & 1.02\phantom{(1)}    & {-}8.{41}(-4)           & 1.41(-3)            & 9.62(-3)  & 3.98(-7)               & 2.73(-6) & \phantom{1}2.43 \\
\enddata
\tablenotetext{}{\textbf{Notes:} Quantities extracted from the dynamical simulations conducted by MF14. The numbers in parentheses indicate the power of $10$ with which the data given must be scaled, e.g.,\ $6.95(-1)$ is $6.95 \times 10^{-1}$. $M_\mathrm{disk}$ is the mass remaining in the disk. $\left<T\right>$ and $\left<Y_e\right>$ are mass-weighted averages of the disk temperature and electron fraction, respectively. $\mathcal{H}_\mathrm{visc}$ is the integrated viscous heating rate. $\mathcal{L}_\mathrm{core}$ is the luminosity of \textit{each} neutrino species emitted from the core. Each of the previous quantities are taken as input to {\tt Sedonu}, and the following can be compared to the {\tt Sedonu} results. $\mathcal{C}_\nu-\mathcal{H}_\nu$ is the net rate of energy loss from the fluid by neutrinos and $\left<dY_e/dt\right>$ is the mass-weighted average of the rate of change of the electron fraction computed by neutrino leakage. $M_\nu$ is the mass in which neutrinos are a larger source of heat than viscosity is, i.e. $\mathcal{H}_\nu-\mathcal{C}_\nu>\mathcal{H}_\mathrm{visc}$. $\mathcal{L}_\mathrm{SI}$ is the luminosity of the disk self-irradiation assumed to be emitted from two rings above and below the equatorial plane. $\left<E_{\nu,\mathrm{SI}}\right>$ is the energy density-weighted average energy of these emitted neutrinos. $1\,\mathrm{B} = 10^{51}\,\mathrm{erg}$.}
\label{tab:flash}
\end{deluxetable*}
Our background fluid snapshots come from the axisymmetric 2D simulations from MF14 of remnant disks modeled after those left behind by a binary neutron star merger. The disks have a mass of $0.03M_\odot$ and circle a $3M_\odot$ BH or HMNS. Table~\ref{tab:flash} lists the times and global properties of the fluid snapshots we use. Figure~\ref{fig:background} shows the background density, temperature, electron fraction, and fluid speed at $3\,\mathrm{ms}$ after the start of the dynamical simulations. To avoid large inconsistencies in using a relativistic transport code with fluid velocities calculated with a Newtonian code, we cap the fluid speeds at a maximum Lorentz factor of 2.

Though the fluid moves around on an orbital timescale of about $3\,\mathrm{ms}$ (at $R=50\,\mathrm{km}$), the time required for the disk to significantly change its structure is set by the viscous timescale of about $1\,\mathrm{s}$ \citep{FM13b}. The diffusion time for the neutrinos can be approximated by $t_\mathrm{diff} \sim \frac{\tau l}{c}\approx10^{-3}\,\mathrm{s}$, where $\tau\sim3$ is a representative average optical depth, $l\sim 10^7\,\mathrm{cm}$ is an approximate characteristic size of the portion of the disk opaque to neutrinos, and $c$ is the speed of light. The neutrino diffusion time is much shorter than the viscous timescale. Hence, although the fluid orbits faster than neutrinos can escape, we can safely assume that the global disk structure is effectively static on the neutrino propagation timescale.

The $3M_\odot$ central object is well above the maximum mass of current observational and theoretical constraints for the maximum mass of a cold neutron star (e.g.,\ \citealt{LP00,SHF13}). The central object in these simulations is likely to become a black hole, but the temporary stability depends on rotation, thermal support against collapse (e.g. \citealt{Kaplan+14}), and magnetic field strength and configuration, though the latter requires magnetic fields on the order of $10^{18}\,\mathrm{G}$ for the effect to be relevant (e.g.,\ \citealt{Cardall+01}). As did MF14, we remain agnostic to how long the mass in the central object is supported against collapse and consider both instant black hole creation and indefinite HMNS stability to bracket the parameter space.


\subsection{Opacities and Emissivities}
\label{sec:opacities}
\begin{deluxetable*}{lccccccccc}
\tablecolumns{10}
\tablecaption{Included Physics}
\tablewidth{7in}
\tablehead{\colhead{Process}          & \colhead{Leakage} & \colhead{Full} & \colhead{Shen} & \colhead{LS220} & \colhead{NoWM} & \colhead{NoScat} & \colhead{NoPair} & \colhead{NoRel} & \colhead{Simple}}
\startdata
$\nu_e/\bar{\nu}_e$ Emis/Abs on $n,p$ & \checkmark        & \checkmark     & \checkmark     & \checkmark     & \checkmark      & \checkmark     & \checkmark       & \checkmark       & \checkmark       \\
Weak Magnetism Correction             &                   & \checkmark     & \checkmark     & \checkmark     &                 & \checkmark     & \checkmark       & \checkmark       &                  \\
Elastic Scatter on $e,n,p,\alpha$     &                   & \checkmark     & \checkmark     & \checkmark     & \checkmark      &                & \checkmark       & \checkmark       &                  \\
$\nu_x$ Emis/Abs/Pair Production      &                   & \checkmark     & \checkmark     & \checkmark     & \checkmark      & \checkmark     &                  & \checkmark       &                  \\
Special Relativity                    &                   & \checkmark     & \checkmark     & \checkmark     & \checkmark      & \checkmark     & \checkmark       &                  &                  \\
Elastic Scatter on Heavy Nuclei       &                   &                & \checkmark     & \checkmark     &                 &                &                  &                  &                  \\
Equation of State                     & Helmholtz         & Helmholtz      & Shen           & LS220          & Helmholtz       & Helmholtz      & Helmholtz        & Helmholtz        & Helmholtz        
\enddata
\tablenotetext{}{\textbf{Notes:} Included physics in each class of neutrino transport simulations. The first column represents the physics included in the leakage scheme in the original dynamical simulations by MF14. Full represents our most complete set of physics, while Simple is designed to replicate the physics used in the leakage calculations as closely as possible. The inclusion of the first four processes and the choice of equation of state go into generating the NuLib opacity tables. Special relativistic physics is turned on or off within {\tt Sedonu}.}
\label{tab:physics}
\end{deluxetable*}
Neutrino-fluid interactions depend on the fluid density $\rho$, temperature $T$, and electron fraction $Y_e$, as well as the neutrino energy $E_\nu$ and species $s_\nu$. The interactions are taken into account via neutrino opacities and emissivities calculated by NuLib\footnote{open source, available at {\tt www.nulib.org}} \citep{Oconnor14} and are output in tabular form for a range of values of $\{\rho,T,Y_e,E_\nu,s_\nu\}$\footnote{input files available at {\tt bitbucket.org/srichers/sedonu}}. We use a table spanning $\rho=10^{6-15}\,\mathrm{g\ cm}^{-3}$ with 82 logarithmically-spaced points, $T=0.05-200\,\mathrm{MeV}$ with 65 logarithmically-spaced points, $Y_e=0.035-0.55$ with 51 linearly-spaced points, and $E_\nu=0.5-200\,\mathrm{MeV}$ with 48 logarithmically-spaced bins. We demonstrate that this table has sufficient resolution in Appendix~\ref{app:resolution}. The opacities and emissivities below the table minima in $\{\rho,T,Y_e,E_\nu\}$ are very low or affect a very small amount of mass. They are hence dynamically unimportant, so we assume them to be zero. Heavy lepton neutrinos play a relatively minor role (see Sections~\ref{sec:BHresults} and \ref{sec:HMNSresults}) since they deposit energy only via neutral current reactions, so we simulate three effective neutrino species $s_\nu = \{\nu_e,\bar{\nu}_e,\nu_x\}$, where $\nu_x$ accounts for $\nu_\mu$, $\bar{\nu}_\mu$, $\nu_\tau$, and $\bar{\nu}_\tau$.

We experiment with excluding various processes and corrections as listed in Table~\ref{tab:physics} to determine how much each approximation affects the resulting neutrino radiation field and fluid source terms. The Full simulations embody our most complete set of physics which will serve as the standard for comparison. The Simple simulations account only for charged-current interactions on free nucleons in order to match as closely as possible the physics assumed in the leakage scheme used in the dynamical simulations of MF14. The Full simulations include weak magnetism and recoil corrections and the opacity for each neutrino species to scatter elastically on electrons, neutrons, protons, $\alpha$ particles, and heavy nuclei, unlike the Simple and Leakage simulations. Opacities for scattering on heavy nuclei are corrected for ion-ion correlations, the heavy-ion form factor, and electron polarization, though heavy nuclei are ignored when using the Helmholtz equation of state (EOS, see below). Each of these processes is implemented as described in \cite{BRT06}.

Absorption opacities are converted into emissivities via Kirchhoff's Law and account for final-state electron and positron blocking. There are additional approximate emissivities calculated for pair processes, namely electron-positron annihilation ($e^- + e^+ \longleftrightarrow \nu_i + \bar{\nu}_i$) and nucleon-nucleon Bremsstrahlung ($n_1 + n_2 \longleftrightarrow n_3 + n_4 + \nu_i + \bar{\nu}_i$ where $n_j$ represents any nucleon). While it is in principle incorrect to apply Kirchhoff's Law to these emissivities to get opacities since the opacities depend on both neutrinos and anti-neutrinos (which are not necessarily in equilibrium), doing so gives correct absorption rates in the optically-thick and trivial neutrino-free limits. In this way, NuLib yields annihilation rates that are correct to an order of magnitude, though it somewhat overestimates the effective opacity. In light of this, we include pair processes only for heavy lepton neutrinos during the transport step (post-processing annihilation calculations are described in Section~\ref{sec:annihilation_methods}).

The opacities are corrected for final-state blocking and in general depend on the chemical potentials of the particles involved in reactions \citep{BRT06}. The nucleon and lepton chemical potentials at a given density, temperature, and electron fraction depend on the details of the equation of state (EOS). To compare as directly as possible with the dynamical simulations, we use the Helmholtz EOS \citep{TS00} including neutrons, protons, and $\alpha$ particles in nuclear statistical equilibrium (NSE). We also use two popular hot nuclear equations of state: those from \cite{Shen+11} and the 220-MeV incompressibility version from \cite{LS91}\footnote{both available in tabular form at {\tt stellarcollapse.org}}.

The opacities calculated by NuLib are given in the rest frame and are transformed into the lab frame according to \cite{Mihalas+99},
\begin{equation}
\kappa_\mathrm{lab} = \kappa_\mathrm{rest} \frac{\gamma}{1-\frac{\mathbf{v} \cdot \mathbf{D}}{c}}\,\,,
\end{equation}
where $\gamma$ is the fluid Lorentz factor, $\mathbf{v}$ is the fluid velocity vector, and $\mathbf{D}$ is the neutrino direction unit vector. To test the effect of neglecting relativistic transformations, we simply set all fluid velocities to 0.

\subsection{Monte Carlo Neutrino Transport}
{\tt Sedonu}\footnote{open source, available at {\tt bitbucket.org/srichers/sedonu}} is a special-relativistic Monte Carlo (MC) neutrino transport code, evolved from the photon transport code {\tt Sedona} \citep{KTN06}. Sedonu simulates neutrinos passing through and interacting with a static background fluid snapshot. We import the fluid grid structure, density $\rho$, temperature $T$, and electron fraction $Y_e$ from grid-based simulation snapshots of any geometry in zero to three dimensions. In a zero-dimensional background (i.e., a one-zone model), symmetries are imposed in all three directions to create a completely isotropic and homogeneous background.

The neutrinos move around in a separate superimposed Cartesian space so that the geometry of the underlying fluid grid comes into play only when the neutrinos require information about the fluid at their current location. The neutrinos are discretized into packets, each of which is specified by the energy $E_\nu$ of each neutrino in the packet, the location $\mathbf{x}$ of the packet, the direction unit vector $\mathbf{D}$ of travel, and the total energy $E_p$ of the packet. When special relativity is included, the grid structure is assumed to be in the lab frame and the fluid properties are given in the rest frame.

In a real merger, general relativity would diminish the energy of outgoing neutrinos and increase the energy of incoming ones. To estimate the magnitude of this effect, we can assume a Schwarzschild metric outside of and sourced only by the $3M_\odot$ central object, which implies that
\begin{equation}
\frac{E_{\nu,1}}{E_{\nu,2}} = \left(\frac{1-2GM/r_2c^2}{1-2GM/r_1c^2}\right)^{1/2}\,\,,
\end{equation}
where $M$ is the mass of the central object and $E_{\nu,1}$ and $E_{\nu,2}$ are the energies of a given neutrino at radii $r_1$ and $r_2$, respectively. The strongest redshift effect we could expect is the difference between the neutrino energy at the inner boundary ($30\,\mathrm{km}$) and the outer edge of the disk ($\sim 250\,\mathrm{km}$), which comes out to be $E_{\nu,250\,\mathrm{km}}/E_{\nu,30\,\mathrm{km}} = 0.86$. The neutrino opacities scale approximately as $\kappa\sim E_\nu^2$ (e.g.,\ \citealt{BRT06}), resulting in about a 25\% effect on the opacities over this distance. However, most of the neutrino energy is emitted and absorbed over distances of tens of kilometers, so errors from excluding gravitational redshift will be necessarily smaller than this. In general relativity, neutrinos would also follow null geodesics rather than straight lab-frame lines, but we defer to other authors to determine the importance of this (see Section~\ref{sec:discussion} for a discussion).

By their very nature, MC simulations output data with random fluctuations that decrease with the number of MC elements. To keep the fluctuations of global quantities in Table~\ref{tab:results} below 0.1\% we propagate $2-4\times10^7$ particles in each simulation. 

\subsubsection{Emission}
\label{sec:emiss}
Neutrinos are either emitted from the fluid itself, or from an off-grid source like a central hypermassive neutron star (HMNS). If the source of neutrinos is the on-grid fluid, the number of neutrino packets spawned in each grid cell is proportional to the cell's total rest-frame emissivity. Within each cell and in the rest frame, the direction and position distribution of spawned neutrinos are isotropically randomly determined. The net comoving luminosity of a grid cell is determined by numerically integrating the emissivity over energy bins according to
\begin{equation}
\label{eq:Lcell}
  \mathcal{L}_\mathrm{cell}=4\pi V \sum\limits_\mathrm{species}\sum\limits_i \varepsilon_i \Delta E_{\nu,i}\,\,,
\end{equation}
where $V$ is the grid cell volume, $\varepsilon_i$ is the emissivity (units of erg s$^{-1}$cm$^{-3}$Hz$^{-1}$sr$^{-1}$), and $\Delta E_{\nu,i}$ is the width of the neutrino energy bin $i$. Each emitted neutrino packet in a given grid cell has the same packet energy, determined by 
\begin{equation}
  E_p = \frac{\mathcal{L}_\mathrm{cell} \Delta t}{{P}_\mathrm{emit}}\,\,,
\end{equation}
where $\Delta t$ is the length of the time step (arbitrary for steady-state calculations like these, as it always cancels out). ${P}_\mathrm{emit}$ is the number of neutrino packets to be emitted from the grid cell and is proportional to the cell's neutrino energy emission rate according to
\begin{equation}
  \label{eq:Nemit}
  {P}_\mathrm{emit} = {P}_\mathrm{total} \frac{\mathcal{L}_\mathrm{cell}}{\Sigma_\mathrm{cells} \mathcal{L}_\mathrm{cell}}\,\,,
\end{equation}
where ${P}_\mathrm{total}$ is the total number of neutrino packets used in the simulation ($2\times10^7$ in our case). Though each cell has a different velocity and hence the lab-frame emissivities are modified, Equation \ref{eq:Nemit} is used only to distribute computational resources through the disk and has no physical meaning.

Neutrino energies fall into one of 48 energy bins matching the NuLib tables we use. The energy bin of any given neutrino is chosen by randomly sampling the local neutrino energy-dependent emissivity. Neutrinos are emitted only from the center of a neutrino energy bin $E_{\nu,i}$. This is to better maintain consistency required by Kirchhoff's Law between the emissivity of a grid cell and the product of the opacity and the neutrino blackbody function, both of which are also evaluated at the bin center. As neutrinos move through fluid cells with different velocities, they are Lorentz transformed away from the bin centers, reducing the level of consistency, but we ignore this minor discrepancy.

Once an MC particle is created, it is Lorentz transformed into the lab frame. Additionally, it should be noted that the fluid in a moving cell is length contracted, such that its rest-frame volume is larger than its lab-frame volume by a factor of the Lorentz factor. Thus, including special relativity increases the rest mass (by at most 4\% in any snapshot), average neutrino energy, and the net luminosity of moving grid cells.

In this paper, the temperature and luminosity of electron neutrinos and electron anti-neutrinos emitted from a central HMNS are taken directly from MF14. There, $T_{\nu_e}=4\,\mathrm{MeV}$ and $T_{\bar{\nu}_e}=5\,\mathrm{MeV}$, and the luminosity of each species obeys
\begin{equation}
  \frac{\mathcal{L}_\mathrm{core}}{20\,\mathrm{B\ s}^{-1}} = 
  \begin{cases}
    \left(\frac{10\,\mathrm{ms}}{30\,\mathrm{ms}}\right)^{-1/2} & t \le 10\,\mathrm{ms}\,\,, \\
    \left(\frac{t}{30\,\mathrm{ms}}\right)^{-1/2} & t > 10\,\mathrm{ms}\,\,. 
  \end{cases}
\end{equation}
The values of the HMNS luminosity at each of our snapshots is also listed in Table~\ref{tab:flash}. When heavy lepton neutrinos are included, we choose their temperature and luminosity to be the same as those of electron anti-neutrinos. The neutrinos emitted from a central HMNS are given an isotropically random position on the inner boundary sphere and an isotropically random direction within the outward $2\pi$ steradians. Neutrino energies are sampled from a Fermi-Dirac blackbody distribution of the appropriate temperature and zero chemical potential. The HMNS emits $2\times10^7$ packets \textit{in addition} to the $2\times10^7$ emitted from fluid in the disk, and the energy of each HMNS-emitted neutrino packet is then chosen such that the total HMNS luminosity is equal to $\mathcal{L}_\mathrm{core}$.

\subsubsection{Propagation}
Once all neutrinos have been created, their motion is in the lab frame along straight lines until they escape or are destroyed, punctuated by moments of scattering. The distance moved along any straight-line segment is the minimum of the following computed distances along the packet's direction of travel: (1) the distance to the simulation outer boundary $d_\mathrm{boundary}$, (2) $d_\mathrm{cell}$, which is 0.4 times the length of the smallest dimension of the cell currently occupied, and (3) the interaction distance $d_\mathrm{interact}$. We use this method of calculating $d_\mathrm{cell}$ rather than computing a geometric distance to the cell boundary for efficiency, but we demonstrate that the factor of 0.4 is small enough to adequately substitute for a more precise geometric calculation in Appendix~\ref{app:resolution}. The interaction distance is randomly sampled such that its probability density obeys
\begin{equation}
  \mathcal{P}(d_\mathrm{interact}) = \rho\kappa e^{-\rho\kappa d_\mathrm{interact}}\,\,,
\end{equation}
where $\kappa=\kappa_s+\kappa_a$ is the sum of the scattering and absorption opacities calculated by NuLib. The particle is then moved a distance $d = \mathrm{min}\{d_\mathrm{boundary},d_\mathrm{interact},d_\mathrm{cell}\}$ and interacts with the fluid if the shortest distance was $d_\mathrm{interact}$.

When a packet interacts with the fluid, it is randomly decided whether it scatters or is absorbed, with probabilities of each proportional to their respective opacities. If the packet scatters on the fluid, it is transformed into the fluid rest frame, given an isotropically random new direction with the same energy, and transformed back into the lab frame. When a packet is absorbed, it is simply destroyed.

Irrespective of which of the three above distances was smallest, the neutrino packet deposits thermal energy continuously into the grid cells it passes through according to $E_\mathrm{dep} = E_p\rho\kappa_a d$ and lepton number according to ${N}_\mathrm{dep} = lE_\mathrm{dep}/E_\nu${, evaluated in the comoving frame}, where $l=1$ for $\nu_e$, $l=-1$ for $\bar{\nu}_e$, and $l=0$ for $\nu_x$. Though these quantities may be more or less than the actual packet energy and lepton number, depending on whether the packet moves more than or less than one optical depth, energy and lepton number conservation are well approximated when averaging over many particles. See Appendix~\ref{app:blackbody} for a test of this algorithm.

The {lab-frame} rate of change of {comoving-frame} internal energy density $\epsilon$ and electron fraction $Y_e$ in a given grid cell is then
\begin{equation}
\begin{aligned}
R_\epsilon & {\equiv} \frac{1}{\epsilon}\frac{d\epsilon}{dt} = \frac{1}{\epsilon V \Delta t}\left({-E_\mathrm{emit}} + \sum\limits_\mathrm{steps}E_\mathrm{dep}\right) + {\frac{\rho q_\mathrm{v}}{\gamma\epsilon}}\,\,,\\
R_{Y_e} &{\equiv} \frac{dY_e}{dt} = \frac{m_n}{\rho V \Delta t}\left({-N_\mathrm{emit}} + \sum\limits_\mathrm{steps}{N}_\mathrm{dep}\right)\,\,.
\end{aligned}
\end{equation}
$V$ is the grid cell's volume in the comoving frame, $\Delta t$ is the (arbitrary) emission time interval {in the lab frame}, $m_n$ is the mass of a neutron, and $E_\mathrm{emit}$ and ${N}_\mathrm{emit}$ are the sum of the emitted {comoving-frame} neutrino energy and lepton number, respectively, from all neutrino species. The sum is over all steps (propagation segments between emission, scattering, absorption, or escape) for all neutrino packets in the cell. {The specific heating rate due to viscosity in the comoving frame $q_\mathrm{v}$ is is taken from the simulations of MF14 and the fluid Lorentz factor $\gamma$ transforms the time derivative in the viscous heating rate to the lab frame.}

Each grid cell has a distribution function consisting of 6144 energy/direction bins composed of $(8\,\,\mathrm{latitudinal\,\,bins})\times(16\,\,\mathrm{longitudinal\,\,bins})\times(48\,\,\mathrm{energy\,\,bins})$. The latitudinal bins have constant size in $\mathrm{cos}(\theta)$, where $\theta$ is the angle from the pole,  so each bin covers the same solid angle. As each neutrino propagates it contributes an energy $E_\mathrm{rad}=E_p d / c \Delta t$ to its corresponding energy and angular distribution function bin $\{\phi_i,\theta_j,\nu_k\}$ in its current grid cell. The energy bins match those of the NuLib table. The total neutrino energy density $\epsilon_\nu$ and average energy $\left<E_p\right>$ in each cell are then
\begin{equation}
  \epsilon_\nu = \frac{1}{V}\sum\limits_{\phi_i}\sum\limits_{\theta_j}\sum\limits_{\nu_k}\sum\limits_\mathrm{steps} E_\mathrm{rad}
\end{equation}
\begin{equation}
  \left<E_p\right> = \frac{1}{\epsilon_\nu V} \sum\limits_{\phi_i}\sum\limits_{\theta_j}\sum\limits_{\nu_k} h \nu_k\sum\limits_\mathrm{steps} E_\mathrm{rad}\,\,.
\end{equation}

\subsubsection{Annihilation}
\label{sec:annihilation_methods}
In a separate post-processing step that does not feed back into the neutrino distribution, we calculate an annihilation rate ($\mathrm{erg\ s}^{-1}\mathrm{cm}^{-3}$) in each cell following \cite{Ruffert+97}. We integrate an annihilation kernel over the distribution according to
\begin{equation}
\begin{split}
\label{eq:annihil}
Q^+_\mathrm{ann} = \sum\limits_{E_{\nu,i}}& \sum\limits_{\bar{E}_{\nu,k}} \sum\limits_{\Omega_j} \sum\limits_{\bar{\Omega}_l} \frac{E_{\nu,i}+\bar{E}_{\nu,k}-2m_e c^2}{E_{\nu,i}\bar{
E}_{\nu,k}} \\ & \times \epsilon_{\nu,ij}\bar{\epsilon}_{\nu,kl} R_{jl}(E_{\nu,i},\bar{E}_{\nu,k},\cos\theta_{jl})\,\,,
\end{split}
\end{equation}
Here, $\epsilon_{\nu,ij}$ and $\bar{\epsilon}_{\nu,ij}$ are the neutrino and anti-neutrino contributions, respectively, to the cell's neutrino energy density ($\mathrm{erg\ cm}^{-3}$) from the neutrino energy bin center $E_{\nu,i}$ and direction bin $j$. The angle between the centers of direction bins $j$ and $l$ is $\theta_{jl}$. This must then be summed over each of the three neutrino-anti-neutrino pairs to get the total annihilation rate. Since we group all four heavy anti/neutrino species together in our simulations, we set $\epsilon_{\nu_\mu} = \bar{\epsilon}_{\nu_\mu} = \epsilon_{\nu_\tau} = \bar{\epsilon}_{\nu_\tau} = \epsilon_{\nu_x}/4$ before performing the annihilation calculations. See Appendix~\ref{app:annihilation} for details on the annihilation kernel $R_{jl}(\nu_i,\bar{\nu}_k,\cos \theta_{kl})$. The derivation of annihilation rates assumes that the sum of the neutrino energies is much larger than the sum of the rest masses of the produced electron and positron. Hence, we subtract the mass energy of the electron-positron pair from the annihilation rate to under-emphasize energy contributed near the minimum-energy limit. Additionally, this causes the annihilation rate to represent only the deposited thermal energy without counting mass energy. To check how large of an effect this has, we calculate the integrated annihilation rate at the $3\,\mathrm{ms}$ snapshot for both the BH and HMNS cases within $45^\circ$ of the axis of symmetry without subtracting the electron rest mass. This caused the energy deposition rate from neutrino annihilation to increase by only $2.5\%$.

\subsubsection{Equilibrium}
After all particles have propagated through the fluid and have left a tally of how much energy and lepton number they deposited in each cell, we can determine what combination of $\{T,Y_e\}$ causes the fluid in each cell to emit the same amount of energy and as many leptons as it absorbed. The equilibrium $T$ and/or $Y_e$ is converged upon using Brent's method \citep{Brent73}, which queries the NuLib tables for emission rates at successive guesses of $\{T,Y_e\}$ until the integrated NuLib emissivities (both energy and lepton number) match the absorption rates calculated during the MC transport. The equilibrium values are physically sensible quantities only where the timescales for such an equilibrium to be reached are short compared with the dynamical timescale. The process of transporting neutrinos and solving for equilibrium can be done iteratively, allowing temperature and electron fraction changes to affect the neutrino sources, until a truly time-independent equilibrium is reached, as is done in the irradiation tests in Appendix~\ref{app:irradiation}. The true time-independent solution of the NS-NS post-merger disk problem we study here is trivially a zero temperature disk, but to evaluate how strongly the fluid and neutrino radiation fields are out of equilibrium we stop after a single iteration to arrive at a local rather than global equilibrium.

\subsection{Neutrino Leakage}

For completeness, we review the neutrino leakage scheme used by \cite{MF14}. Throughout the following, only electron neutrino and anti-neutrinos are included.

The central HMNS emits neutrinos with the same temperature and luminosity as described in Section~\ref{sec:emiss}, and the neutrino flux due to the HMNS at any given location is attenuated by the (grey) optical depth integrated radially from the HMNS.

\cite{MF14} determine the rate of energy loss at any location in the torus by interpolating between the optically thin free-streaming limit and the optically thick diffusion limit, given by the effective luminosity\footnote{Note the typographical error in \cite{MF14} that reverses the order of the timescales.}
\begin{equation}
\label{eq:Leff}
\mathcal{L}_\mathrm{cell}^\mathrm{eff} = \sum\limits_i \frac{1}{1+t_\mathrm{diff}/t_\mathrm{loss}} 4\pi V\int\limits_0^\infty\varepsilon_i dE_{\nu,i}\,\,.
\end{equation}
Here, $\varepsilon_i$ is again the the neutrino emissivity for species $i$, $t_\mathrm{loss} = \epsilon V/\mathcal{L}_\mathrm{cell}^\mathrm{eff}$ is the characteristic time for the fluid to lose its internal energy via neutrino emission, and $t_\mathrm{diff} = (\kappa d)\times(d/c)$ is the characteristic time for neutrinos diffusing over a characteristic escape distance. The first term in the expression for $t_\mathrm{diff}$ represents a typical optical depth through which neutrinos would need to diffuse to escape, where $\kappa$ is the energy-averaged neutrino absorption coefficient due to charged-current reactions at the given location (see \citealt{FM13b} for details). The second term is the time required for an unimpeded neutrino to cross the same distance. The escape distance is taken to be $d = \min\{r,H_\perp,H_\parallel\}$, where $H_\perp$ and $H_\parallel$ are the vertical and horizontal scaleheights, respectively.

The disk self-irradiation scheme assumes that neutrinos are emitted from two rings, one above and one below the midplane. The location of the ring is at the effective luminosity-weighted average radius and polar angle in each hemisphere, and the luminosity of each ring is half of the volume-integrated effective neutrino luminosity. Neutrinos are emitted from both rings with a zero chemical potential blackbody spectrum, the temperature of which is the effective luminosity-weighted average fluid temperature. The fluxes of each neutrino species at a given location are independently attenuated by an optical depth $\tau_\mathrm{irr} = \max(\kappa_i d,\kappa_{i,\mathrm{ring}}d_\mathrm{ring})$, where $d_\mathrm{ring}$ is $d$ evaluated at the ring's location and $\kappa_{i,\mathrm{ring}}$ is the absorption coefficient for species $i$ at the ring's location. For details, see \cite{FM13b} and \cite{MF14}. 

Comparing Equation~\ref{eq:Lcell} (used in computing $\mathcal{L}_\mathrm{emit}$ in Tables \ref{tab:results} and \ref{tab:res_study}) with Equation~\ref{eq:Leff} (used in computing the self-irradiation luminosity $\mathcal{L}_\mathrm{SI}$ in Table~\ref{tab:flash}), it is clear that $\mathcal{L}_\mathrm{SI} < \mathcal{L}_\mathrm{emit}$. The quantities both represent neutrino radiation coming from the disk itself in some capacity, but $\mathcal{L}_\mathrm{SI}$ is diminished by optical depth effects, preventing direct comparison with $\mathcal{L}_\mathrm{emit}$.


\section{Results (Central BH)}
\label{sec:BHresults}

In this section, we present the results for the simulations where it is assumed that a black hole forms immediately upon merger and is present in every snapshot. Table~\ref{tab:results} lists the times at which we simulate MC neutrino transport and the corresponding global fluid and neutrino radiation properties. In what follows, we will probe the neutrino radiation field and its interaction with the fluid, and try to explain differences between the Full MC simulations, the Simple MC simulations, and the leakage data of MF14. We do most of our comparisons with snapshots from a time of $3\,\mathrm{ms}$ after the start of the dynamical simulation as differences between the methods are most striking then, though the composition and amount of ejecta are determined by the long-term evolution.

Mentioning one caveat is in order. The fluid snapshots were evolved in the dynamical simulations using the leakage neutrino treatment of MF14, and we simply perform our MC transport on snapshots of the evolved fluid. Though the Simple MC transport employs the same set of neutrino interactions, the geometry and spectral shape of the neutrino radiation field are different from what is assumed in the leakage scheme. Fluid may be near thermal and weak equilibrium with neutrinos in the leakage scheme, but this is not necessarily true after switching to MC transport. This potential discrepancy can easily cause artificially large or small heating and leptonization rates. If the fluid were evolved with an MC treatment instead, it would likely be much closer to equilibrium with the MC neutrinos, and the rates might not be as high. Because of this, our comparisons between leakage and MC results indicate the qualitative effects, such as faster cooling and leptonization rates, but the magnitudes of the differences are likely not reliable. The effects of MC transport on the end results of dynamical simulations are thus difficult to determine. The dynamical simulations also begin with a disk of uniform electron fraction $Y_e=0.1$, which is not initially in equilibrium with either leakage or MC neutrinos. Addressing this out-of-equilibrium issue requires the dynamical simulations to begin before the merger and to be coupled to MC neutrino transport, which we leave to future work.

\subsection{Neutrino Radiation Field (Central BH)}
\label{sec:BHnufield}
\begin{figure}
  \includegraphics[width=\linewidth]{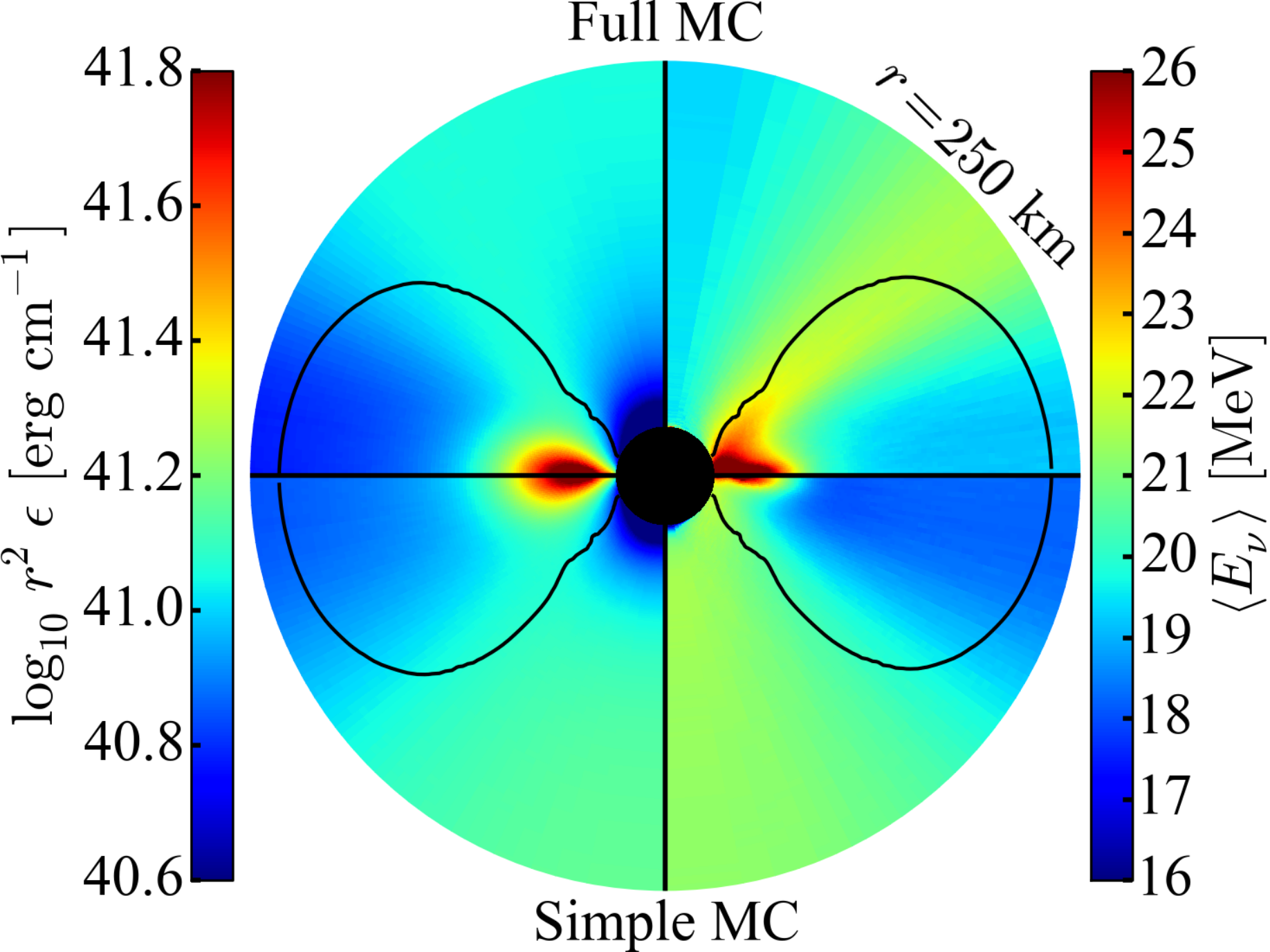}
  \caption{\textbf{Neutrino energy density and average energy (Central BH)} at $t=3\,\mathrm{ms}$. \textit{Left hemisphere}: Neutrino energy density, summed over all species and multiplied by $r^2$ to remove effects of distance from the center. \textit{Right hemisphere}: energy density-weighted average neutrino energy, averaged over all species. \textit{Bottom hemisphere}: Simple MC results. \textit{Top hemisphere}: Full MC results. The black curve is the $\rho=10^6\,\mathrm{g\ cm}^{-3}$ contour, below which {\tt Sedonu} opacities and emissivities are set to zero. The outer radius on the plot is at $250\,\mathrm{km}$ and the inner radius is $30\,\mathrm{km}$. The neutrino radiation field is very asymmetric and sensitive to the included physics. The disk casts a shadow as higher-energy neutrinos are preferentially absorbed. Much more asymmetry is present when the Full suite of physics is included.}
  \label{fig:BHnufield}
\end{figure}

In Figure~\ref{fig:BHnufield} we show the spatial distribution of the neutrino radiation field at $t=3\,\mathrm{ms}$ for both Simple and Full neutrino physics. Though the plot includes all neutrino species, the radiation is dominated by electron anti-neutrinos. Most of the neutrino energy comes from the inner regions of the disk close to the central object, and the dense disk casts a shadow that reduces the neutrino luminosity and energy density at large radii near the equator. The inner boundary also blocks neutrinos from moving to the other side of the disk and creates a polar shadow. The elastic electron scattering in the Full simulations results in higher opacities, which in turn deepens the equatorial and polar shadows. The Lorentz transformation of neutrinos in fluid moving at around $0.6c$ near the inner boundary increases the average energy of neutrinos emitted from the hot inner disk by $\sim 30\%$. Additionally, the neutrinos are beamed in the azimuthal direction, causing fewer of the higher-energy neutrinos coming from the inner parts of the disk to be present in the polar regions and more to be present along the $45^\circ$ radial. With either set of physics, this is different from the neutrino radiation field described by \cite{FM13b}, which becomes spherical at large distances.

\begin{figure}
  \includegraphics[width=\linewidth]{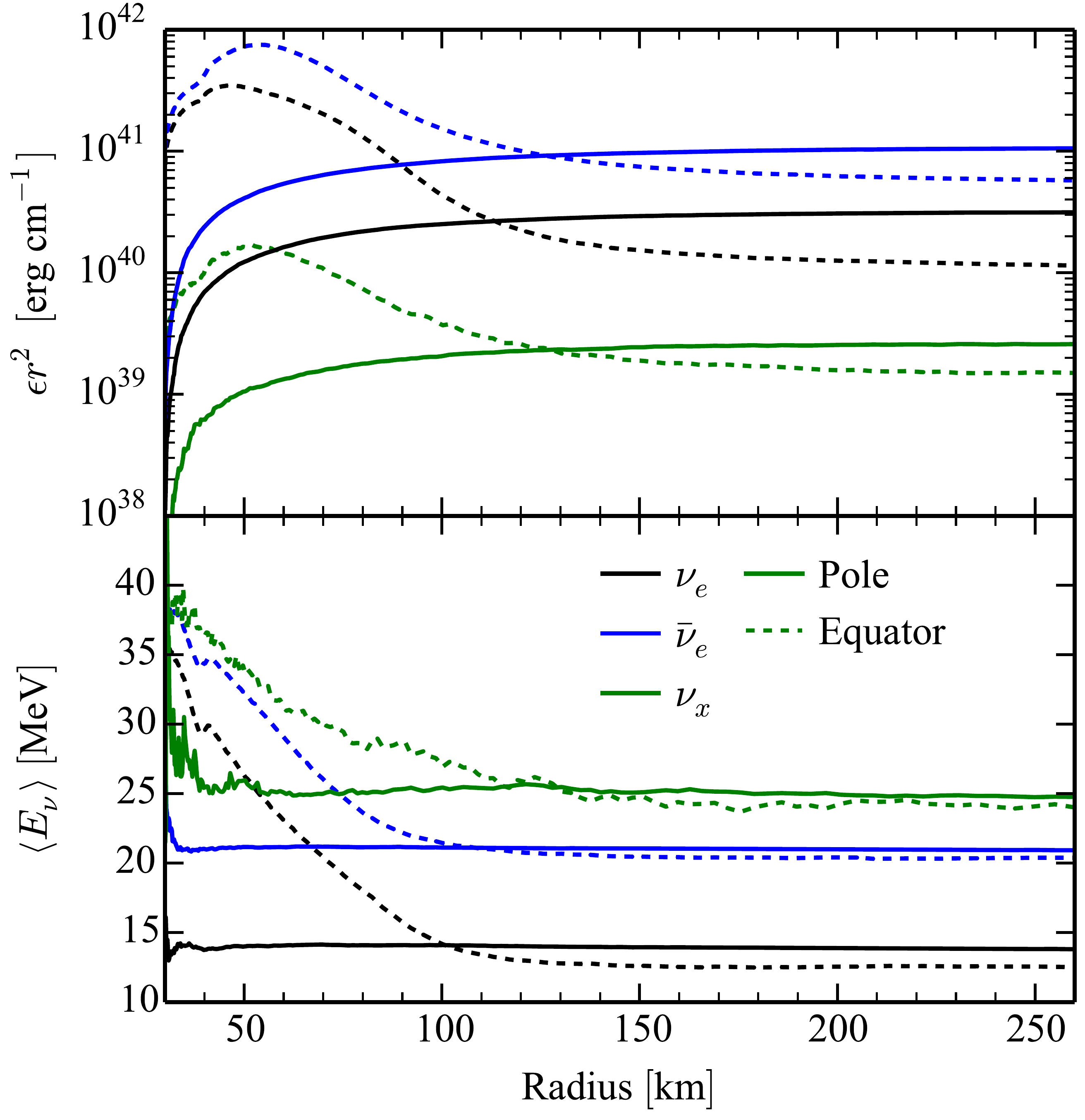}
  \caption{\textbf{Neutrino Radiation Profile (Central BH)} from the Full MC simulation at $t=3\,\mathrm{ms}$ for all three simulated neutrino species. The neutrino radiation field is asymmetric and dominated by electron anti-neutrinos. \textit{Top}: neutrino energy density along the pole (solid lines) and the equator (dashed lines), multiplied by $r^2$ to remove effects of distance from the center. \textit{Bottom}: energy density-weighted average neutrino energy along radial lines. The green $\nu_x$ curves represent the sum of all four heavy lepton neutrino species.}
  \label{fig:BHprofile}
\end{figure}
\begin{figure}
  \includegraphics[width=\linewidth]{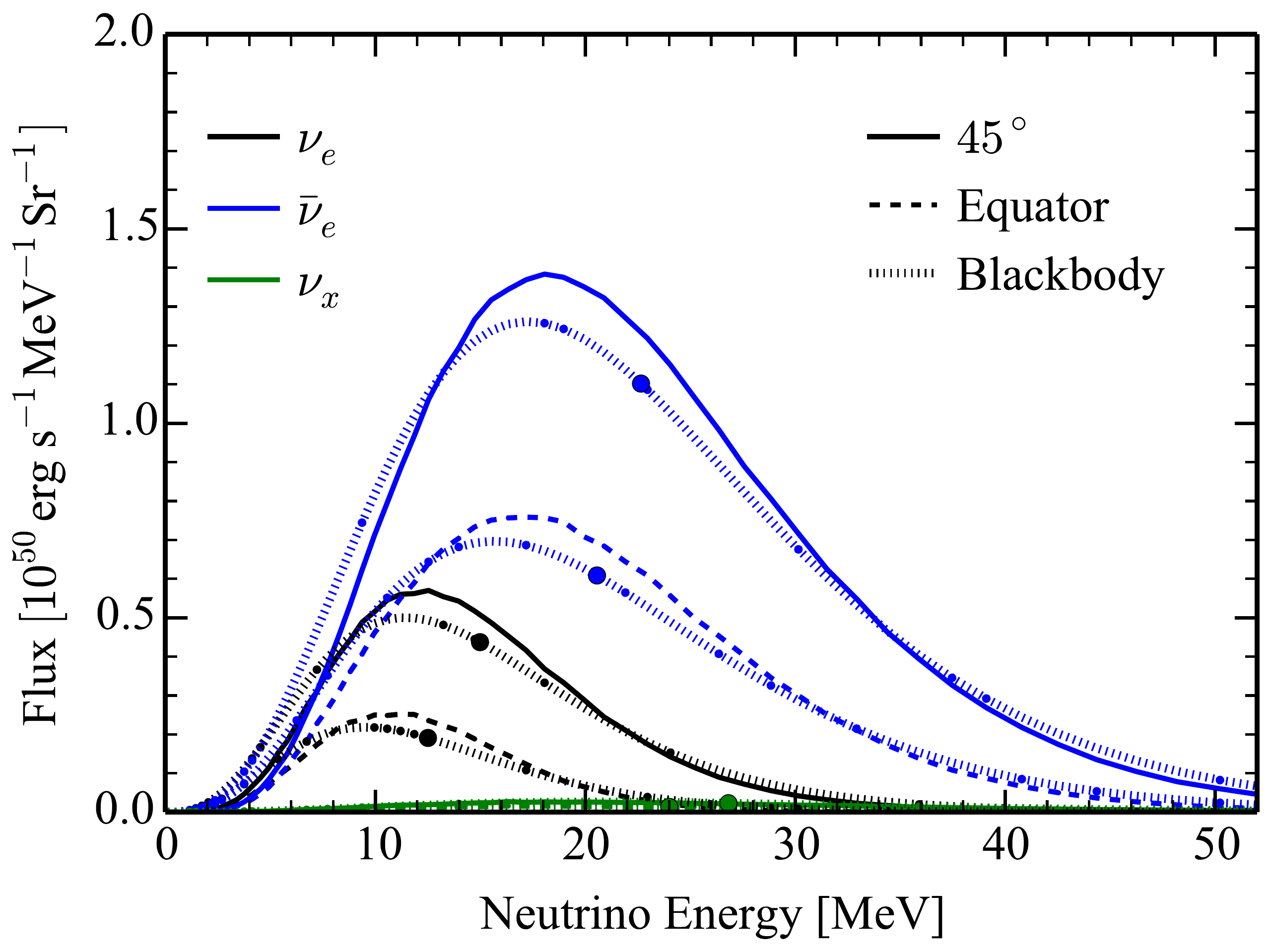}
  \caption{\textbf{Neutrino Spectra (Central BH)} from the Full MC simulation at $t=3\,\mathrm{ms}$. Dashed lines are spectra of each neutrino species escaping from within $10^\circ$ of the equator, while the solid lines are those from within $10^\circ$ of the $45^\circ$ cones, normalized by the solid angle covered by the respective regions. Overplotted for both directions (distinguished by proximity to the data curves) are dotted zero-chemical potential blackbody curves with the same total flux and average energy as the measured spectrum. The large dot on the blackbody curve indicates this average energy. For smoothness, the spectra are taken from the 2xEnergy run in Appendix~\ref{app:resolution}. The escaping neutrino radiation is somewhat nonthermal and asymmetric.}
  \label{fig:BHspectra}
\end{figure}

Profiles of the neutrino radiation field split into the different neutrino species are shown in Figure~\ref{fig:BHprofile}. There is some noise at small radii due to a relatively small number of simulated neutrino packets present there. Along the equator, normalized neutrino energy density and average energy are much higher close to the black hole than farther out in the disk, since the higher-energy neutrinos are preferentially absorbed by the disk. Moving radially along the pole, the energy density increases quickly as more of the disk becomes visible. However, the average energy is always close to the average energy escaping from the disk listed in Table~\ref{tab:results}. In all directions, the neutrino species follow the same hierarchy, such that electron anti-neutrinos everywhere contribute most to the energy density and heavy lepton neutrinos everywhere have the highest average energy.

$\mathcal{L}_\mathrm{emit}$, $\mathcal{L}_\mathrm{escape}$, $\langle E_{\nu,\mathrm{emit}}\rangle$, and $\langle E_{\nu,\mathrm{escape}}\rangle$ in Table~\ref{tab:results} describe the global lab-frame properties of the neutrinos that are emitted and of those that escape through the outer boundary. In the snapshot at $3\,\mathrm{ms}$, there is more energy emitted as electron neutrinos than as electron anti-neutrinos, but this is before the initial data in the simulations of MF14 has had any time to come into a quasi-equilibrium with the viscous heating and neutrino interactions. At all other times, electron anti-neutrino emission is stronger, reflecting a tendency of the fluid to relax to a higher electron fraction. The subsequent emission rates of all species decrease with time as the disk loses mass and cools. Heavy lepton neutrinos interact much more weakly with the fluid than do electron neutrinos and anti-neutrinos both in absorption and emission since they participate only in neutral-current reactions. This, combined with the low optical depths that prevent a blackbody distribution from building up, causes the heavy lepton neutrinos to always be subordinate to electron neutrinos and anti-neutrinos in energy density and in fluid heating and cooling.

If we assume the neutrinos form a zero-chemical potential blackbody distribution as is done in MF14, we can relate temperature to average energy through
\begin{equation}
  \left<E_\nu\right> = \frac{\int_0^\infty E_\nu B_{E_\nu}(0,T)\,dE_\nu}{\int_0^\infty B_{E_\nu}(0,T)\,dE_\nu} = 4.11 k_b T\,\,,
  \label{eq:avgE}
\end{equation}
where $E_\nu$ is the neutrino energy, $k_b$ is the Boltzmann constant, and $B_{E_\nu}(\mu,T)$ is the neutrino blackbody function at temperature $T$ and chemical potential $\mu$ (Equation~\ref{eq:blackbody}). While the density-weighted average fluid temperature at $t=3\,\mathrm{ms}$ is around $3\,\mathrm{MeV}$, the average emitted neutrino energy is between 26 and 29 MeV for all species. Thus, most of the neutrinos are created in the hottest regions of the disk very close to the black hole. The opacity to neutrinos scales approximately like $E_\nu^2$ (e.g.,\ \citealt{BRT06}) causing more higher-energy neutrinos to be absorbed and the average energy of escaping neutrinos to be much smaller than that of the emitted neutrinos. The heavy lepton neutrinos have the coolest emission temperature but the hottest escape temperature, since electron neutrinos and anti-neutrinos have much larger opacities and higher-energy neutrinos are preferentially absorbed.

The disk's self-irradiation in the leakage scheme is calculated as in MF14, and the global properties are summarized in Table~\ref{tab:flash}. The temperature of the radiation in the leakage scheme is determined by an emissivity-weighted average and is the same for both electron neutrinos and anti-neutrinos. The average energy in Table~\ref{tab:flash} is computed from a zero-chemical potential blackbody of this temperature using Equation \ref{eq:avgE}. Both the average energy and volume-integrated emission luminosities from the leakage data are much lower than those computed by {\tt Sedonu}, since the ``emission'' in the leakage scheme approximately accounts for immediate re-absorption in the same grid cell. The leakage and MC emission quantities then do not represent the same physics, but the difference further illustrates that the higher-energy neutrinos are re-absorbed locally while the lower-energy neutrinos are able to escape.

Using instead the Simple set of physics described in Table~\ref{tab:physics} does little to bring the leakage (Table~\ref{tab:flash}) and {\tt Sedonu} (Table~\ref{tab:results}) results closer together. However, it does result in significant deviations from the Full set of physics. In the following, we exclude individual pieces of physics to pinpoint the origin of the differences in the $t=3\,\mathrm{ms}$ snapshot. The exclusion of scattering predictably does nothing to the properties of the created neutrinos in the disk, but by decreasing the optical depth, allows more of the higher-energy neutrinos to escape. Ignoring special relativity results in a decrease of the emission luminosity and average energy of all species. Neglecting to correct for weak magnetism and recoil effects causes the electron anti-neutrino emission rate to increase by about 20\%, but since most of the opacity is also increased by a similar amount, the escaping luminosity increases by only about 3\%. The other species are minimally affected.

The low electron fraction throughout the disk causes electron neutrinos to be very likely to absorb onto neutrons, allowing few of them to escape. Their escape luminosity in Table~\ref{tab:results} is around an order of magnitude lower than their emitted luminosity, indicating an average optical depth of $\tau\sim3$ for electron neutrinos, $\tau\sim1$ for electron anti-neutrinos, and $\tau\ll 1$ for heavy lepton species. Figure~\ref{fig:BHspectra} shows neutrino spectra that further demonstrate the asymmetry of the escaping neutrino radiation. The leakage scheme does not yield data with which we can directly compare our escape spectra. However, we show that the spectra appear qualitatively similar in shape to zero-chemical potential blackbody spectra with the same average energy and total flux (dotted lines in Figure~\ref{fig:BHspectra}), though the MC spectra are somewhat pinched with peak energies higher by a few MeV.

\subsection{Neutrino-Fluid Interaction (Central BH)}
\label{sec:BHinteraction}
\begin{figure*}
  \includegraphics[height=0.25\linewidth]{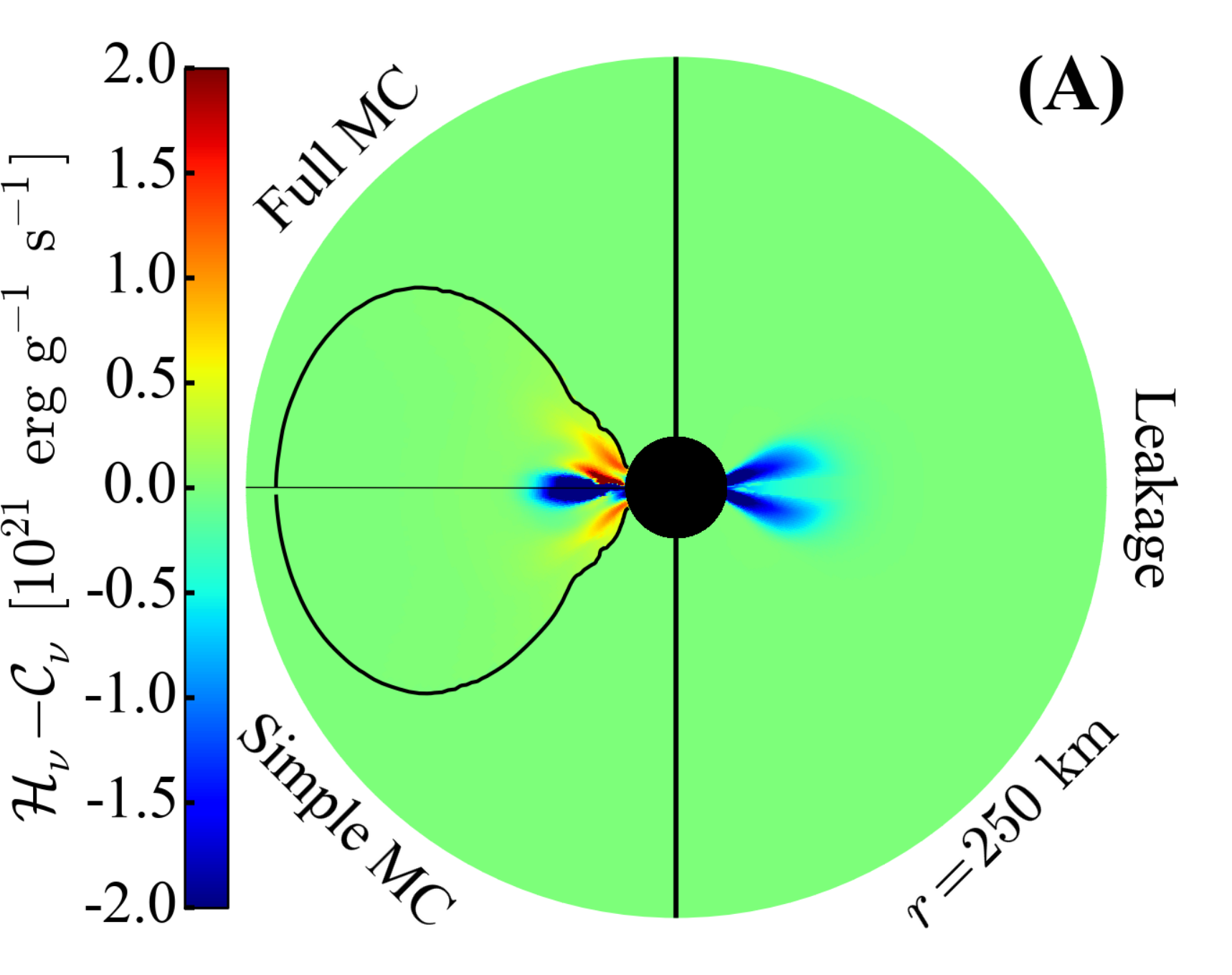}
  \includegraphics[height=0.25\linewidth]{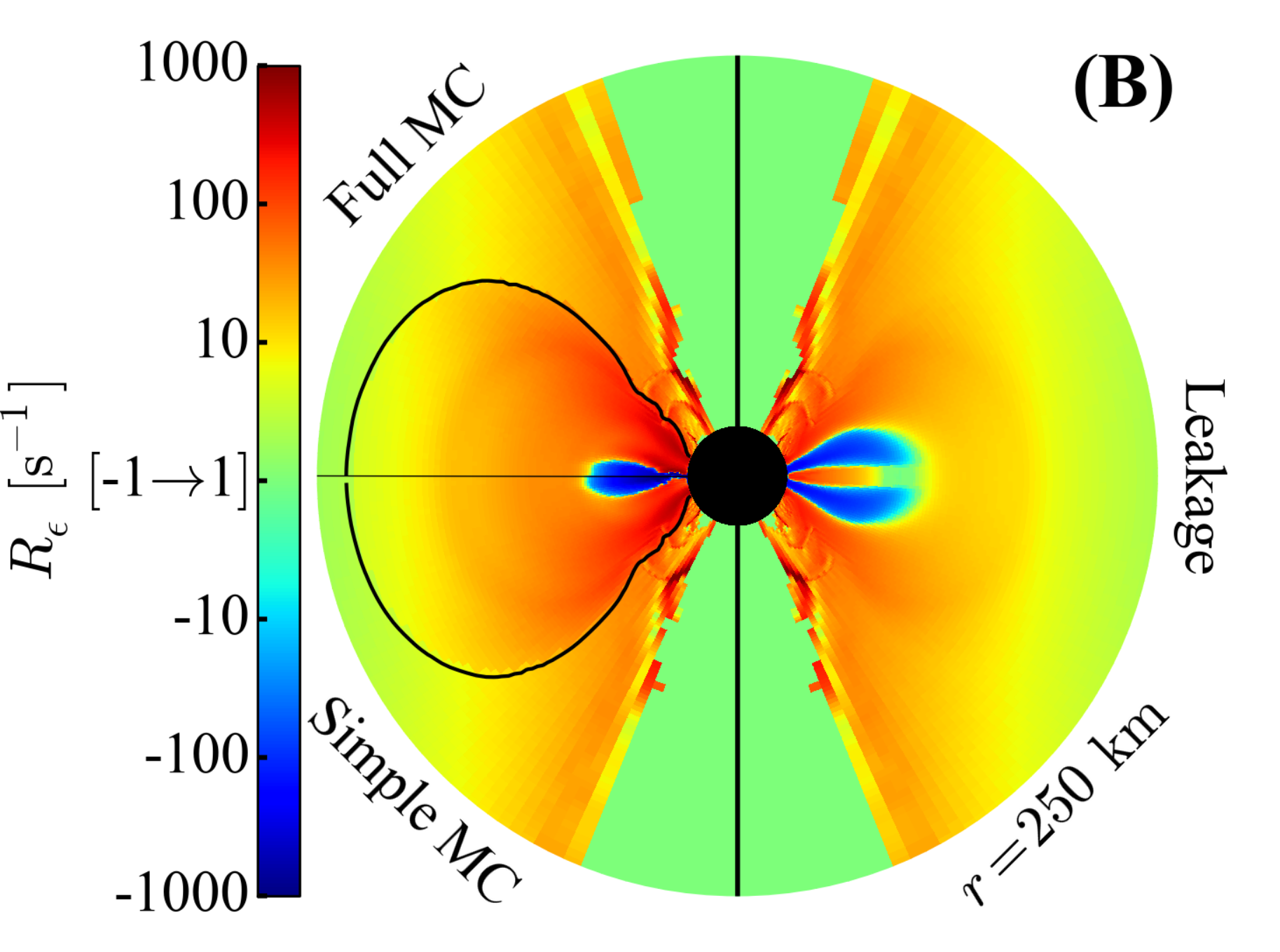}
  \includegraphics[height=0.25\linewidth]{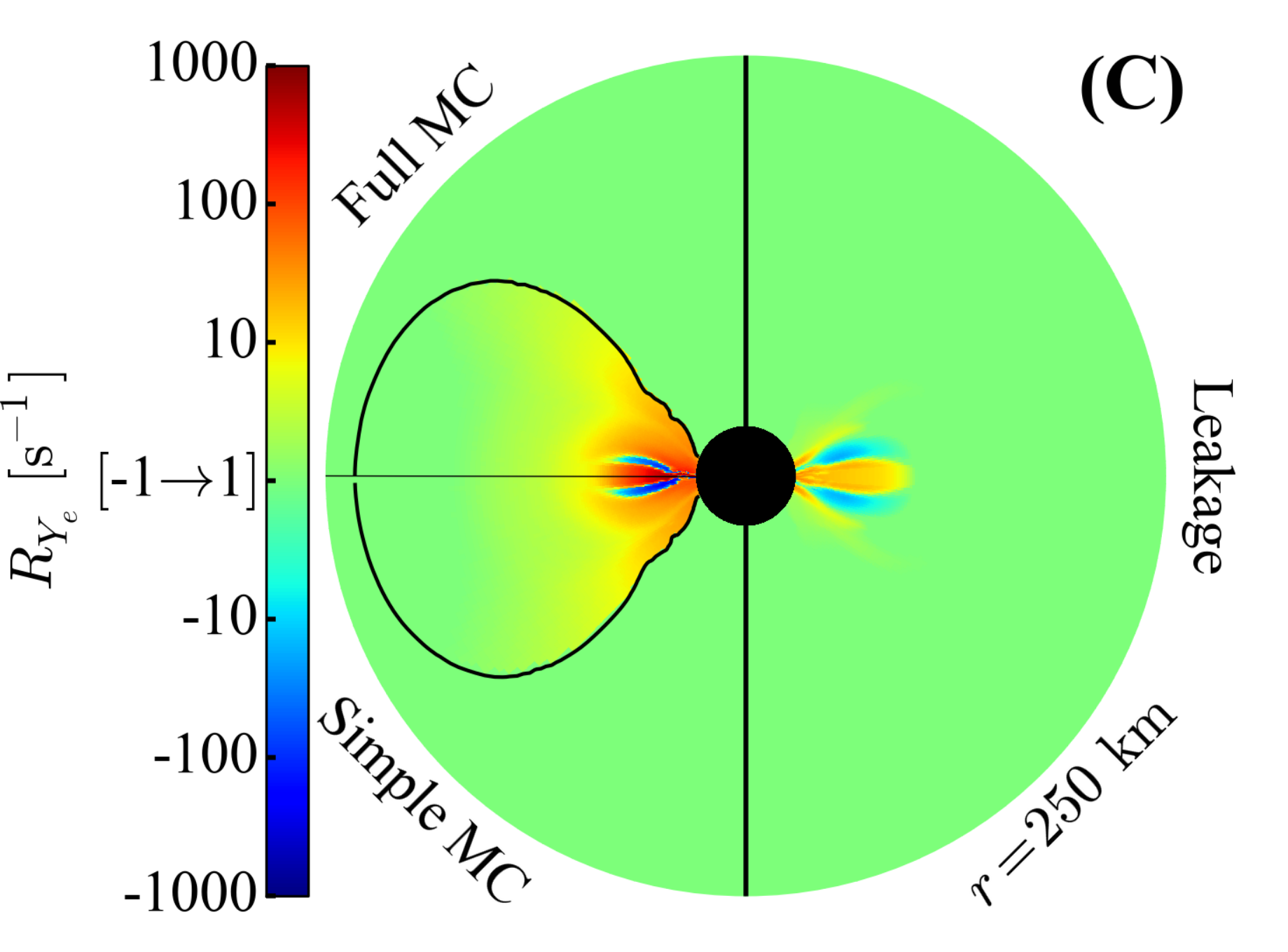}
  \caption{\textbf{Neutrino-Fluid Interaction (Central BH)} at $t=3\,\mathrm{ms}$. In each plot, the right half shows results calculated with neutrino leakage in the dynamical simulations of MF14, while the left half is calculated by {\tt Sedonu}. Each quadrant of Sedonu results depicts only half of the simulation domain. The top left quadrant uses the Full set of physics, while the bottom left uses the Simple set of physics. Panel A shows the difference between absorptive heating and emissive cooling. Panels B and C depict the relative rate of change of internal energy (including viscous heating) divided by internal energy and of electron fraction, respectively. Red represents a large positive rate of change while blue represents a large negative rate of change. Any rate of change whose magnitude is smaller than $1\,\mathrm{s}^{-1}$ is plotted as 0. The outer radius on each plot is at $250\,\mathrm{km}$ and the inner radius is $30\,\mathrm{km}$. The black curve is the $\rho=10^6\,\mathrm{g\ cm}^{-3}$ contour, below which {\tt Sedonu} opacities and emissivities are set to zero. Using MC neutrino transport would likely significantly affect the thermal and compositional evolution of the disk.}
  \label{fig:BHinteraction}
\end{figure*}

Different treatments of neutrino effects can have a significant impact on the fluid evolution. Panel~A of Figure~\ref{fig:BHinteraction} shows rapid cooling in the densest parts of the disk and net heating above and below the equatorial plane in both the MC and leakage results at $t=3\,\mathrm{ms}$. However, MC transport results in faster heating directly above the disk by more than an order of magnitude, a smaller cooling region, and much faster cooling on the equator than leakage, making the leakage heating nearly invisible in Panel A of Figure~\ref{fig:BHinteraction}. The differences are largely due to the approximate nature of the disk self-irradiation and leakage scheme of MF14, the accuracy of which suffers especially at the midplane of the disk. For efficiency, the leakage scheme calculates the optical depth at a given point to be $\tau=\kappa\ \mathrm{min}\{r,H_\perp,H_\parallel\}$, where $\kappa$ is the opacity, $r$ is the radius, and $H_\perp$ and $H_\parallel$ are the vertical and horizontal pressure scale heights, respectively. This optical depth calculation is only accurate to within a factor of a few. MC transport allows neutrinos to escape in any direction rather than just vertically and radially, increasing the amount of escaping neutrinos. Using Full neutrino physics dramatically increases the heating rate just above the hottest part of the disk due to special relativistic effects boosting the luminosity and average energy of neutrinos, and the larger global heating rate is also reflected in a smaller value of $\mathcal{C}_\nu - \mathcal{H}_\nu$ in Table~\ref{tab:results}.

Viscosity, neutrino heating, and neutrino cooling all affect the thermal evolution of the disk. The relative importance of neutrinos and viscosity can be seen in the amount of mass for which neutrino heating is larger than viscous heating in Table~\ref{tab:results}. At the $3\,\mathrm{ms}$ Full MC snapshot, $3.15\times10^{-3}M_\odot$ is heated more strongly by neutrinos than viscosity, though after this time the number drops very quickly. Simple neutrino physics causes this mass to be $60\%$ smaller, and in the Leakage simulation this mass is almost zero. In Panel~B of Figure~\ref{fig:BHinteraction} we show $R_\epsilon=(1/\epsilon)d\epsilon/dt$, the relative rate of change of internal energy including the viscous heating rate used in the dynamical simulations of MF14, scaled by each point's current energy density. The extra heated wings above the disk in the MC simulation cause the internal energy to change several times faster than leakage would suggest. More dramatically, the MC simulations show neutrino cooling dominating viscous heating along the equator, while the opposite is true in the leakage results. From this we would expect a stronger neutrino-driven wind, a thinner disk, and much faster disk cooling, though dynamical simulations would be required to investigate this quantitatively.

\begin{figure}
  \includegraphics[width=\linewidth]{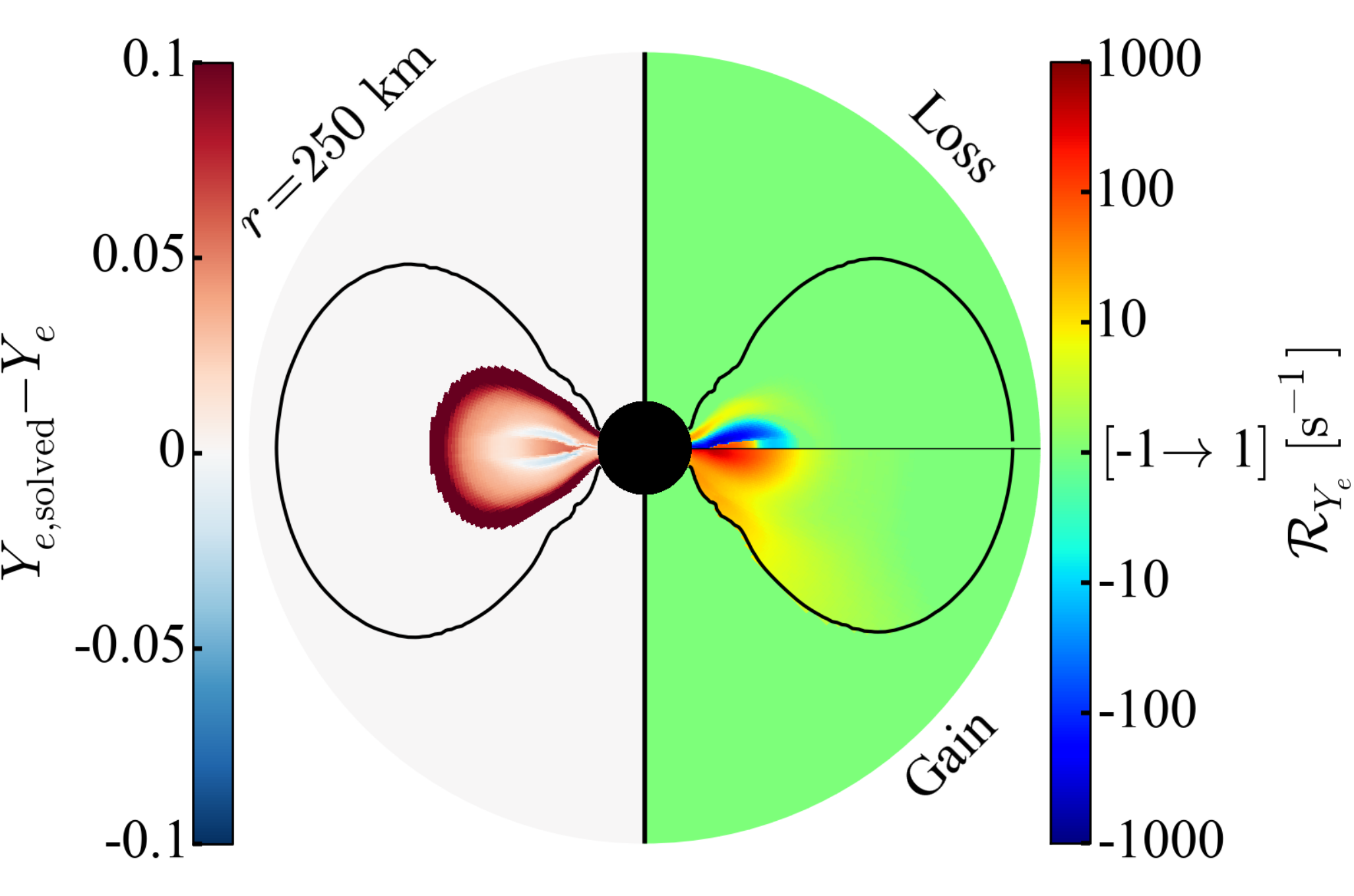}
\caption{\textbf{Equilibrium electron fraction (Central BH)} at $t=3\,\mathrm{ms}$. \textit{Left Hemisphere:} equilibrium electron fraction at which the net lepton number absorption rate is equal to the net lepton number emission rate. The equilibrium solver is unreliable below $\sim1.5\,\mathrm{MeV}$ as the energy grid ceases to be able to resolve the neutrino distributions, so $Y_{e,\mathrm{solved}}-Y_e$ at locations with a temperature less than this is plotted as zero. \textit{Right Hemisphere:} rate of change of electron fraction, as in Figure~\ref{fig:BHinteraction}, but separately depicting that caused by emission (top) and absorption (bottom). The black curve is the $\rho=10^6\,\mathrm{g\ cm}^{-3}$ contour, below which {\tt Sedonu} opacities and emissivities are set to zero. The outer radius on the plot is at $250\,\mathrm{km}$ and the inner radius is $30\,\mathrm{km}$.}
\label{fig:BHequilibriumYe}
\end{figure}

In a similar manner, we show $R_{Y_e}=dY_e/dt$ in Panel~C of Figure~\ref{fig:BHinteraction}. There is very little difference between the Simple and Full MC simulations, though both represent a significant departure from the leakage data. In all cases, electron fraction is increasing near the equator as the low-electron fraction fluid is emitting primarily electron anti-neutrinos. Above the equator, there is a pattern of regions of both increasing and decreasing electron fraction. The right half of Figure~\ref{fig:BHequilibriumYe} shows that this pattern of increasing and decreasing electron fraction is caused by neutrino emission rather than absorption, indicating that variations in density and temperature cause the equilibrium electron fraction to vary. In the left half of Figure~\ref{fig:BHequilibriumYe}, we show the difference between the current electron fraction and the equilibrium electron fraction (including neutrino interactions, see Section~\ref{sec:methods}). The initial conditions of the dynamical simulation began with an electron fraction of $Y_e=0.1$, leaving the center of the disk far from equilibrium, and the slower rates in the leakage scheme prevent it from coming into equilibrium more quickly.

As the disk spreads, cools, and accretes onto the central BH, neutrinos affect the evolution of the disk and outflow more slowly. However, MC transport differs significantly from leakage for at least several tens of milliseconds. The volume-integrated neutrino cooling minus heating {through \sout{in}} the $3\,\mathrm{ms}$ snapshot is {up to} $\sim3{8}\%$ larger in the Full MC simulations than in the Leakage ones. At all later times, though, the Leakage simulations cool faster by {up to 22\% \sout{a similar percentage}}. Throughout the disk's evolution MC results in a higher leptonization rate, up to 7.7 times that of the Leakage simulation at $3\,\mathrm{ms}$. This is due to the way the leakage scheme treats optical depths, as discussed in Section~\ref{sec:discussion}.

\begin{figure}
  \includegraphics[width=\linewidth]{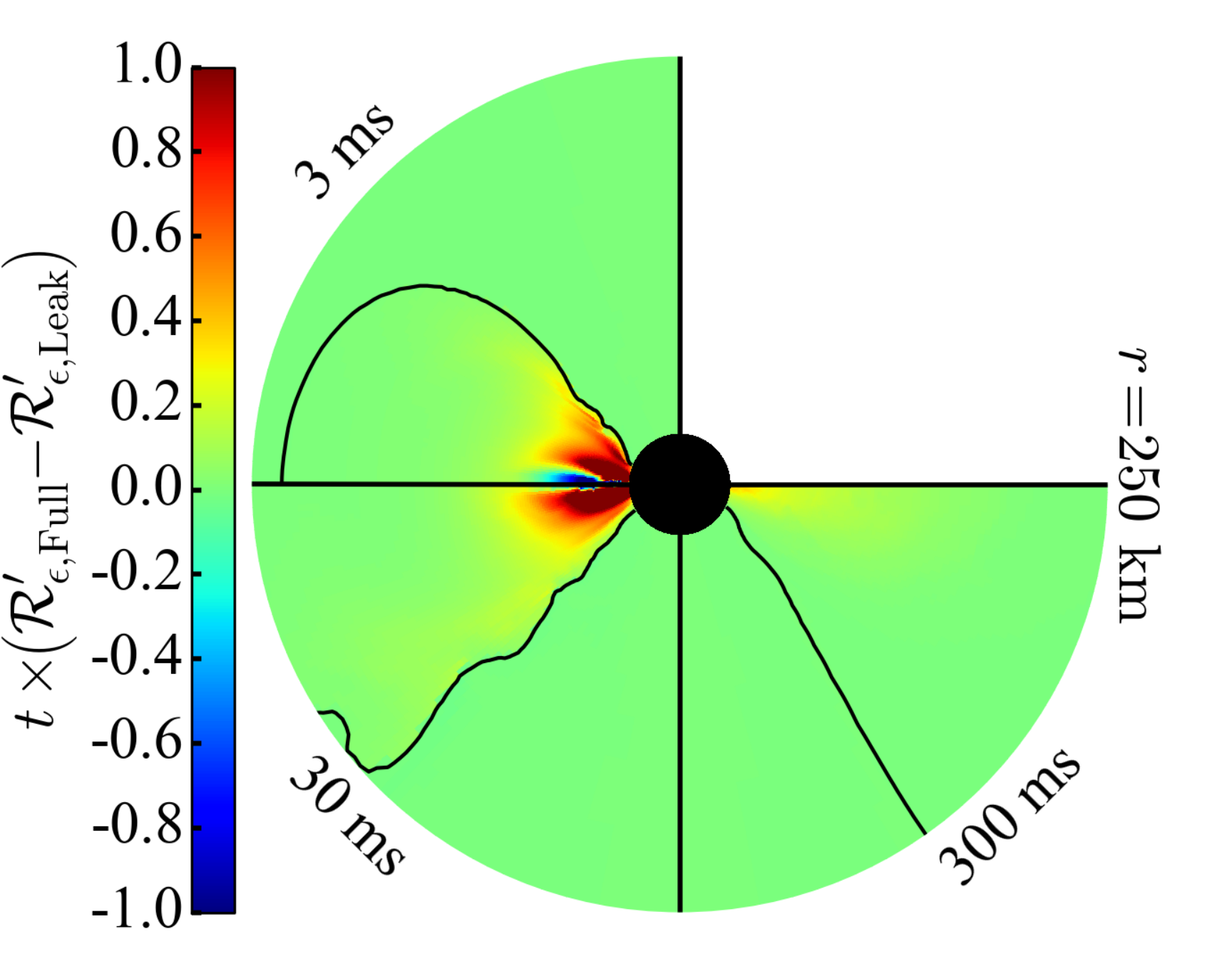}
  \includegraphics[width=\linewidth]{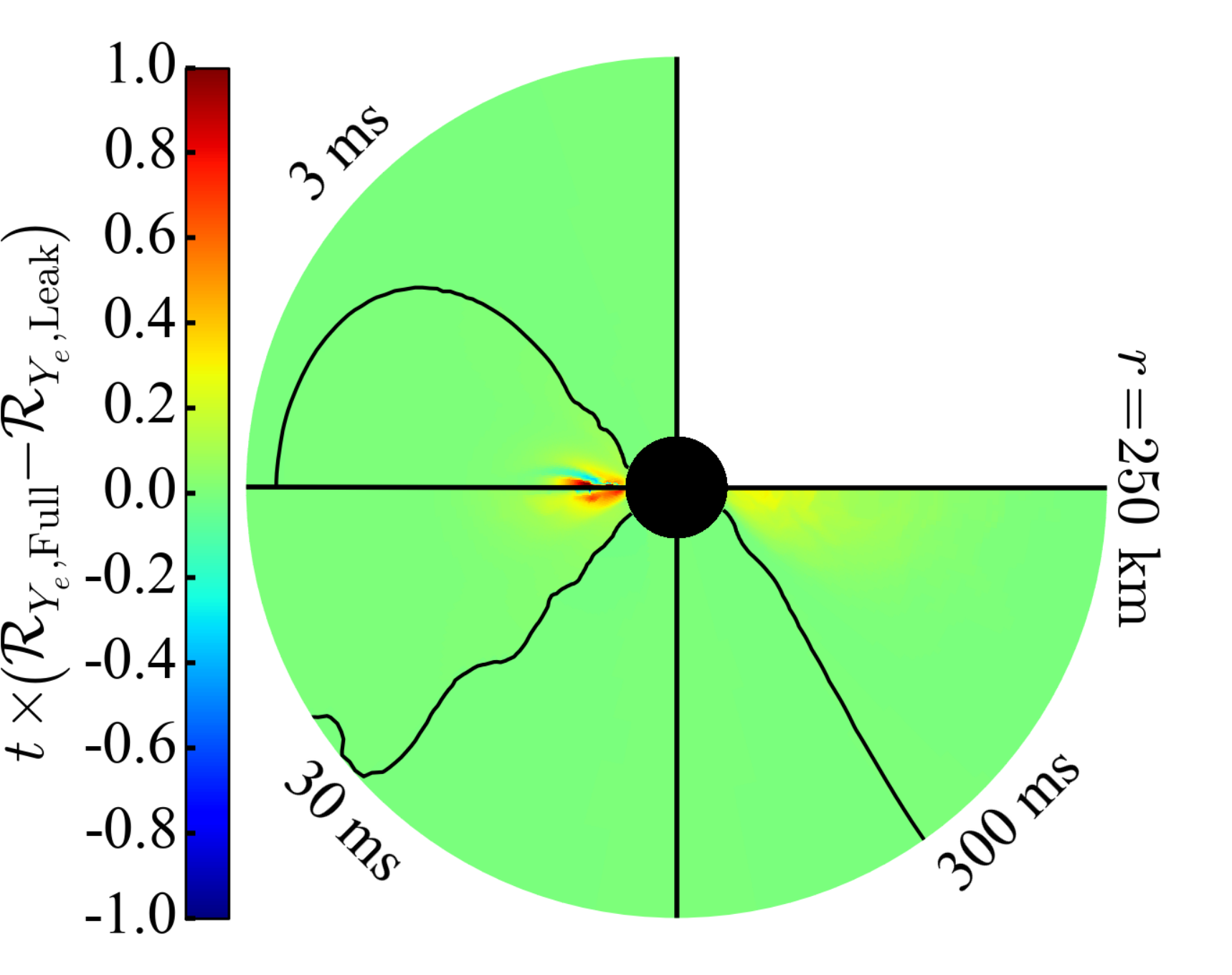}
  \caption{\textbf{Leakage-MC Difference (Central BH)}. Full MC transport results differ significantly from leakage results for at least several tens of milliseconds. Difference between the Full and Leakage relative rate of change of internal energy ignoring viscosity (top panel) and electron fraction (bottom panel), multiplied by the snapshot time in order to estimate the potential impact of improved neutrino transport on dynamical simulations. The black curve is the $\rho=10^6\,\mathrm{g\ cm}^{-3}$ contour, below which {\tt Sedonu} opacities and emissivities are set to zero. The outer radius on the plot is at $250\,\mathrm{km}$ and the inner radius is $30\,\mathrm{km}$. In both plots, values larger than unity imply that if such rates continued for a time equal to the snapshot time, the difference between MC and leakage would be dynamically important. For this plot, data is taken from the 2xEnergy run in Appendix~\ref{app:resolution} for increased solution accuracy.}
  \label{fig:BHLateTime}
\end{figure}

In Figure~\ref{fig:BHLateTime}, we show the difference between the rates of change of internal energy due only to neutrinos ($\mathcal{R}^\prime_\epsilon$) and electron fraction ($\mathcal{R}_{Y_e}$) calculated using leakage and using MC transport. Though the differences are most striking at $3\,\mathrm{ms}$, we still see a significantly larger $\mathcal{R}^\prime_\epsilon$ near the $45^\circ$ radials and a larger $\mathcal{R}_{Y_e}$ near the midplane in the MC results at $30\,\mathrm{ms}$. In these simulations, {\tt Sedonu} takes the opacity at densities lower than $10^6\,\mathrm{g\ cm}^{-3}$ to be zero, so the leakage scheme shows more neutrino heating for a very small amount of mass outside the disk. However, the left and right halves of Panel~B of Figure~\ref{fig:BHinteraction} outside of the region covered by {\tt Sedonu} appear almost identical because in this region viscous heating is completely dominant. Comoving frame viscous heating is identical in all cases, but time dilation slightly modifies the heating rates in all but the Simple and NoRel cases. At $300\,\mathrm{ms}$ neutrino cooling by leakage is more efficient than Monte Carlo cooling, though the differences are only apparent at the densest part of the disk and are anyway dominated by viscous heating.

For calculating opacities and interaction rates, NuLib requires as input an EOS to determine the chemical potentials of the particles involved in each interaction. Thus, different EOS result in different interaction rates. In addition to the fiducial Helmholtz EOS \citep{TS00}, we repeat the calculations using the LS220 \citep{LS91} and the H. Shen \citep{Shen+11} EOS. We find that the choice of EOS has no significant effect, as is demonstrated by the results summarized in Table~\ref{tab:results}. This is reassuring, since all of the neutrino emission and absorption occurs at sub-nuclear densities where the details of the treatment of the strong force are less significant.

\section{Results (Central HMNS)}
\label{sec:HMNSresults}
Following the merger of two neutron stars, the central object that forms may be a black hole, a stable neutron star, or an only temporarily-stable hypermassive neutron star (HMNS), depending on details of the equation of state and the object's mass (e.g.,\ \citealt{Kaplan+14}). In this section, we bracket the parameter space by repeating the analysis of the previous section, but with simulations including an HMNS assumed to be permanently stable. The inner boundary, which models an HMNS by reflecting matter, prevents mass from accreting through it and leads to a disk that stays hot and massive for a much longer time. As did MF14, we assume the HMNS emits neutrinos with a zero-chemical potential blackbody distribution and the average energies listed in Table~\ref{tab:flash}. Heavy lepton neutrinos, when present, have the same luminosity and average energy as electron anti-neutrinos. Table~\ref{tab:results} lists the simulations we run and the corresponding global properties of the fluid and radiation field.

\subsection{Neutrino Radiation Field (Central HMNS)}
\label{sec:HMNSnufield}
\begin{figure}
  \includegraphics[width=\linewidth]{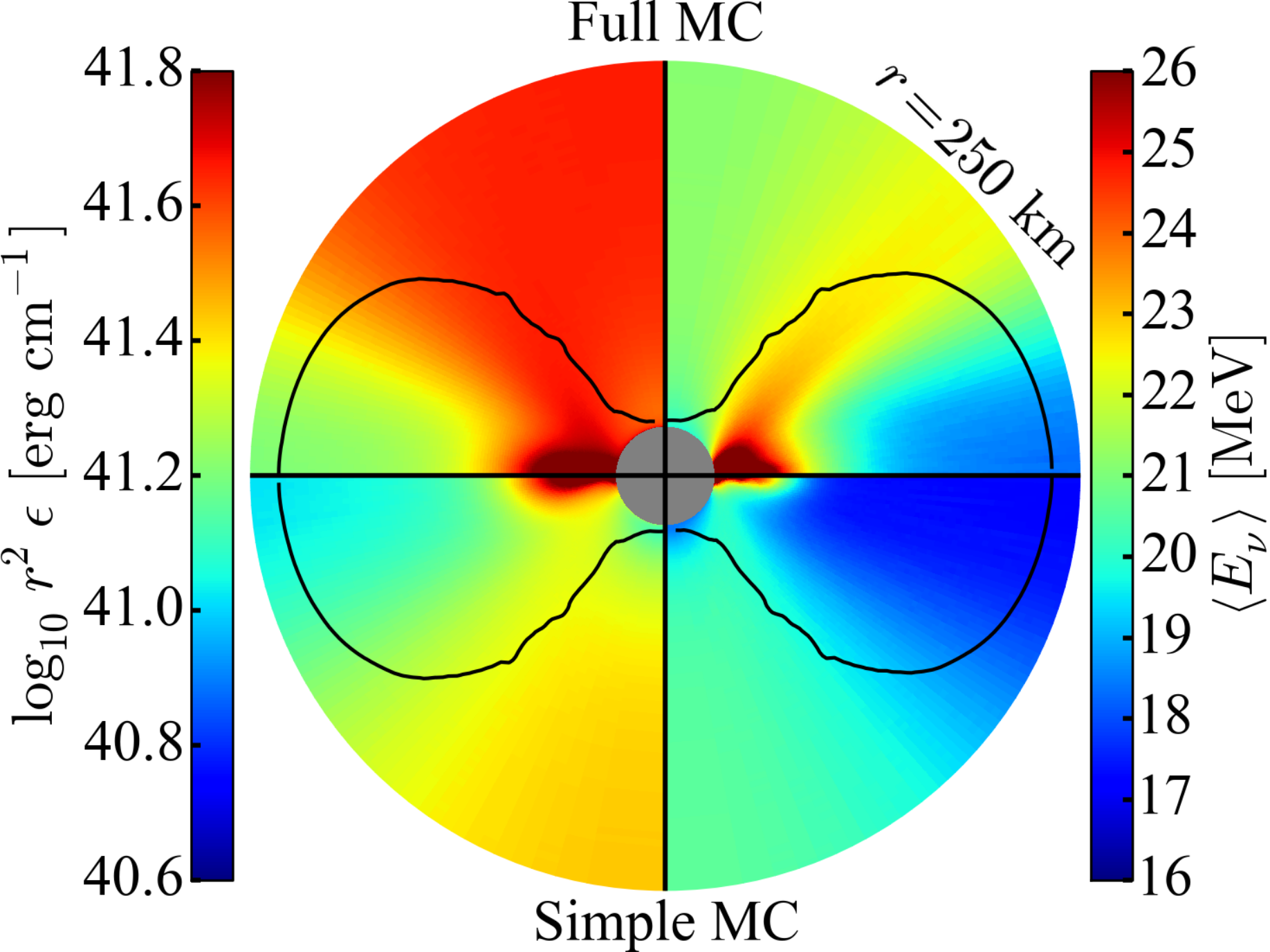}
  \caption{\textbf{Neutrino energy density and average energy (Central HMNS)} at $t=3\,\mathrm{ms}$. The quantities shown are the same as in Figure~\ref{fig:BHnufield}, but using the HMNS snapshot. \textit{Left hemisphere}: Neutrino energy density, summed over all species and multiplied by $r^2$ to remove effects of distance from the center. \textit{Right hemisphere}: Neutrino energy density-weighted average energy, averaged over all species. \textit{Bottom hemisphere}: Simple MC results. \textit{Top hemisphere}: Full MC results. The black curve is the $\rho=10^6\,\mathrm{g\ cm}^{-3}$ contour, below which {\tt Sedonu} opacities and emissivities are set to zero. The outer radius on the plot is at $250\,\mathrm{km}$ and the inner radius is $30\,\mathrm{km}$. The neutrino radiation field is very asymmetric and sensitive to the included physics. The disk casts a shadow as higher-energy neutrinos are preferentially absorbed. Much more asymmetry is present when the Full suite of physics is included. Both the energy density and average neutrino energy are higher than in the BH case. At this point in time, the luminosities and average energies of $\{\nu_e,\bar{\nu}_e,\nu_x\}$ from the central HMNS are set to $\{10.2,10.2,40.8\}\,\mathrm{B\ s}^{-1}$ and $\{16.4,20.5,20.5\}\,\mathrm{MeV}$, respectively.}
  \label{fig:HMNSnufield}
\end{figure}
As seen in Figure~\ref{fig:HMNSnufield}, in the presence of an HMNS the neutrino radiation field at $t=3\,\mathrm{ms}$ shows the same disk and polar shadows and relativistic beaming as when a BH is present, though the neutrino energy densities and average energies are somewhat higher due to the hotter fluid and extra irradiation from the HMNS. The Full physics neutrino radiation field shows higher energy densities by a factor of $\sim1.5$ in the free streaming regions outside the disk than the Simple physics neutrino radiation field. This is due mostly to the production of copious amounts of heavy lepton neutrinos in the HMNS, which have comparatively small cross sections and so are able to pass much more easily through the disk. As in the BH case, special relativity increases the average neutrino energy by up to $\sim30\%$, especially 45 degrees from the pole.

$\mathcal{L}_\mathrm{emit}$, $\mathcal{L}_\mathrm{escape}$, $\langle E_{\nu,\mathrm{emit}}\rangle$, and $\langle E_{\nu,\mathrm{escape}}\rangle$ in Table~\ref{tab:results} describe the global lab-frame properties of the neutrinos in the simulations with an HMNS. Since the initial disk conditions are the same for both dynamical simulations, the properties of the neutrinos emitted from the disk at $t=0\,\mathrm{ms}$ are also identical. The extra irradiation from the HMNS results in a slower net cooling rate of the disk at early times. After the first $3\,\mathrm{ms}$, the large disk mass around the HMNS causes the net cooling rate to be much larger, though the higher temperatures and amount of irradiation by nearly equal numbers of electron neutrinos and anti-neutrinos imposed from the HMNS boundary condition (see Section~\ref{sec:emiss}) cause a slower change in electron fraction than in the BH case. The high densities and viscously-amplified temperatures near the HMNS cause 1.9 times as much neutrino energy to be emitted from the disk, though the vast majority is immediately re-absorbed. A combination of the HMNS's extra radiation and the higher disk luminosity causes the escaping neutrino luminosity to be larger by factors of 2.8, 1.5, and 90 for $\nu_e$, $\bar{\nu}_e$, and $\nu_x$, respectively. The energies of the escaping neutrinos are similar to those in the BH case, but the $\nu_x$ average energy is decreased due to dilution from the HMNS. Note that in simulations that include the HMNS, the HMNS is much hotter than the disk (e.g., \citealt{Dessart+09}), but we parameterize the neutrinos being emitted from the HMNS for consistency with the dynamical simulations of MF14.

\begin{figure}
  \includegraphics[width=\linewidth]{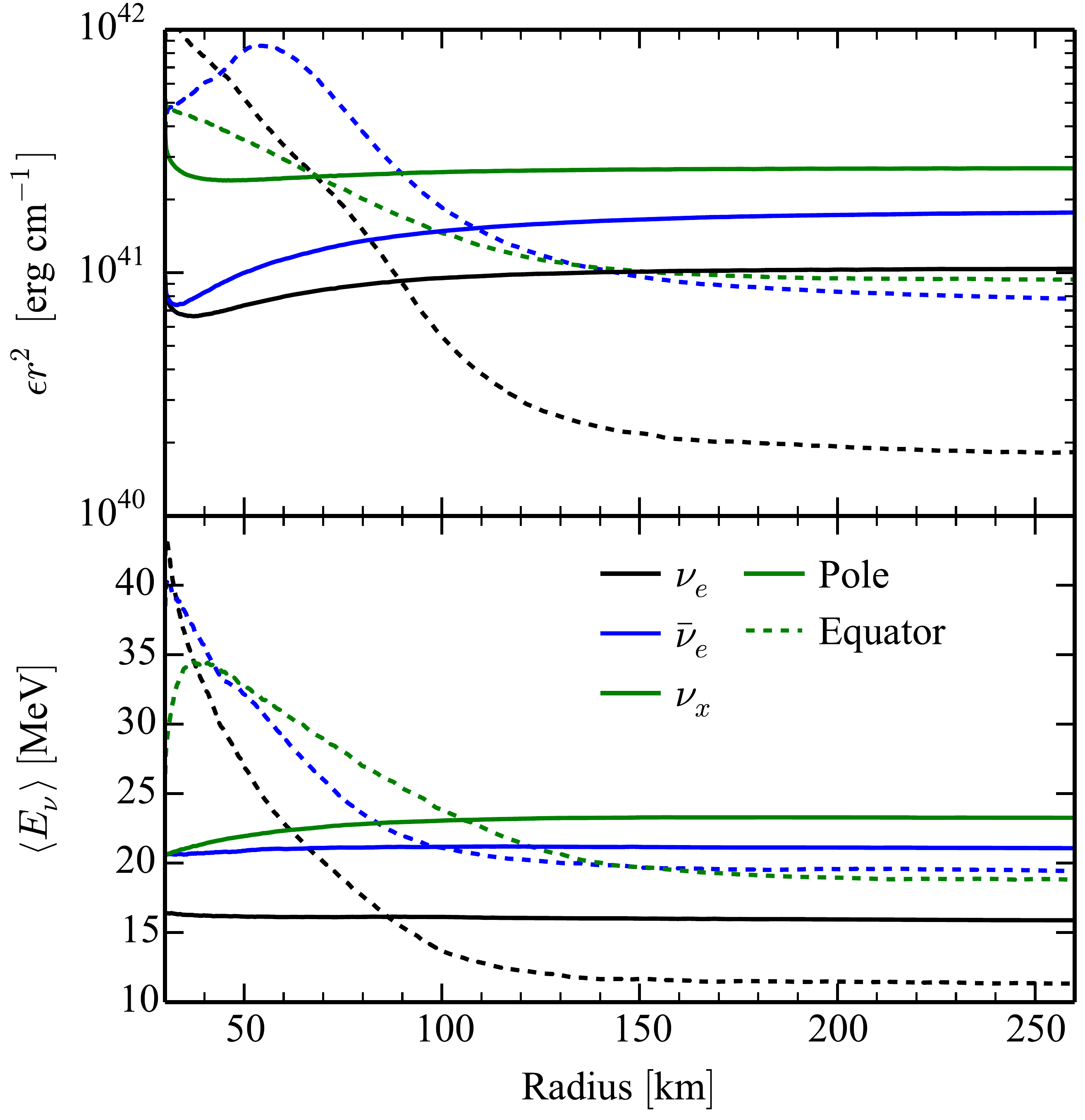}
  \caption{\textbf{Neutrino Radiation Profile (Central HMNS)} from the Full MC simulation at $t=3\,\mathrm{ms}$ for all three simulated neutrino species. The quantities shown are the same as in Figure~\ref{fig:BHprofile}. The neutrino radiation field is asymmetric and dominated by electron anti-neutrinos. \textit{Top}: neutrino energy density along the pole (solid lines) and the equator (dashed lines), multiplied by $r^2$ to remove effects of distance from the center. \textit{Bottom}: energy density-weighted average neutrino energy along radial lines. The green $\nu_x$ curves represent the sum of all four heavy lepton neutrino species. Note the difference in the y-axis scale compared with Figure~\ref{fig:BHprofile}. The hierarchy between neutrino species is shuffled at small radii due to competing emission from the HMNS and the disk, and from disk absorption.}
  \label{fig:HMNSprofile}
\end{figure}

\begin{figure}
  \includegraphics[width=\linewidth]{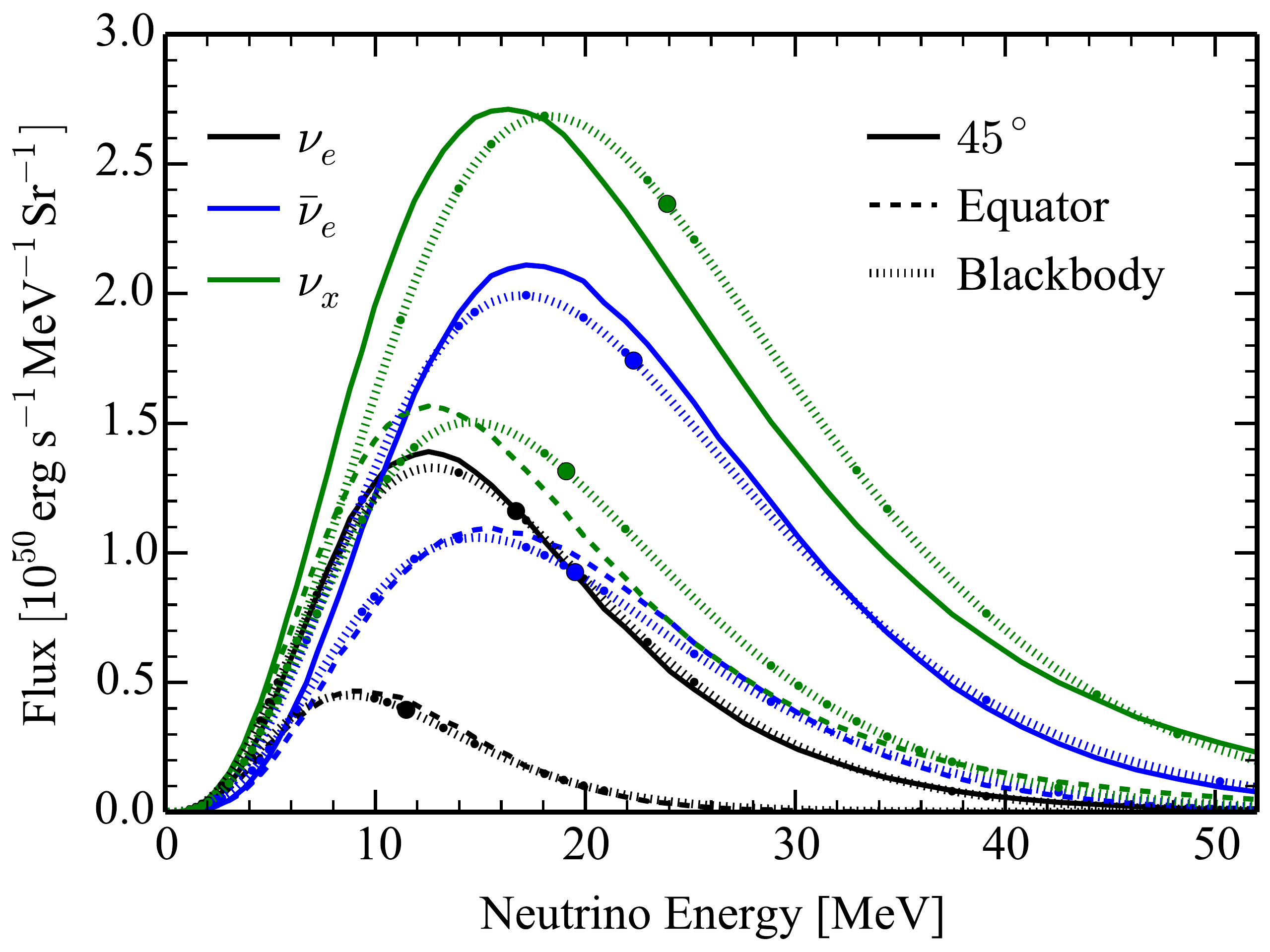}
  \caption{\textbf{Neutrino Spectra (Central HMNS)} from the Full MC simulation at $t=3\,\mathrm{ms}$. The quantities shown are the same as in Figure~\ref{fig:BHspectra}, though note that the vertical axis differs. The escaping neutrino radiation is somewhat nonthermal and asymmetric. Dashed lines are spectra of each neutrino species escaping from within $10^\circ$ of the equator, while the solid lines are those from within $10^\circ$ of the $45^\circ$ cones, normalized by the solid angle covered by the respective regions. Overplotted for both directions (distinguished by proximity to the data curves) are dotted zero-chemical potential blackbody curves with the same total flux and average energy as the measured spectrum. The large dot on the blackbody curve indicates this average energy. For smoothness, the spectra are taken from the 2xEnergy run in Appendix~\ref{app:resolution}. The heavy lepton neutrinos are near blackbodies since most come from the central HMNS.}
  \label{fig:HMNSspectra}
\end{figure}

Figure~\ref{fig:HMNSprofile} shows a complicated interaction between radiation from the HMNS and that from the disk. Along the poles there is an initial dip in intensity as neutrinos are absorbed by a layer of matter just outside of the HMNS. Moving outward along the pole, as the disk comes into view, the electron neutrinos and anti-neutrinos emitted from the disk bump their respective intensities up again. Though they interact more weakly than electron neutrinos and anti-neutrinos, heavy lepton neutrinos are scattered by the disk, as seen in the divergence of the pole and equatorial energy density. In fact, the scattering combined with Doppler boosting in the Full simulation cause the average heavy lepton neutrino energy peak in the inner regions of the disk, and the scattered neutrinos even cause radially increasing average neutrino energy along the poles. Below $150\,\mathrm{km}$ along the equator, electron and heavy lepton neutrino intensities decline as their emission from the HMNS is absorbed by the disk. The dense part of the disk below $60\,\mathrm{km}$ is such a strong emitter of electron anti-neutrinos, however, that the intensity rises before falling again farther out. In all cases, the neutrino radiation field becomes free-streaming after a radius of $\sim150\,\mathrm{km}$, indicated by horizontal lines in Figure~\ref{fig:HMNSprofile}.

Figure~\ref{fig:HMNSspectra} demonstrates the asymmetry of the neutrino radiation field and its departure from a zero-chemical potential blackbody. At all times, the irradiation from the HMNS makes the net escaping flux from heavy lepton neutrinos comparable to that from electron anti-neutrinos, in contrast to the BH case.

Just as with the BH snapshots, using Full physics adds a scattering opacity which prevents neutrinos from escaping as easily as with Simple physics, causing Simple physics to allow for a larger cooling and leptonization rate, as well as naturally larger escape luminosities. Full physics also increases the neutrino creation rate through weak magnetism corrections and Lorentz transformations. Even though a large number of heavy lepton neutrinos are produced by the HMNS, Figure~\ref{fig:HMNSprofile} shows that they contribute much less to disk heating than electron anti-neutrinos do. This is because they deposit energy into the fluid only through NuLib's approximate treatment of inverse Bremsstrahlung and neutrino pair annihilation. Excluding heavy lepton neutrinos only results in the disk cooling 3\% more quickly. Otherwise, the exclusion of various elements of physics has the same effect as in the BH snapshots.

\subsection{Neutrino-Fluid Interaction (Central HMNS)}

\begin{figure*}
  \includegraphics[height=0.25\linewidth]{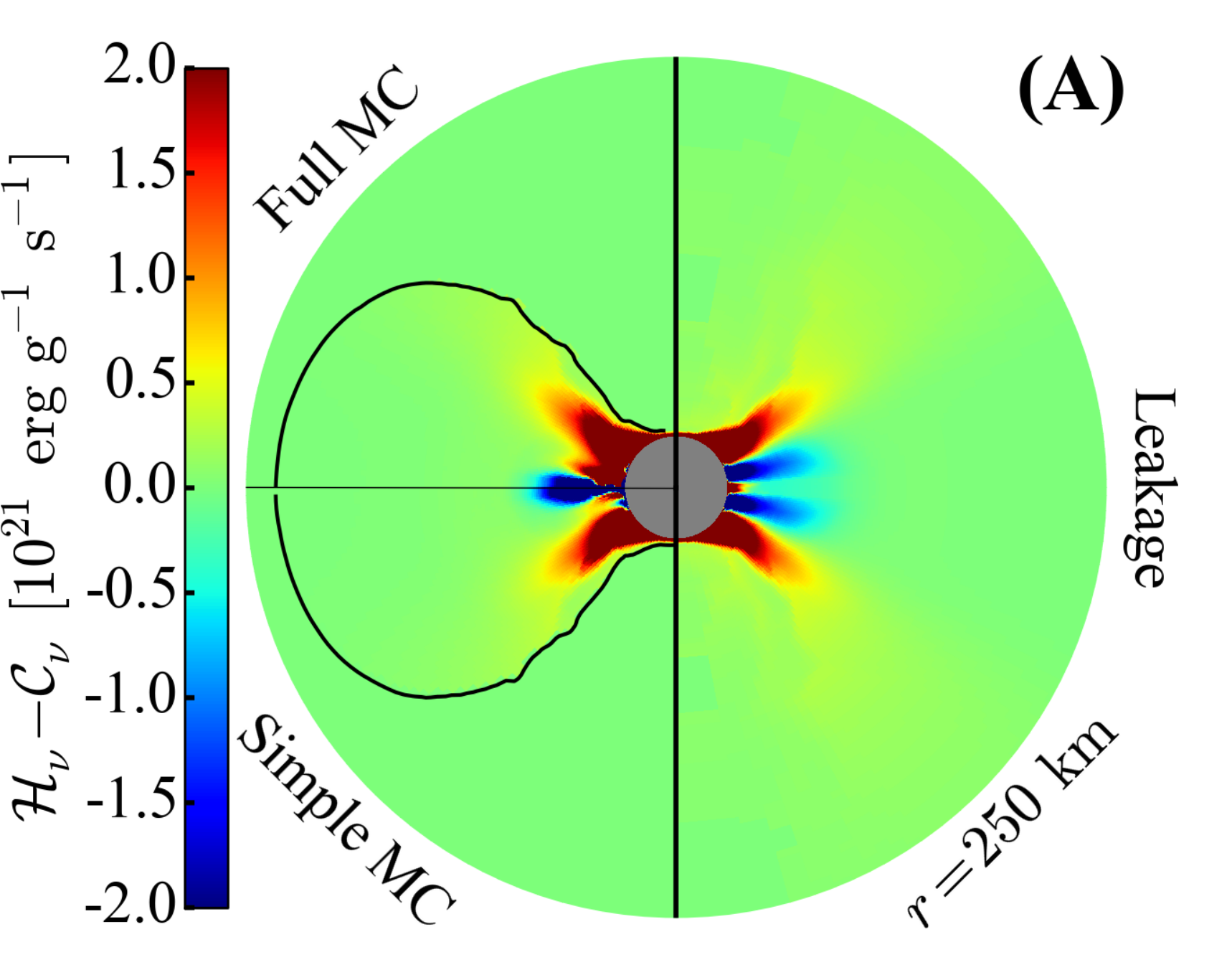}
  \includegraphics[height=0.25\linewidth]{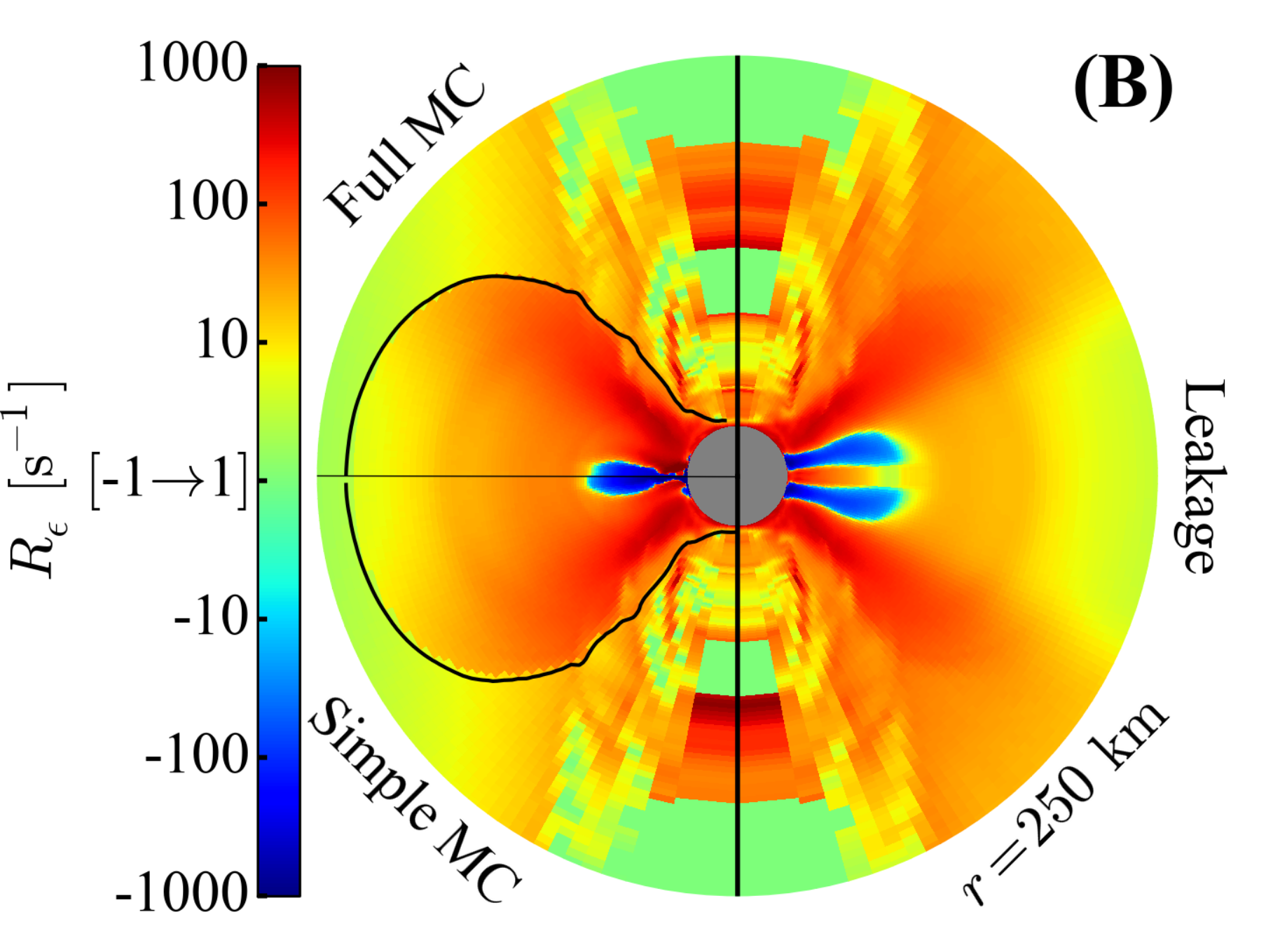}
  \includegraphics[height=0.25\linewidth]{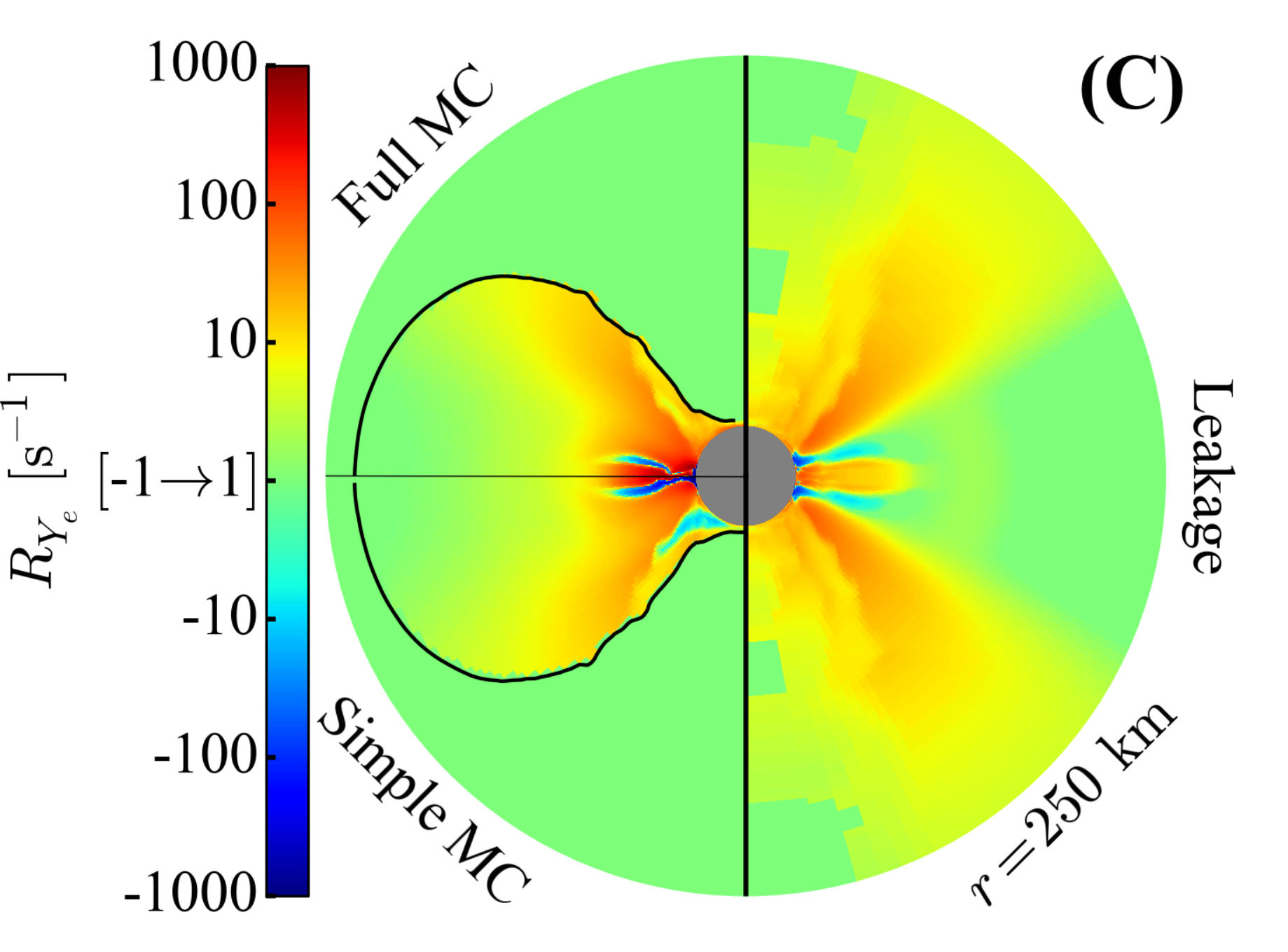}
  \caption{\textbf{Neutrino-Fluid Interaction (Central HMNS)}  at $t=3\,\mathrm{ms}$. The quantities shown are the same as in Figure~\ref{fig:BHinteraction}. In each plot, the right half shows results calculated with neutrino leakage in the dynamical simulations of MF14, while the left half is calculated by {\tt Sedonu}. Each quadrant of Sedonu results depicts only half of the simulation domain. The top left quadrant uses the Full set of physics, while the bottom left uses the Simple set of physics. Panel A shows the difference between absorptive heating and emissive cooling. Panels B and C depict the relative rate of change of internal energy (including viscous heating) divided by internal energy and of electron fraction, respectively. Red represents a large positive rate of change while blue represents a large negative rate of change. Any rate of change whose magnitude is smaller than $1\,\mathrm{s}^{-1}$ is plotted as 0. The outer radius on each plot is at $250\,\mathrm{km}$ and the inner radius is $30\,\mathrm{km}$. The black curve is the $\rho=10^6\,\mathrm{g\ cm}^{-3}$ contour, below which {\tt Sedonu} opacities and emissivities are set to zero. Using MC neutrino transport would likely significantly affect the thermal and compositional evolution of the disk.}
  \label{fig:HMNSinteraction}
\end{figure*}

Figure~\ref{fig:HMNSinteraction} describes the interaction of neutrinos with the background fluid. Near the equator, the structures of the heating rates, $R_\epsilon$, and $R_{Y_e}$ are very similar to the BH case. Panel~A shows the local neutrino heating rates, the volume integrals of which are displayed in Table~\ref{tab:results}. Full MC, as in the BH case, shows slower integrated neutrino cooling than the Simple MC simulation (factor of $\sim0.91$) but much faster cooling than the Leakage simulation (factor of $\sim{2.2}$). The relative effects of neutrinos and viscosity can be seen in the amount of mass for which neutrino heating is larger than viscous heating. This mass is $\sim26\%$ larger in the Full MC simulations than in the Simple MC simulation and $\sim5.6$ times larger than in the Leakage simulation. This is visible in the relative sizes of the neutrino-heated regions above and below the disk.

The rates of change of internal energy in Panel~B of Figure~\ref{fig:HMNSinteraction} demonstrate that both Full and Simple MC neutrinos escape from and pass through the densest regions of the disk more easily than leakage allows, causing visibly faster heating near the equator beyond the disk. Similar to the heating rates, the volume-integrated leptonization rate indicated by the Full MC simulation is faster than that predicted by the Leakage simulation. The slight difference between the Simple and Full MC simulations is also reflected in Table~\ref{tab:results}. Simple physics decreases opacities and results in a faster increase in electron fraction, mostly near the HMNS above the disk. Monte Carlo allows neutrinos to escape easier than they could in the Leakage simulation and hence has higher cooling and leptonization rates.

Although the differences stem from the same effects, the Full MC net neutrino cooling rate is at times a factor of {8} higher than the leakage net neutrino cooling rate (at $t=30\,\mathrm{ms}$), though this is likely artificially high due to the out-of-equilibrium effects discussed in Section~\ref{sec:BHresults}. Unlike in the BH case, MC net neutrino cooling minus heating is larger than that from leakage calculations through $300\,\mathrm{ms}$ since the disk retains its mass, and the greater ease of escape for MC neutrinos allows the disk to absorb less energy from the HMNS neutrinos and to cool more quickly. In addition, the HMNS emits nearly equal numbers of electron neutrinos and anti-neutrinos, but the easier escape allowed to MC neutrinos means the HMNS neutrinos are not as effective at bringing the electron fraction up. At $t=3\,\mathrm{ms}$, MC leads to faster cooling than leakage does by a factor of $\sim2.{2}$ and leptonization by a factor of $\sim2.7$. However, unlike the BH case, after the $3\,\mathrm{ms}$ snapshot MC leptonization is slower than leakage. By the $3\,\mathrm{s}$ snapshot, the dynamical simulation has overshot MC's equilibrium and so the MC disk is slowly heating rather than cooling. 

\begin{figure}
  \includegraphics[width=\linewidth]{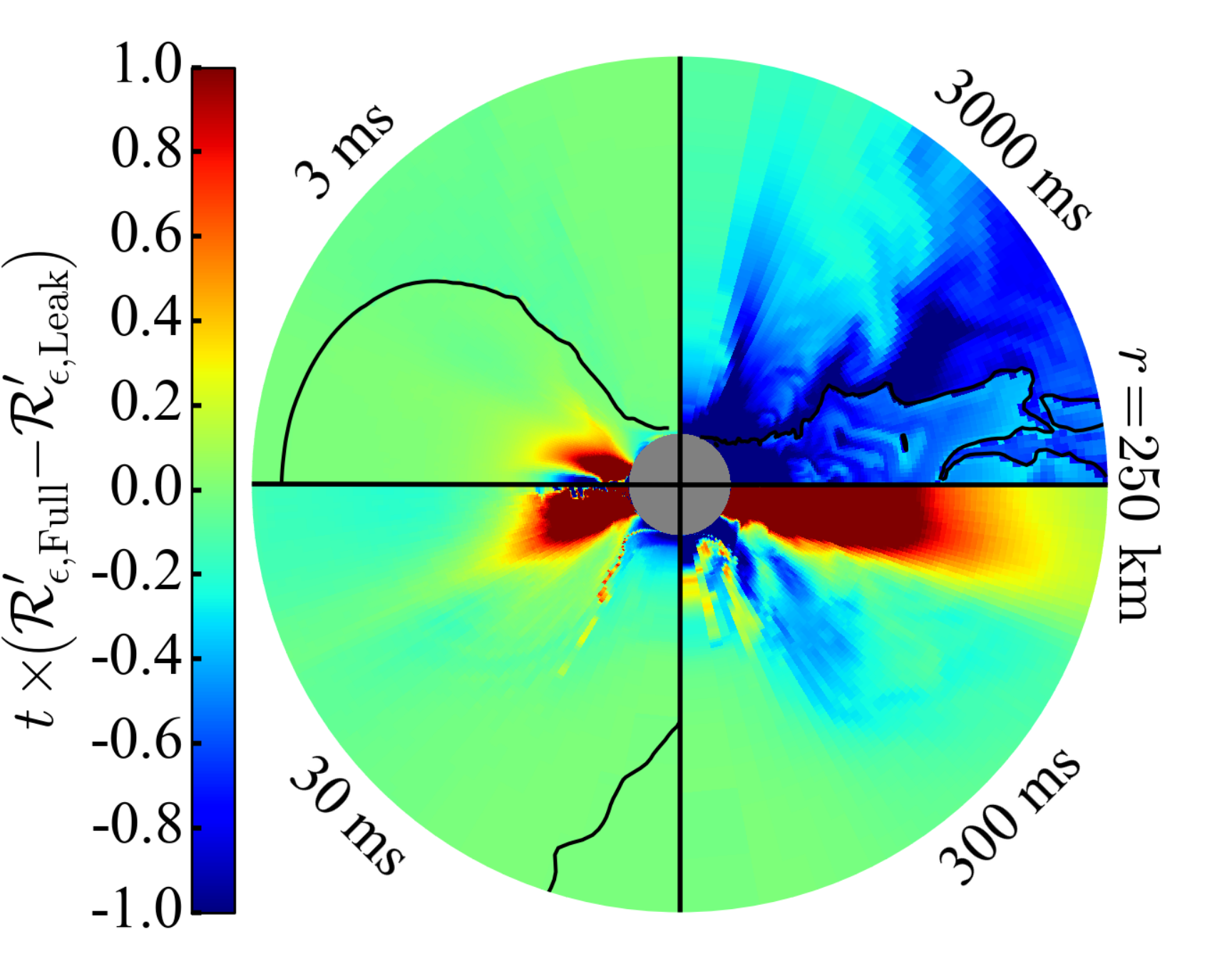}
  \includegraphics[width=\linewidth]{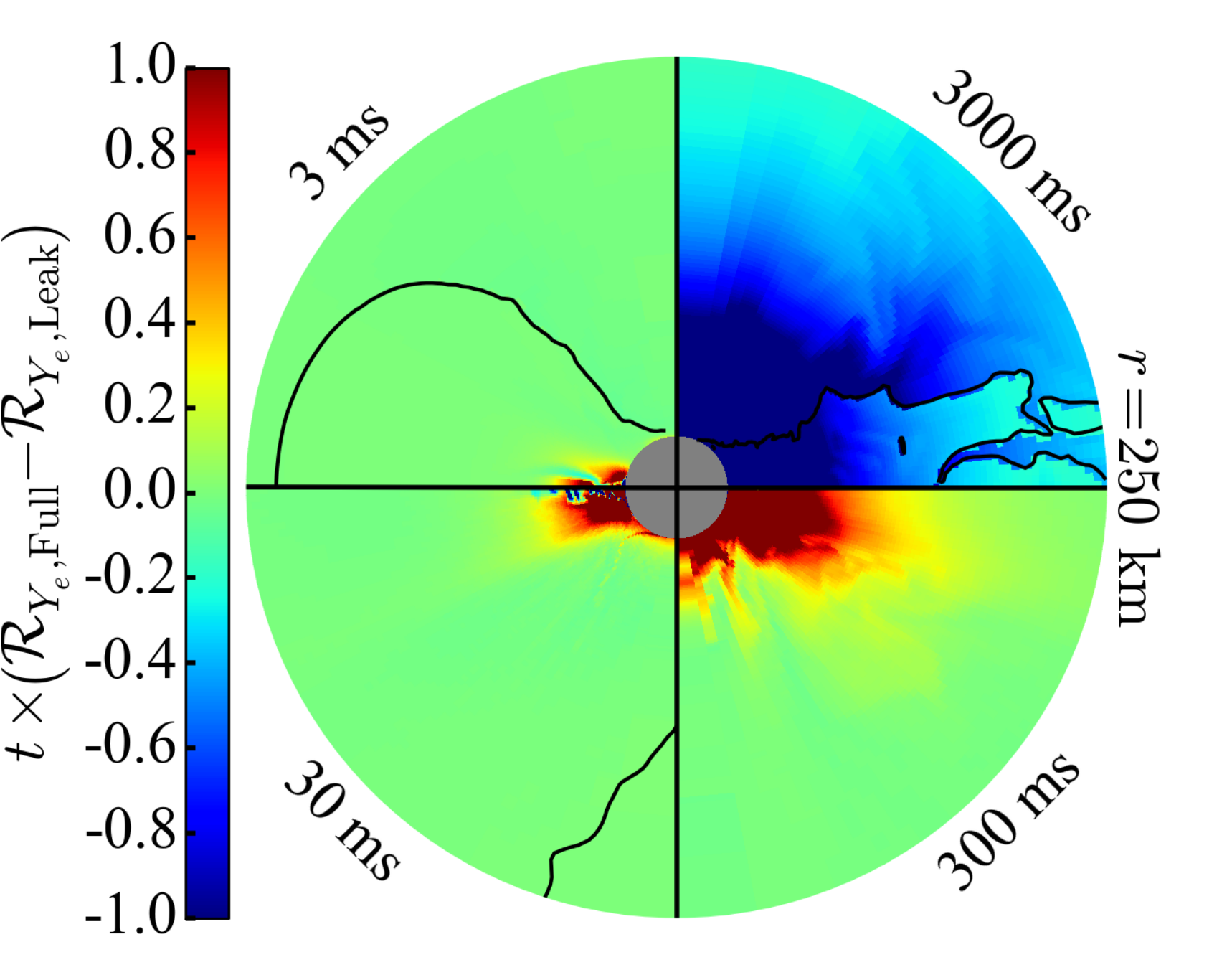}
  \caption{\textbf{Leakage-MC Difference (Central HMNS)}. The quantities shown are the same as in Figure~\ref{fig:BHLateTime}. Difference between the Full and Leakage relative rate of change of internal energy ignoring viscosity (top panel) and electron fraction (bottom panel), multiplied by the snapshot time in order to estimate the potential impact of improved neutrino transport on dynamical simulations. The black curve is the $\rho=10^6\,\mathrm{g\ cm}^{-3}$ contour, below which {\tt Sedonu} opacities and emissivities are set to zero. At $300\,\mathrm{ms}$ everything in the image is above the density cutoff. The outer radius on the plot is at $250\,\mathrm{km}$ and the inner radius is $30\,\mathrm{km}$. In both plots, values larger than unity imply that if such rates continued for a time equal to the snapshot time, the difference between MC and leakage would be dynamically important. Full MC transport results differ significantly from leakage results at all times. For this plot, data is taken from the 2xEnergy run in Appendix~\ref{app:resolution} for increased solution accuracy.}
  \label{fig:HMNSLateTime}
\end{figure}

The disk mass remains in the HMNS simulation for almost a hundred times as long as it does in the BH snapshots, and neutrinos from both disk and HMNS emission play an important role for at least ten times as long as in the BH snapshots. Figure~\ref{fig:HMNSLateTime} compares the leptonization rates and the difference between cooling and heating rates due only to neutrinos through the $3\,\mathrm{s}$ snapshot. The data in the $3\,\mathrm{ms}$ quadrants effectively replicates information conveyed in Figure~\ref{fig:HMNSinteraction} by showing that MC transport allows the disk to cool faster, heats the regions above the disk faster, and allows the disk $Y_e$ to change more quickly. In the $30$ and $300\,\mathrm{ms}$ quadrants, leakage predicts that neutrinos are unable to cool the matter in the equatorial plane, but are able to cool the disk above and below the equator. MC transport, on the other hand, predicts some cooling in isolated domains on the equator and next to the HMNS at high latitudes, but neutrino heating balances neutrino cooling in the mid-latitude disk regions. In addition, leakage predicts much stronger heating of the low-density polar regions, a trend which continues to be evident at later times. In the $3\,\mathrm{s}$ quadrants, very little mass remains in the disk and the predicted leakage rates are far in excess of the MC ones, just as they were in the low-density regions in previous snapshots. This is all consistent with the above statement that MC allows neutrinos to escape more easily than leakage does.

\begin{figure}
  \includegraphics[width=\linewidth]{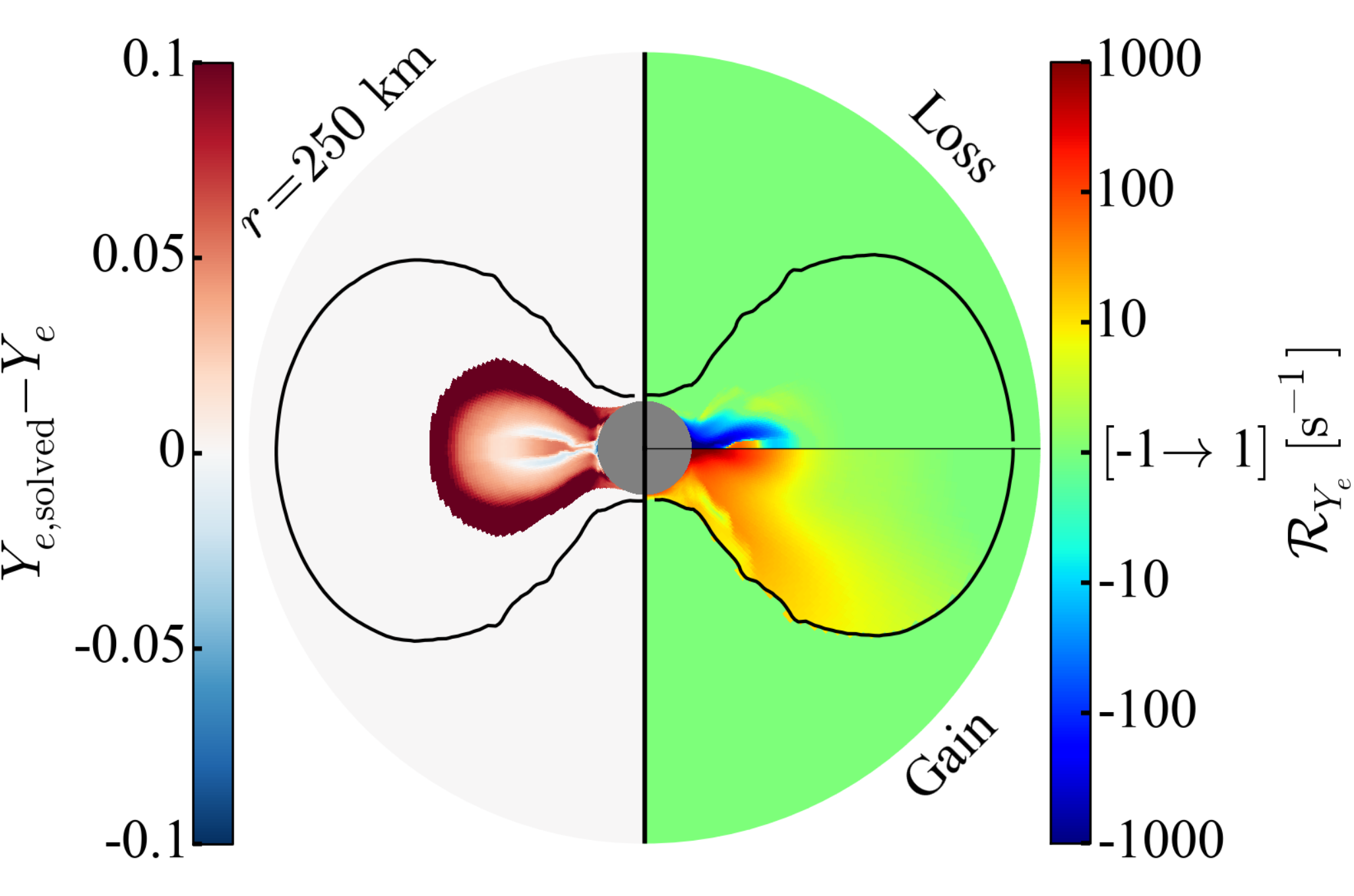}
\caption{\textbf{Equilibrium Electron Fraction (Central HMNS)} at $t=3\,\mathrm{ms}$. The dense region of the disk is far from equilibrium, as is the fluid in contact with the HMNS at high latitudes. The quantities shown are the same as in Figure~\ref{fig:BHequilibriumYe}. \textit{Left Hemisphere:} equilibrium electron fraction at which the net lepton number absorption rate is equal to the net lepton number emission rate. The equilibrium solver is unreliable below $1\,\mathrm{MeV}$ as the energy grid ceases to be able to resolve the neutrino distributions, so $Y_{e,\mathrm{solved}}-Y_e$ at locations with a temperature less than this is plotted as zero. \textit{Right Hemisphere:} rate of change of electron fraction, as in Figure~\ref{fig:BHinteraction}, but separately depicting that caused by emission (top) and absorption (bottom). The black curve is the $\rho=10^6\,\mathrm{g\ cm}^{-3}$ contour, below which {\tt Sedonu} opacities and emissivities are set to zero. The outer radius on the plot is at $250\,\mathrm{km}$ and the inner radius is $30\,\mathrm{km}$.}
\label{fig:HMNSequilibriumYe}
\end{figure}

We solve for the equilibrium electron fraction in Figure~\ref{fig:HMNSequilibriumYe}. The results are very similar to the BH case. The leakage neutrinos effectively interact more strongly than MC neutrinos do, which causes the fluid in the low-density polar regions to an increase in electron fraction through absorption of electron neutrinos from the HMNS in the dynamical simulation, bringing it closer to equilibrium. In the main region of the disk, the fluid is below the equilibrium electron fraction since there is a sufficiently large amount of mass that neutrinos have not yet been able to significantly raise the electron fraction. If significant neutrino processing occurs before the disk is formed (i.e., before the initial conditions of the dynamical simulations of MF14), the electron fraction would be higher and not as far from equilibrium.

\section{Neutrino Pair Annihilation}
\label{sec:annihil}
\begin{figure}
  \includegraphics[width=\linewidth]{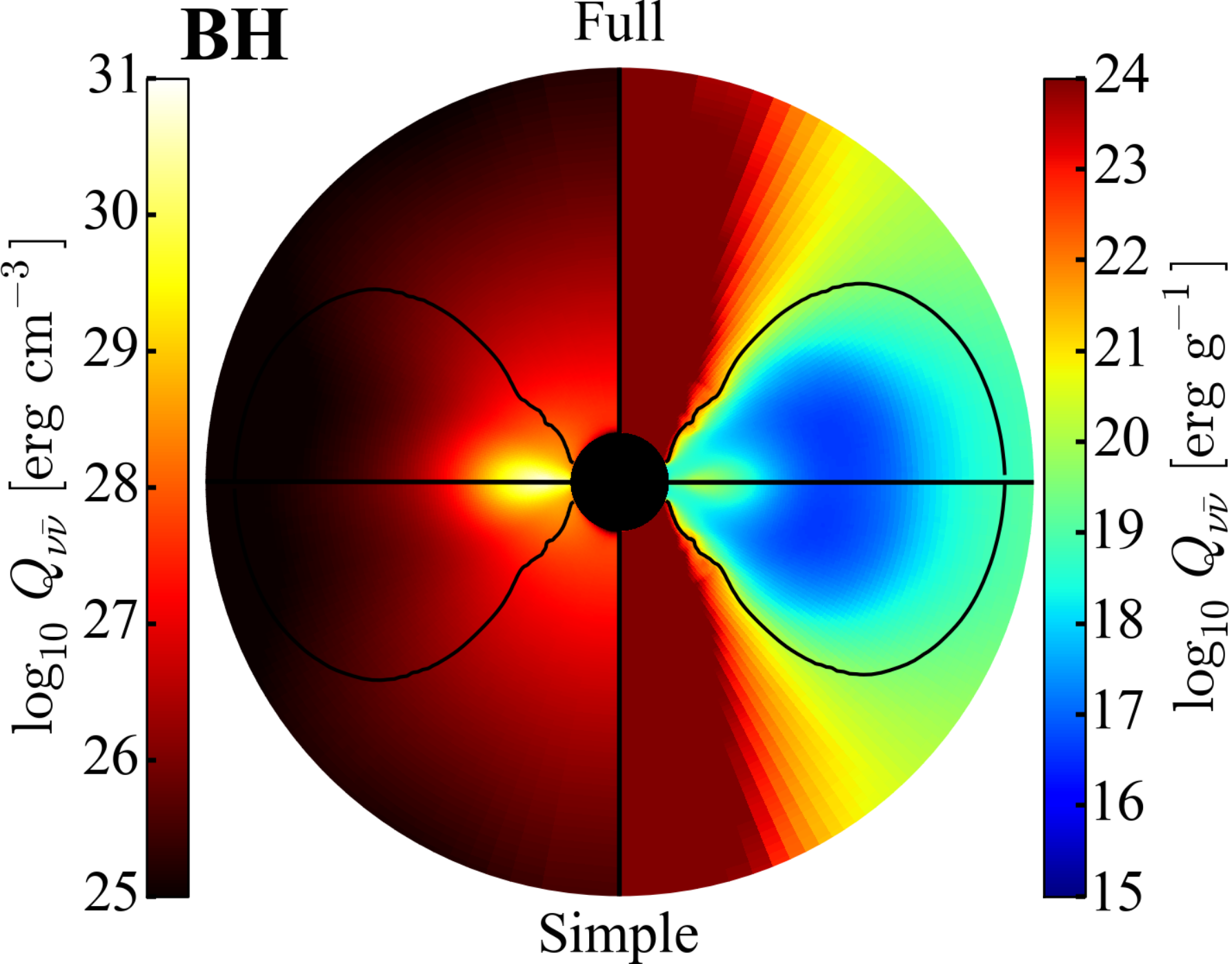}
  \includegraphics[width=\linewidth]{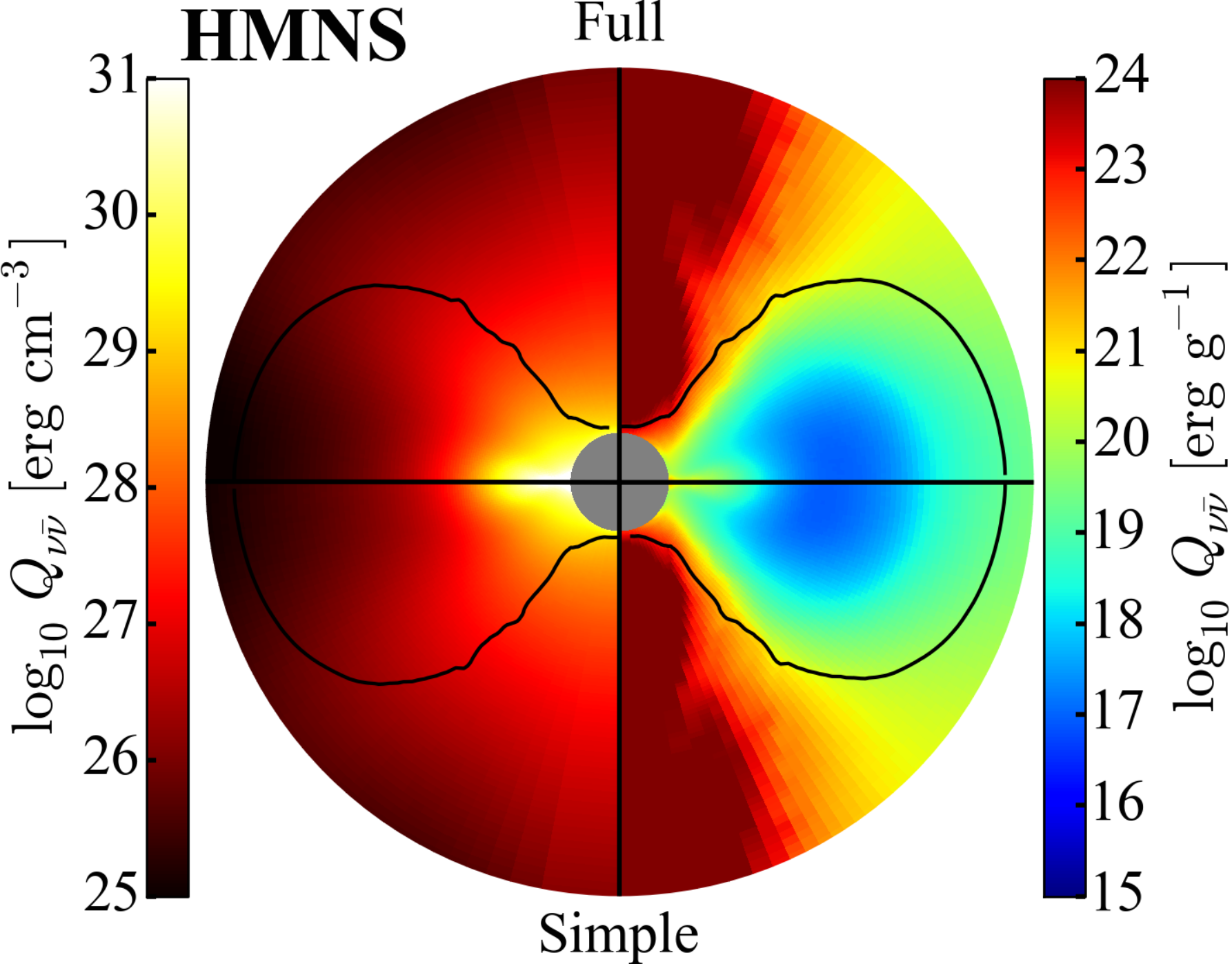}
  \caption{\textbf{Neutrino Annihilation Rates} at $t=3\,\mathrm{ms}$ for the BH case (top panel) and the HMNS case (bottom panel). \textit{Left hemisphere}: annihilation rate per unit volume. \textit{Right hemisphere}: the same annihilation rate per unit mass. \textit{Top hemisphere}: Full MC simulations. \textit{Bottom hemisphere}: Simple MC simulations. The outer curve is the $\rho=10^6\,\mathrm{g\ cm}^{-3}$ contour, below which {\tt Sedonu} opacities and emissivities are set to zero. The outer radius on the plot is at $250\,\mathrm{km}$ and the inner radius is $30\,\mathrm{km}$. Most annihilation occurs in the dense disk, but is most significant per unit mass along the poles.}
  \label{fig:annihilation}
\end{figure}

We calculate neutrino pair annihilation rates in a post-processing step after neutrinos have finished propagating through the disk. The resulting rates are plotted in Figure~\ref{fig:annihilation} for the $3\,\mathrm{ms}$ snapshots with both BH and HMNS backgrounds, and volume-integrated rates are given for every snapshot in Table~\ref{tab:results}. The NoPair simulations do not include pair processes in the NuLib tables, but the annihilation post-processing requires only neutrino distribution functions and does not rely on the NuLib tables. In the sparse polar regions, the density is low enough that annihilation would rapidly increase the temperature and entropy, which has the potential to generate a rapid outflow. However, annihilation accounts for at most $\sim4\%$ of the global energy gain/loss rate in any snapshot with either central object, and so will not dramatically affect the dynamics of most of the disk mass. In order to better estimate the role annihilation would have in driving a relativistic jet, we integrate only over regions less than $45^\circ$ from the poles. This excludes annihilation in the bulk of the disk and in regions far from the poles, which are either too dense or too far from the pole to contribute to acceleration along the poles. The ratio of the total annihilation to that just within $45^\circ$ of the poles can be as high as 120 in the BH snapshots (at $3\,\mathrm{ms}$) and 220 in the HMNS snapshots (at $30\,\mathrm{ms}$). Determining how much mass and energy is driven by annihilation, and whether this can actually produce a jet, would require including annihilation rates in the dynamical simulation.

An order-of-magnitude estimate of the total energy deposited in polar regions from neutrino pair annihilation can be found by time interpolating the volume-integrated values of annihilation rate given in Table~\ref{tab:results} and integrating assuming that $\mathcal{H}_{\nu\bar{\nu}}=C t^{k}$, where $k$ and $C$ are parameters set to create a piecewise-continuous interpolation. Integrating this interpolation for the Full physics simulations, we find that the total amount of energy deposited is $E_{\nu\bar{\nu},\mathrm{net}}=2.2\times10^{48}\,\mathrm{erg}$ after $300\,\mathrm{ms}$ for the BH case and $E_{\nu\bar{\nu},\mathrm{net}}=1.8\times10^{50}\,\mathrm{erg}$ after $3\,\mathrm{s}$ for the HMNS case. If for the reasons above we include only the volume within $45^\circ$ of the poles as in Table~\ref{tab:results}, the integrated deposited energy becomes $E_{\nu\bar{\nu},\mathrm{net}}=2.8\times10^{46}\,\mathrm{erg}$ for the BH case and $E_{\nu\bar{\nu},\mathrm{net}}=1.9\times10^{48}\,\mathrm{erg}$ for the HMNS case. However, $10^{48}-10^{50}\,\mathrm{erg\,s}^{-1}$ is required to launch a GRB jet (e.g.,\ \citealt{LR07}). Given the assumptions used in both the MC neutrino transport and the dynamical simulations of MF14, this calculation is certainly not accurate enough to definitively rule out the possibility of an annihilation-driven $e^+-e^-$ jet, but it is on the lower end of this energy requirement.

The effects of various approximations on the annihilation rate at $3\,\mathrm{ms}$ for both the BH and HMNS cases are also summarized in Table~\ref{tab:results}. Though special relativity increases the average neutrino energy, it also beams the neutrinos along similar trajectories in the azimuthal direction, which decreases the relative angle between neutrinos and causes the NoRel annihilation rate to be higher than the Full rate by  $\sim44\%$ in the BH snapshot and $\sim13\%$ in the HMNS snapshot. Scattering off of rapidly moving fluid boosts neutrino energies and so the annihilation rate in the NoScat simulation of the BH snapshot is $\sim5\%$ lower than in the Full simulation. Additionally, scattering causes neutrinos that would otherwise have passed straight through the disk to be deflected up toward the polar regions, and together with the increased neutrino energy this results in a $\sim18\%$ increase in the annihilation rate in the HMNS snapshot. The weak magnetism correction results in a smaller escape luminosity of electron anti-neutrinos, which decreases the annihilation rate by $\sim5\%$ in the BH snapshot and $\sim2\%$ in the HMNS snapshot. In the BH snapshots, so few heavy lepton neutrinos are produced that they provide essentially no additional contribution to pair annihilation rates, but the heavy lepton neutrinos emitted from an HMNS can increase the global annihilation rate by $\sim20\%$. These numbers apply only to the $3\,\mathrm{ms}$ snapshots, but indicate the direction and approximate relative magnitude of the effect each piece of physics has on the instantaneous annihilation rate.

\section{Discussion}
\label{sec:discussion}
Alhough MC neutrino transport is a less approximate treatment of neutrinos than leakage, it should be noted that there are still many approximations being made. The largest is the neglect of general relativity, which would red/blueshift neutrinos moving outward/inward, respectively, and would bend the neutrino trajectories along geodesics. Previous works have indicated that this following of geodesics causes neutrino trajectories to intersect at higher angles, increasing the annihilation rates by at most a factor of two \citep{AF00,AF01,Miller+03,Birkl+07,Harikae+10}. Moreover, the step between depositing neutrino energy and building a jet cannot be determined by our stationary simulations. Neutrino annihilation in the very sparse regions would also result in a very large entropy per baryon and thus a very efficient r-process even in matter that is barely neutron-rich, but the amount of mass in the polar regions is so small that the amount of r-process elements would be insignificant \citep{Fernandez+15a}.

Second, the annihilation kernels we use do not account for final-state electron and positron blocking, and are valid only for neutrinos with energies much larger than the electron rest mass (see Appendix~\ref{app:annihilation}). Third, neutrinos are fermions, and Pauli exclusion in regions where neutrinos are degenerate should affect their trajectories (e.g.,\ \citealt{Janka+92}). We account for the fermionic nature of neutrinos only in their interactions with matter and not in their propagation, but the very low degeneracy of the neutrinos makes this a good approximation. Fourth, scattering kernels are actually inelastic and anisotropic, but we treat them as elastic and isotropic, and include a correction factor to approximate ``effective'' anisotropic scattering (e.g.,\ \citealt{BRT06}), and we ignore inelasticity for the sake of simplicity. Since scattering opacities can have a significant impact on the neutrino radiation field, proper treatment of inelastic and anisotropic scattering could become an important source of energy deposition as, e.g.,\ in the context of core-collapse supernovae \citep{Lentz+12}. Finally, the opacities in the outskirts of the disk where $T<\sim0.5\,\mathrm{MeV}$ depend on the composition, which is likely not in NSE. However, addressing these approximations is beyond the current capabilities of {\tt Sedonu} and we defer to future work to evaluate their importance more carefully.

We see very significant differences in the cooling and leptonization rates between MC transport and leakage. The rate of energy loss predicted by MC transport is consistently larger than that predicted by leakage, but some systematic error is not unexpected from such an approximation and we expect the MC neutrinos to be out of equilibrium with the fluid evolved with the leakage scheme. In the BH snapshots, the global leptonization rates calculated by MC transport are $\sim7$ times larger than those calculated by leakage through the $30\,\mathrm{ms}$ snapshot. Since neutrino opacities scale roughly with the square of the neutrino energy (e.g.,\ \citealt{BRT06}), the leakage scheme used by MF14 attempts to account for the energy loss rate by calculating the optical depth based on the opacity of the fluid at the neutrino energy 
\begin{equation}
\begin{split}
\left<E_\nu\right> & = \sqrt{\frac{\int_0^\infty E_\nu^2 B_\nu(0,T)\,dE_\nu}{\int_0^\infty B_\nu(0,T)\,dE_\nu}} \\
 & = \sqrt{\mathcal{F}_5(0) / \mathcal{F}_3(0)} = 4.56\,k_BT\,\,,
\end{split}
\end{equation}
where $\mathcal{F}_n(\mu)$ are Fermi integrals of order $n$ Equation~\ref{eq:fermi}). Note that here we assume the chemical potential is zero, as do MF14. When applied to energy escape, this accounts for the fact that low-energy neutrinos are able to escape more easily than higher-energy neutrinos due to the scaling of the opacity with neutrino energy. However, this choice of energy is designed to properly account for energy loss, not lepton number change. If we repeat the same exercise, but replace $B_\nu(T)$ with $B_\nu(T)/E_\nu$ to represent number escape rather than energy escape, we get
\begin{equation}
\begin{split}
\left<E_\nu\right> & = \sqrt{\frac{\int_0^\infty E_\nu^2 B_\nu(0,T)\,dE_\nu/E_\nu}{\int_0^\infty B_\nu(0,T)\,dE_\nu/E_\nu}} \\
 & = \sqrt{\mathcal{F}_4(0) / \mathcal{F}_2(0)} = 3.59\,k_BT\,\,.
\end{split}
\end{equation}
Thus, the average energy to use when calculating opacities to account for lepton number loss is about 11\% smaller than that used when accounting for energy loss rate. Since the opacity for lower energy neutrinos is lower, using the same mean opacity for number and energy escape causes the leakage scheme to underestimate the number of escaping leptons. The leakage scheme could be made more consistent by calculating separate optical depths for neutrino energy and number escape \citep{Ruffert+96} or by calculating separate optical depths for each energy bin (e.g.,\ \citealt{Perego+14} in the context of the isotropic diffusion source approximation). Additionally, since this issue is a spectral effect rather than a geometric one, it could be accounted for in many of the more sophisticated energy-dependent transport schemes, such as spectral two-moment transport (e.g.,\ \citealt{Shibata+11,Cardall+13,Just+15a}).

One of the main differences between the effects of neutrinos calculated by leakage versus those calculated by MC transport is the amount of heating above the disk where densities are relatively low. MF14 argued that neutrinos are unable to drive a significant wind, but can affect the composition, especially above the disk, which increases the electron fraction of the viscously driven ejecta. To estimate the potential impact of the increased heating in our MC simulations, we look at the amount of mass $M_\nu$ for which neutrino heating dominates over viscous heating. $M_\nu$ is listed in Tables \ref{tab:flash} and \ref{tab:results}. In the BH snapshots, the leakage results indicate that essentially no mass is heated more strongly by neutrinos than by viscosity. However, for the first several tens of milliseconds, MC transport shows neutrinos being dynamically important in 11-50\% of the mass in same snapshots (though quickly approaching zero after that). In the HMNS snapshots, $M_\nu$ increases with time according to leakage ($\sim{8}\%$ of the disk mass at ${0}\,\mathrm{ms}$ to $\sim73\%$ at $3\,\mathrm{s}$), but decreases according to MC transport ($\sim4{2}\%$ at ${3}\,\mathrm{ms}$ to essentially none at $3\,\mathrm{s}$). This is largely due to the disk spreading out in the HMNS simulations such that much of the disk mass is below the minimum density for which neutrino interactions are accounted for in {\tt Sedonu}. Though essentially all of the mass is still above the minimum density at $30\,\mathrm{ms}$, by $300\,\mathrm{ms}$ 75\% of the disk mass is above the minimum density (average density is $10^{10}\,\mathrm{g\ cm}^{-3}$), and by $3\,\mathrm{s}$ only 0.09\% is above the minimum density (average density is $10^4\,\mathrm{g\ cm}^{-3}$).

There is still much to be done before predictive simulated kilonova light curves become available, but the differences we see between the leakage and MC results could have dramatic implications for the elements formed in the ejecta and the resulting light curve. Previous studies indicate that the production of heavy r-process elements requires electron fractions below $Y_e\sim0.2-0.3$ (e.g.,\ \citealt{Wanajo+14,Kasen+15}). Our Figures~\ref{fig:BHLateTime} and \ref{fig:HMNSLateTime} suggest that significant increases to the electron fraction of the disk ejecta are possible with Monte Carlo neutrino transport, since it results in the matter outside of the disk being more strongly neutrino processed. A weak r-process would still make elements up to $A\sim90$ in electron fractions up to $Y_e\sim0.4$ if the entropy is sufficiently high  \citep{Wanajo+14}. However, the lack of a strong r-process in the disk wind would imply a stronger early blue peak in the kilonova light curve if the merger is observed from a polar direction where the disk is not obscured by the lanthanide-rich dynamical ejecta (MF14;\citealt{Kasen+15}).

We can apply the interpolation and integration scheme we used in Section~\ref{sec:annihil} to the annihilation rates listed in Table~1 of \cite{Dessart+09} (assuming no BH spin) to estimate the total annihilation energy deposited for the $100\,\mathrm{ms}$ simulation to be $5.1\times10^{48}\,\mathrm{erg}$. This is about two orders of magnitude smaller than our estimate in Section~\ref{sec:annihil} using the entire domain in the HMNS case, but is very similar to the estimate using only the regions within $45^\circ$ of the poles. However, a direct comparison is somewhat difficult for the following reasons. (1) \cite{Dessart+09} do not have an inner boundary condition, (2) their HMNS is $\sim0.5M_\odot$ less massive than the one we assume, (3) the HMNS luminosity is an order of magnitude more luminous at $30\,\mathrm{ms}$, (4) the disk is $\sim7$ times more massive (with correspondingly higher densities), (5) they neglect viscous heating, and (6) they use a density cutoff of $\rho=10^{11}\,\mathrm{g\,cm}^{-3}$ for calculating annihilation rates rather than an angle from the poles. This cutoff serves to exclude the HMNS and dense inner disk from the annihilation calculations, since energy deposited there is effectively trapped and unable to contribute to outflows. Since our disk is so much less massive, all of our disk mass has density $\rho<10^{11}\,\mathrm{g\,cm}^{-3}$. Given the vast differences in the background fluid with which the annihilation calculations were performed, the differences in the annihilation rates are reasonable and to be expected.

The annihilation rate, however, depends on the product of the neutrino and anti-neutrino intensities, and so is sensitive to changes in the neutrino luminosity. Though \cite{Dessart+09} see neutrino luminosities similar to ours, other studies show somewhat higher luminosities of all species. \cite{Foucart+15} simulate a NS-BH merger and find electron neutrino luminosities of $\lesssim100\,\mathrm{B\ s}^{-1}$, electron anti-neutrino luminosities of $\lesssim300\,\mathrm{B\ s}^{-1}$ and collective heavy-lepton neutrino luminosities of $\lesssim100\,\mathrm{B\ s}^{-1}$ for a few tens of milliseconds. \cite{Sekiguchi+15} simulate a NS merger including the HMNS and see see similar luminosities over a similar time. In both cases, the luminosities are larger than ours by a factor of a few, which has the potential to increase the annihilation rate by an order of magnitude. If these luminosities are closer to those in nature than those computed by us and \cite{Dessart+09} and the geometry of the emission favors increased annihilation rates, there may yet be hope for the neutrino annihilation-powered GRB model.

Dynamical simulations with MC neutrino transport (or other methods more sophisticated than leakage) are required to determine the true long-term effects of the increased cooling and leptonization rates. For instance, although our results show that MC results in faster global cooling at all times, what may happen in a full simulation is faster cooling at early times and slower cooling at late times, since the disk will have become cold much faster. However, MC transport is currently too computationally expensive to be used at every timestep in a three-dimensional dynamical calculation. Other transport methods like energy-dependent two-moment transport (e.g.,\ \citealt{Thorne81,Shibata+11,Just+15a}) will be able to account for the spectral effects and many of the geometric ones, but in this approximation one must choose an otherwise undetermined closure relation to close the system of equations. Several physically-motivated analytic closures and variable Eddington factor methods (e.g.,\ \citealt{Cardall+13}) have been proposed, but any method with a local closure (i.e., one that is determined only by the radiation in the current grid cell) introduces a nonlinearity into the transport equation that leads to unphysical radiation shocks (e.g.,\ \citealt{Olson+00}). In the future, it may be possible to find a closure treatment that maintains the efficiency of the two-moment transport scheme while accurately reproducing Monte Carlo results.

\section{conclusions}
\label{sec:conclusions}
We present the new open-source, steady-state, special relativistic Monte Carlo neutrino transport code {\tt Sedonu}. It efficiently calculates the energy- and angle-dependent neutrino distribution function on multi-dimensional fluid backgrounds and solves for the equilibrium fluid temperature and electron fraction. We have simulated neutrino transport through snapshots of post-merger disks using Monte Carlo (MC) techniques with various elements of physics. We compare the results to the leakage scheme used in the original dynamical simulations by MF14. Since the Monte Carlo neutrinos are out of equilibrium with the fluid evolved with the leakage scheme in the dynamical calculations, we believe that the qualitative trends we indicate are robust. However, determining the magnitudes of the differences between the two methods would require MC transport to be coupled to the fluid evolution. In light of this, we summarize our findings below.

\begin{enumerate}
  \item Compared with leakage, MC transport results in global cooling and leptonization rates that are higher than those predicted by leakage during the optically-thick disk stage. If the disk is optically thin with a central BH, MC cooling is slower and leptonization is faster than leakage. If the disk is optically thin with a central HMNS, MC cooling is faster and leptonization is slower than leakage. This suggests a stronger blue component of the kilonova light curve if viewed from polar angles.
  \item MC exhibits up to an order of magnitude stronger neutrino heating above the disk that could increase the strength of a neutrino-driven wind.
  \item In the disk midplane, cooling via MC neutrinos dominates viscous heating through the $30\,\mathrm{ms}$ snapshot with either central object, in constrast to the leakage results.
  \item The neutrino radiation field at large radii is very asymmetric, with most of the radiation escaping around $45^\circ$ from the equator.
  \item The peak energies of the neutrino distribution functions can be shifted by a few MeV higher ($\nu_e,\bar{\nu}_e$) or lower ($\nu_x$) from the peak of a zero-chemical potential blackbody with the same average energy and flux.
  \item Neutrino pair annihilation deposits an order of magnitude more energy with a central HMNS ($\sim1.9\times10^{48}\,\mathrm{erg}$) than with a central BH ($\sim2.8\times10^{46}\,\mathrm{erg}$), though this is still unlikely to be sufficient to drive a jet, though higher neutrino luminosities could make it plausible.
  \item Special relativity increases the average energies of escaping neutrinos by around $1-3\,\mathrm{MeV}$ and beams higher-energy neutrinos away from the poles. The inclusion of heavy lepton neutrinos, pair processes, scattering, weak magnetism, and variations in the equation of state have together at most a 10\% effect on the integrated cooling and leptonization rates.
\end{enumerate}

\acknowledgements We thank A.~Burrows, M.~Duez, F.~Foucart,
L.~Roberts, J.~Lippuner, E.~Murchikova, and T.~Urbatsch for helpful
discussions and Rollin Thomas for an interface to the Lua library.
SR is supported by a DOE Computational Science Graduate
Fellowship under grant number DE-FG02-97ER25308.  SR and CDO
acknowledge support by the National Science Foundation under awards
AST-1205732, AST-1333520, and PHY-1151197, by the Sherman Fairchild
Foundation, and by the Los Alamos National Laboratory Institute for
Geophysics, Planetary Physics and Signatures. EO acknowledges support
from NASA through Hubble Fellowship grant \#51344.001-A awarded by the
Space Telescope Science Institute, which is operated by the
Association of Universities for Research in Astronomy, Inc., for NASA,
under contract NAS 5-26555. RF acknowledges support from the
University of California Office of the President, and from NSF award
AST-1206097. This research used computing and storage resources (repo
m2058) provided by the National Energy Research Scientific Computing
Center (NERSC), which is supported by the Office of Science of the
U.S. Department of Energy under Contract No. DE-AC02-05CH11231. Parts
of the computations were also performed on the Caltech compute cluster
Zwicky (NSF MRI-R2 award PHY-0960291), on the NSF XSEDE network
under allocation TG-PHY100033, and on NSF/NCSA Blue Waters under NSF
PRAC award ACI-1440083.


\appendix
\section{Microphysics Resolution Study}
\label{app:resolution}
To ensure that our results are robust against our choices of numerical discretization, we repeat transport simulations for the $3\,\mathrm{ms}$ snapshots with different discretizations, all using the Full set of physics, and compare to the original simulations in Table~\ref{tab:res_study}. The discretization of the NuLib tables was tested by in turn doubling the neutrino energy, matter density, matter temperature, and electron fraction grid resolution. We also in turn double the angular resolution of the neutrino distribution functions in each cell, which only has an effect on the annihilation rates. Finally, we double the number of steps each neutrino packet must take by changing $d_\mathrm{cell}$ to be $0.2$ rather than $0.4$ times the cell's smallest dimension (see Section~\ref{sec:methods} for details). 

The differences between simulations with enhanced microphysics resolution in Table~\ref{tab:res_study} and the originals in Table~\ref{tab:results} are all small, indicating that our discretization is sufficiently fine. Increasing the angular resolution of the annihilation kernels causes annihilation rates to drop slightly, supporting the supposition that most of the annihilation is from small incident angles.

Increasing the number of Monte Carlo neutrino packets does not introduce any systematic change, but rather only reduces the size of random fluctuations in results when the same simulation is run multiple times. We use $2-4\times10^7$ packets in order to keep random fluctuations of the results in Table~\ref{tab:results} at $\sim0.1\%$.

\section{code tests}
We perform a pair of related tests to ensure that our transport and equilibrium finding methods arrive at the correct answer. In the first test, we demonstrate that we are able to reach a blackbody distribution function at high optical depth. In the second, we demonstrate that our equilibrium solver settles to the correct values of temperature and electron fraction.

\subsection{Blackbody Generation}
\label{app:blackbody}
In this test we use a single unit-volume fluid cell with periodic boundary conditions. We give the cell a density, temperature, and electron fraction, and observe the resulting neutrino radiation that builds up within the cell, similar to \cite{Tubbs78}. The neutrinos should settle into a Fermi blackbody distribution given by
\begin{equation}
\label{eq:blackbody}
  B_{E_\nu}(\mu,T) = \frac{E_\nu^3/h^3c^2}{e^{(E_\nu-\mu)/kT} + 1}
\end{equation}
in CGS units of $\mathrm{erg\,s}^{-1}\mathrm{cm}^{-2}\mathrm{MeV}^{-1}\mathrm{sr}^{-1}$, where $E_\nu$ is the neutrino energy (different from the packet energy $E_p$ mentioned in the main text), $\mu$ is the neutrino chemical potential, and $T$ is the fluid temperature. This is identical to the photon blackbody, with two exceptions. The sign in the denominator originating from the Fermi-Dirac distribution, and there is no factor of 2 in the numerator because only left-handed neutrinos have been observed. Integrating over neutrino energy and angle gives the blackbody energy density
\begin{equation}
\label{eq:BBendens}
  \epsilon_{\nu,i} = \frac{4\pi}{c}\int_0^\infty B_{E_\nu}(\mu,T)\,dE_\nu = \frac{4\pi}{(hc)^3} \mathcal{F}_3(\mu)\,\,,
\end{equation}
where
\begin{equation}
\label{eq:fermi}
\mathcal{F}_n(\mu) = \int_0^\infty \frac{E_\nu^n\,dE_\nu}{e^{(E_\nu-\mu)/kT}+1}
\end{equation}
are Fermi integrals of order n. In the special case of $\mu=0$, this has a simple analytical solution akin to the Stefan-Boltzmann law,
\begin{equation}
  \epsilon_{\nu,i} = \frac{7 \pi^5 (kT)^4}{30 (hc)^3}\,\,.
\end{equation}

\begin{figure}
  \includegraphics[width=\linewidth]{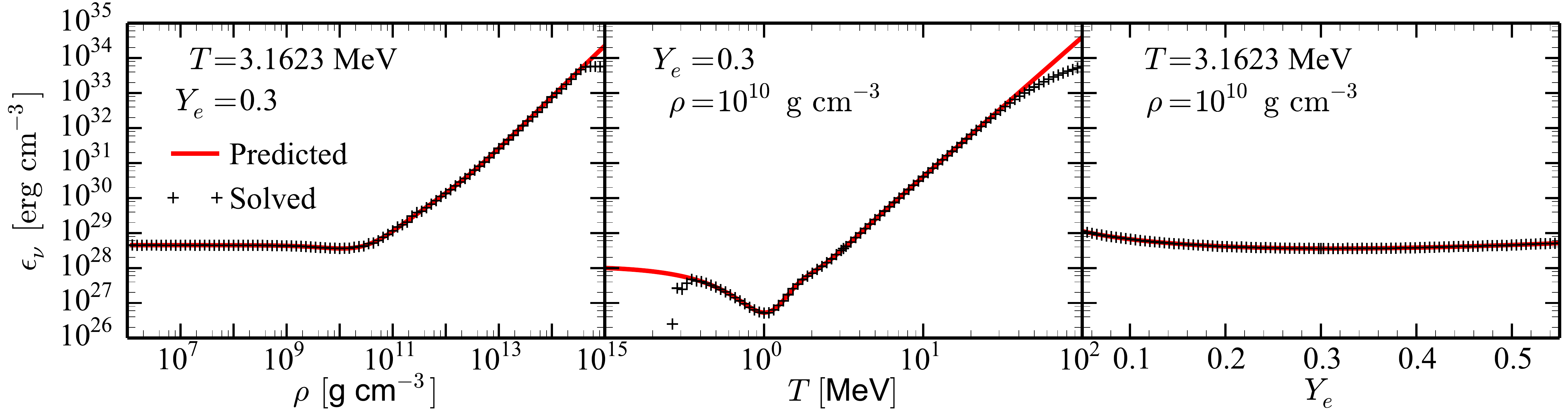}
  \caption{\textbf{Blackbody Generation Test:} The equilibrium total neutrino energy density $\epsilon_\nu$ as a function of fluid density $\rho$, temperature $T$, and electron fraction $Y_e$. The fluid variables are varied around $\rho=10^{10}\,\mathrm{g\ cm}^{-3}$, $T=3.1623\,\mathrm{MeV}$, and $Y_e=0.3$. The {\tt Sedonu}-calculated energy densities (black crosses) match the theoretical ones (red lines) in a wide range of fluid conditions. Regions of mismatch at low and high temperatures occur as the peak of the neutrino distribution approaches the low and high neutrino energy limits of the NuLib tables.\\}
  \label{fig:test_blackbody}
\end{figure}

We let the fluid emit neutrinos and allow them to propagate until they are absorbed and compare the resulting neutrino energy density to Equation~\ref{eq:BBendens}. Because scattering opacities become much larger than absorption opacities at high densities, we use the NoScat physics for efficiency. We expect the net energy density to be the sum of contributions from each of the 6 neutrino species, constrained by $\mu_{\bar{\nu}_e} = -\mu_{\nu_e}$ and $\mu_{\nu_x} = 0$, where $\nu_x$ represents any of the four heavy lepton neutrino/anti-neutrino species. The equilibrium $\mu_{\nu_e}$ can be taken directly from the EOS for a given $\{\rho,T,Y_e\}$. Figure~\ref{fig:test_blackbody} demonstrates the match between the predicted energy density and that calculated by {\tt Sedonu} for many values in each direction of $\{\rho,T,Y_e\}$. The plots appear very similar for all equations of state, but we use the Helmholtz EOS to complement the results plotted in the main text. The computed and theoretical results disagree at low and high temperatures, where the neutrino distribution functions extend past the energy limits in the NuLib tables.

\subsection{Blackbody Irradiation}
\label{app:irradiation}
Rather than determining what radiation field is in equilibrium with the input fluid properties as in the previous test, we determine what fluid properties are in equilibrium with the input radiation. That is, we solve for the equilibrium properties of fluid that is allowed to relax in a bath of blackbody neutrinos. We set up a thin shell of fluid ($dr/r = 10^{-4}$), apply a reflective outer boundary condition, and emit blackbody neutrinos from an absorbing inner boundary specified by a neutrino temperature $T_\mathrm{input}$ and electron neutrino chemical potential $\mu_{\nu_e,\mathrm{input}}$. The chemical potentials of all other neutrino species satisfy the constraints detailed in the previous test, namely that $\mu_{\bar{\nu}_e,\mathrm{input}}=-\mu_{\nu_e,\mathrm{input}}$ and $\mu_{\nu_x,\mathrm{input}}=0$. Unlike in the main text, the luminosity of each species $s$ is determined by the input temperature and chemical potential of the blackbody neutrinos being emitted from the inner boundary, such that
\begin{equation}
\mathcal{L}_s=4\pi^2 r^2 \sum\limits_{E_{\nu,i}}B_{E_{\nu,i}}(\mu_{s,\mathrm{input}},T_\mathrm{input}) \Delta E_{\nu,i}\,\,,
\end{equation}
where $r$ is the radius of the inner boundary, $E_{\nu,i}$ is the center of energy bin $i$, and $\Delta E_{\nu,i}$ is the width of energy bin $i$. 

We then iteratively relax the fluid to its equilibrium temperature and electron fraction as described in the following. Each iteration is done as described in Section~\ref{sec:methods}. If we are solving for temperature, we then set the temperature of each grid cell to $T_{i+1}=T_i + d\Delta T_i$, where $T_{i+1}$ and $T_i$ are the temperatures for the iterations $i+1$ and $i$, respectively, $d=0.3$ is a somewhat arbitrary damping factor, and $\Delta T_i$ is the difference between the temperature and the equilibrium temperature. The same process is also applied to the electron fraction if we are solving for it. Then the new fluid properties are used in another transport and solve iteration. The results presented here are the result of 20 such iterations.

\begin{figure}
  \includegraphics[width=\linewidth]{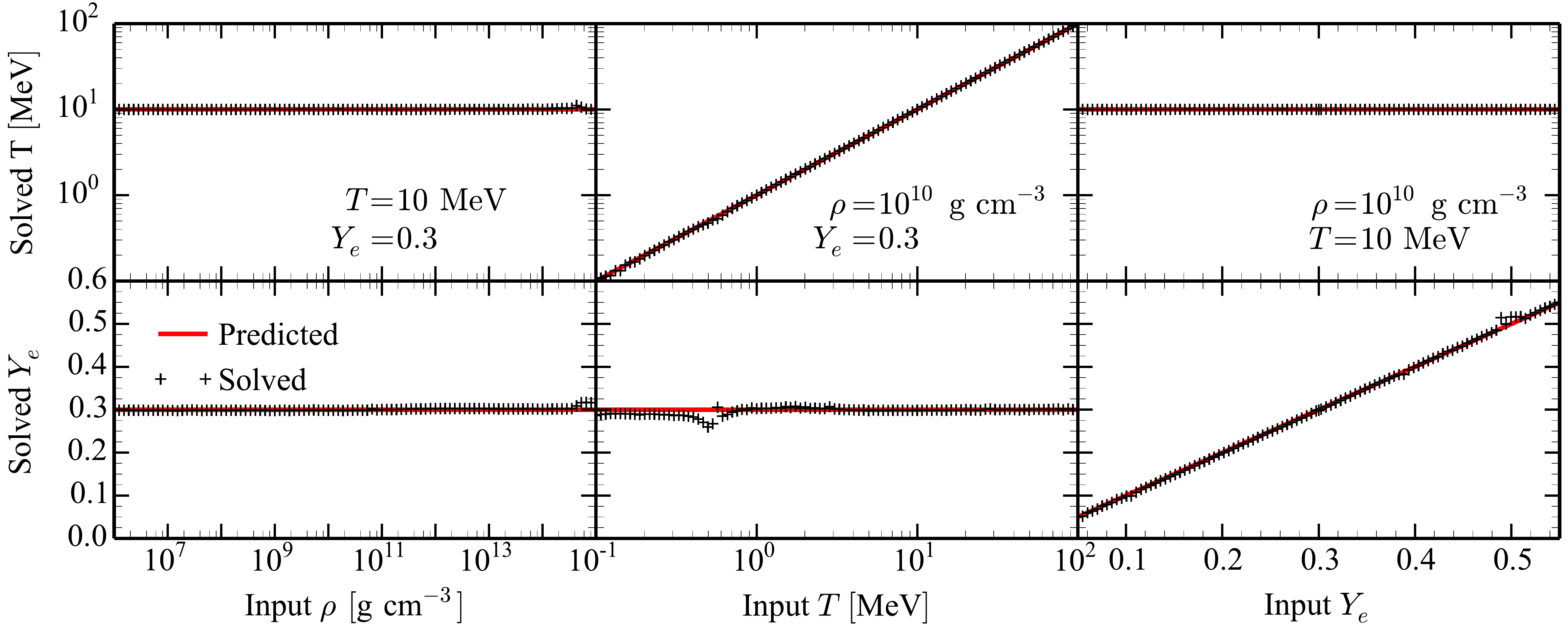}
  \caption{\textbf{Blackbody Irradiation Test:} The equilibrium fluid temperature ($\mathrm{Solved}\,T$) and electron fraction ($\mathrm{Solved}\,Y_e$) as a function of fluid density ($\mathrm{Input}\,\rho$), neutrino temperature ($\mathrm{Input}\,T$), and neutrino chemical potential (via the proxy $\mathrm{Input}\,Y_e$). The three top panels show results where only fluid temperature is solved for, and the three bottom panels show results where only fluid electron fraction is solved for. The {\tt Sedonu}-calculated energy densities (black crosses) match the theoretical ones (red lines) in a wide range of fluid conditions. The equilibrium solver has difficulty converging on an electron fraction when the chemical potential is large relative to the temperature (i.e. high $\rho$ or low $T$), since the neutrino Fermi-Dirac distributions become too sharp to be resolved by the NuLib tables.\\}
  \label{fig:test_irradiation}
\end{figure}

We expect that the equilibrium temperature $T_\mathrm{eq}$ and electron fraction $Y_{e,\mathrm{eq}}$ should settle to values such that $\mu_{\nu_e,\mathrm{EOS}}(\rho,T_\mathrm{eq},Y_{e,\mathrm{eq}})=\mu_{\nu_e,\mathrm{input}}$ and $T_\mathrm{eq}=T_\mathrm{input}$, where $\mu_{\nu_e,\mathrm{EOS}}=\mu_e+\mu_p-\mu_n$ is given by the EOS. Figure~\ref{fig:test_irradiation} demonstrates the estimated and calculated equilibrium temperature and electron fraction at several values of $\{\rho,T_\mathrm{input},Y_{e,\mathrm{input}}\}$, where $Y_{e,\mathrm{input}}$ is a proxy for the chemical potential inputs, such that $\mu_{\nu_e,\mathrm{EOS}}(\rho,T_\mathrm{input},Y_{e,\mathrm{input}})=\mu_{\nu_e,\mathrm{input}}$. The equilibrium values of $T$ and $Y_e$ are each determined with independent iterative calculations. There are regions at low temperatures and high densities where the correct solution is not found, as in the previous test. This is a combination of inadequate energy resolution and range in the NuLib tables, as the peaks of the emissivity spectra approach either end of the energy range or become too sharp to resolve. When equilibrium temperature and electron fraction are calculated simultaneously (not plotted), the now two-dimensional solver is less robust and only consistently reaches the correct solution when $\rho\lesssim3\times10^{13}\,\mathrm{g\,cm}^{-3}$ and $T\gtrsim 2\,\mathrm{MeV}$. 

\section{Neutrino Pair Annihilation}
\label{app:annihilation}
In each grid cell, {\tt Sedonu} records the neutrino distribution function by accumulating neutrino energy density in bins of neutrino species, neutrino energy, and direction. We use this information to calculate neutrino annihilation rates. The derivation here assumes that the resulting electrons and positrons are extremely relativistic, i.e., that the incoming neutrino energies are much larger than the sum of the electron and positron rest masses. This assumption is well justified for most of the neutrino energy range we consider.

The general neutrino annihilation rate is given by \cite{Ruffert+97}, equation 1, representing the rate of energy deposition from the annihilation of neutrinos and anti-neutrinos, which we reproduce here for completeness:
\begin{equation}
\label{eq:Q+vac}
\begin{split}
Q^+_\mathrm{ann} = \frac{\sigma_0 c}{4(m_e c^2)^2}\int_0^\infty dE_\nu \int_0^\infty d\bar{E}_\nu \oint_{4\pi}d\Omega \oint_{4\pi}d\bar{\Omega} & \frac{fE_\nu^3}{(hc)^3} \frac{\bar{f}\bar{E}_\nu^3}{(hc)^3} \left(E_\nu+\bar{E}_\nu\right) \\
& \times \left[\frac{C_1+C_2}{3} (1-\cos\theta)^2 + C_3\frac{(m_e c^2)^2}{E_\nu\bar{E}_\nu} (1-\cos\theta)\right]\,\,.
\end{split}
\end{equation}
Barred quantities refer to the anti-neutrino species, $E_\nu$ is the neutrino energy, $m_e$ is the electron mass, $c$ is the speed of light, $\theta$ is the angle between the two incoming neutrinos, and $\sigma_0=1.76\times 10^{-44}\,\mathrm{cm^2}$ is the fiducial weak interaction cross section. The weak coupling constants depend on the neutrino species that is annihilating. For electron neutrino and electron anti-neutrino (other species) annihilation, $C_1+C_2\approx2.34\,(0.50)$ and $C_3\approx1.06\,(-0.16)$. $f$ is the phase space distribution function (values lie between 0 and 1) and $f\nu^3/c^3$ is the neutrino energy density per unit neutrino energy per steradian of direction. The latter quantity only differs by a factor of c from the specific intensity $I_\nu$ used in \cite{Dessart+09}. For simplicity, we perform a first-order numerical integral, and assume the energies and directions are confined to delta functions at the energy/direction bin centers. Applying this assumption, we arrive at
\begin{equation}
Q^+_\mathrm{ann} = \frac{\sigma_0 c}{4(m_e c^2)^2}\sum\limits_{E_{\nu,i}} \sum\limits_{\bar{E}_{\nu,k}} \sum\limits_{\Omega_j} \sum\limits_{\bar{\Omega}_l} \epsilon_{\nu,ij}\bar{\epsilon}_{\nu,kl} (E_{\nu,i} + \bar{E}_{\nu,k}) \left[\frac{C_1+C_2}{3} (1-\cos\theta_{jl})^2 + C_3\frac{(m_e c^2)^2}{E_{\nu,i}\bar{E}_{\nu,k}} (1-\cos\theta_{jl})\right],
\end{equation}
where $E_{\nu,i}$ and $\bar{E}_{\nu,k}$ are energy bin centers for the neutrino and anti-neutrino species, respectively. $\epsilon_{\nu,ij}$ and $\bar{\epsilon}_{\nu,kl}$ are the integrated (i.e. total measured in the transport simulation) energy density in the corresponding energy/direction bin. $\theta_{jl}$ is the angle between the centers of direction bins $j$ and $l$. This must then be summed over all three neutrino species--anti-species pairs to get the total energy deposition rate.

Recall that we differ slightly from this formalism in that we subtract the outgoing lepton masses from the incoming neutrino energies to de-emphasize low-energy neutrino annihilations that do not fit the assumption of high-energy neutrinos used in the original derivation. One can subtract the electron masses from the incoming neutrino energies and match the result to Equation~\ref{eq:annihil} to see that the annihilation kernel is
\begin{equation}
  R_{jl}(E_{\nu,i},\bar{E}_{\nu,k},\theta_{jl}) = \frac{\sigma_0 c}{4(m_e c^2)^2}\left[E_{\nu,i}\bar{E}_{\nu,k}\frac{C_1+C_2}{3} (1-\cos\theta_{jl})^2 + C_3(m_e c^2)^2 (1-\cos\theta_{jl})\right]\,\,.
\end{equation}

\clearpage
\begin{turnpage}
\begin{deluxetable*}{cccccccccccccccccc}
\tablecaption{Results - Global Quantities}
\tablecolumns{18}
\tablewidth{9.5in}
\startdata
\tablehead{\colhead{Physics} & \colhead{Time} & \colhead{$\mathcal{C}_\nu-\mathcal{H}_\nu$} & \colhead{$\left<\frac{dY_e}{dt}\right>$} & \colhead{$M_\nu$}            & \multicolumn{3}{c}{$\mathcal{L}_\mathrm{emit}\,(\mathrm{B\ s}^{-1})$} & \multicolumn{3}{c}{$\mathcal{L}_\mathrm{escape}\,(\mathrm{B\ s}^{-1})$} & \multicolumn{3}{c}{$\left<E_{\nu,\mathrm{emit}}\right>\,\mathrm{(MeV)}$} & \multicolumn{3}{c}{$\left<E_{\nu,\mathrm{escape}}\right>\,\mathrm{(MeV)}$} & \colhead{$\mathcal{H}_{\nu\bar{\nu}}$} \\
                             & \colhead{(ms)} & \colhead{$(\mathrm{B\ s}^{-1})$}            & \colhead{$(\mathrm{s}^{-1})$}            & \colhead{$(M_\odot)$} & \colhead{$\nu_e$} & \colhead{$\bar{\nu}_e$} & \colhead{$\nu_x$}         & \colhead{$\nu_e$} & \colhead{$\bar{\nu}_e$} & \colhead{$\nu_x$}           & \colhead{$\nu_e$} & \colhead{$\bar{\nu}_e$} & \colhead{$\nu_x$}            & \colhead{$\nu_e$} & \colhead{$\bar{\nu}_e$} & \colhead{$\nu_x$}      & \colhead{$(\mathrm{B\ s}^{-1})$}     }
\cutinhead{BH Disk Global Quantities}\\
       & 0   & \phantom{-}8.{39}(1)\phantom{-} & \phantom{-}5.6{6}(1)\phantom{-} & {7.56(-3)} & 5.4{6}(1)\phantom{-} & 1.32(2)\phantom{-} & 1.06\phantom{(-1)} & 4.9{6}\phantom{(-1)} & 7.1{3}(1)\phantom{-} & 9.{76}(-1) & 27.3\phantom{3} & 27.6\phantom{3} & 23.8\phantom{3} & 14.5\phantom{3} & 22.8\phantom{3} & 23.8\phantom{3} & 6.2{8}(-3)\phantom{3} \\
       & 3   & \phantom{-}4.54(1)\phantom{-} & \phantom{-}1.89(1)\phantom{-} & 3.15(-3)           & 1.10(2)\phantom{-} & 7.15(1)\phantom{-} & 8.70(-1)             & 7.63\phantom{(-1)} & 3.12(1)\phantom{-} & 7.74(-1)             & 29.1\phantom{3} & 29.0\phantom{3} & 26.1\phantom{3} & 14.2\phantom{3} & 21.9\phantom{3} & 26.0\phantom{3} & 3.85(-3)\phantom{3} \\ 
Full   & 30  & \phantom{-}4.50\phantom{(-1)} & \phantom{-}3.27\phantom{(-1)} & 2.17(-8)           & 3.33\phantom{(-1)} & 3.54\phantom{(-1)} & 2.34(-2)             & 1.16\phantom{(-1)} & 2.81\phantom{(-1)} & 2.08(-2)             & 19.0\phantom{3} & 18.8\phantom{3} & 16.0\phantom{3} & 13.9\phantom{3} & 17.6\phantom{3} & 15.9\phantom{3} & 2.82(-5)\phantom{3} \\
       & 300 & \phantom{-}1.70(-2)           & \phantom{-}1.90(-1)           & 0.00\phantom{(-1)} & 2.16(-3)           & 1.44(-2)           & 5.46(-4)             & 2.03(-3)           & 1.38(-2)           & 5.15(-4)             & \phantom{3}9.22 & \phantom{3}9.13 & \phantom{3}8.03 & \phantom{3}9.13 & \phantom{3}9.07 & \phantom{3}7.98 & 7.85(-11)           \\ \\ \hline \\
       & 0   & \phantom{-}9.6{2}(1)\phantom{-} & \phantom{-}6.9{3}(1)\phantom{-} & {8.07(-3)} & 5.0{6}(1)\phantom{-} & 1.5{1}(2)\phantom{-} & \nodata\phantom{(-)} & 5.2{5}\phantom{(-2)} & 8.2{3}(1)\phantom{-} & \nodata\phantom{(-)} & 24.8\phantom{2} & 26.0\phantom{2} & \nodata & 14.2\phantom{2} & 22.1\phantom{2} & \nodata & 8.9{3}(-3)\phantom{2} \\
       & 3   & \phantom{-}5.12(1)\phantom{-} & \phantom{-}2.40(1)\phantom{-} & 1.17(-3)           & 1.01(2)\phantom{-} & 8.16(1)\phantom{-} & \nodata\phantom{(-)} & 8.35\phantom{(-1)} & 3.63(1)\phantom{-} & \nodata\phantom{(-)} & 25.7\phantom{3} & 26.9\phantom{3} & \nodata         & 13.9\phantom{3} & 21.1\phantom{3} & \nodata         & 6.06(-3)\phantom{3} \\
Simple & 30  & \phantom{-}4.65\phantom{(-1)} & \phantom{-}3.98\phantom{(-1)} & 0.00\phantom{(-1)} & 3.05\phantom{(-1)} & 3.74\phantom{(-1)} & \nodata\phantom{(-)} & 1.15\phantom{(-1)} & 2.94\phantom{(-1)} & \nodata\phantom{(-)} & 16.7\phantom{3} & 17.0\phantom{3} & \nodata         & 13.0\phantom{3} & 16.0\phantom{3} & \nodata         & 3.77(-5)\phantom{3} \\
       & 300 & \phantom{-}1.69(-2)           & \phantom{-}2.04(-1)           & 0.00\phantom{(-1)} & 2.05(-3)           & 1.48(-2)           & \nodata\phantom{(-)} & 1.92(-3)           & 1.42(-2)           & \nodata\phantom{(-)} & \phantom{3}8.49 & \phantom{3}8.66 & \nodata         & \phantom{3}8.40 & \phantom{3}8.58 & \nodata         & 9.90(-11)           \\ \\  \hline \\
NoPair & 3   & \phantom{-}4.45(1)\phantom{-} & \phantom{-}1.89(1)\phantom{-} & 3.15(-3)           & 1.10(2)\phantom{-} & 7.15(1)\phantom{-} & \nodata\phantom{(-)} & 7.62\phantom{(-1)} & 3.12(1)\phantom{-} & \nodata\phantom{(-)} & 29.1\phantom{3} & 29.0\phantom{3} & \nodata         & 14.2\phantom{3} & 21.9\phantom{3} & \nodata         & 3.85(-3)\phantom{3} \\
NoScat & 3   & \phantom{-}5.03(1)\phantom{-} & \phantom{-}2.07(1)\phantom{-} & 6.13(-3)           & 1.10(2)\phantom{-} & 7.15(1)\phantom{-} & 8.71(-1)             & 8.27\phantom{(-1)} & 3.54(1)\phantom{-} & 7.90(-1)             & 29.1\phantom{3} & 29.1\phantom{3} & 26.2\phantom{3} & 14.6\phantom{3} & 23.3\phantom{3} & 26.3\phantom{3} & 3.65(-3)\phantom{3} \\
NoRel  & 3   & \phantom{-}4.47(1)\phantom{-} & \phantom{-}1.96(1)\phantom{-} & 5.30(-4)           & 1.03(2)\phantom{-} & 6.70(1)\phantom{-} & 8.16(-1)             & 7.66\phantom{(-1)} & 3.01(1)\phantom{-} & 7.06(-1)             & 25.8\phantom{3} & 26.0\phantom{3} & 23.3\phantom{3} & 13.5\phantom{3} & 20.3\phantom{3} & 22.9\phantom{3} & 5.54(-3)\phantom{3} \\
NoWM   & 3   & \phantom{-}4.68(1)\phantom{-} & \phantom{-}2.05(1)\phantom{-} & 3.83(-3)           & 1.08(2)\phantom{-} & 8.71(1)\phantom{-} & 8.68(-1)             & 7.65\phantom{(-1)} & 3.23(1)\phantom{-} & 7.69(-1)             & 28.9\phantom{3} & 30.1\phantom{3} & 26.1\phantom{3} & 14.3\phantom{3} & 21.2\phantom{3} & 25.8\phantom{3} & 4.04(-3)\phantom{3} \\
Shen   & 3   & \phantom{-}4.53(1)\phantom{-} & \phantom{-}1.89(1)\phantom{-} & 2.98(-3)           & 1.08(2)\phantom{-} & 7.07(1)\phantom{-} & 8.71(-1)             & 7.58\phantom{(-1)} & 3.12(1)\phantom{-} & 7.75(-1)             & 29.1\phantom{3} & 29.0\phantom{3} & 26.1\phantom{3} & 14.2\phantom{3} & 22.0\phantom{3} & 25.9\phantom{3} & 3.82(-3)\phantom{3} \\
LS220  & 3   & \phantom{-}4.52(1)\phantom{-} & \phantom{-}1.88(1)\phantom{-} & 3.50(-3)           & 1.09(2)\phantom{-} & 7.07(1)\phantom{-} & 8.71(-1)             & 7.58\phantom{(-1)} & 3.11(1)\phantom{-} & 7.73(-1)             & 29.1\phantom{3} & 29.0\phantom{3} & 26.2\phantom{3} & 14.2\phantom{3} & 21.9\phantom{3} & 25.9\phantom{3} & 3.84(-3)\phantom{3} \\
\cutinhead{HMNS Disk Global Quantities}\\
       & 0    & \phantom{-}{7.28}(1)\phantom{-} & \phantom{-}{5.78}(1)\phantom{-} & 1.{06}(-2) & 5.4{5}(1)\phantom{-} & 1.32(2)\phantom{-} & 1.06\phantom{(-1)} & 1.49(1) & 8.5{7}(1) & 7.08(1) & 27.3\phantom{3} & 27.6\phantom{3} & 23.9\phantom{3} & 15.0 & 22.4 & 22.5 & 5.9{4}(-2) \\
       & 3    & \phantom{-}6.14(1)\phantom{-} & \phantom{-}1.78(1)\phantom{-} & 1.25(-2) & 2.50(2)\phantom{-} & 1.02(2)\phantom{-} & 2.10\phantom{(-1)}   & 2.12(1)           & 4.80(1)           & 6.93(1)              & 36.0\phantom{3} & 32.4\phantom{3} & 32.6\phantom{3} & 15.6 & 21.4 & 22.7    & 4.65(-2)\\
Full   & 30   & \phantom{-}1.37(2)\phantom{-} & \phantom{-}3.83(-1)           & 9.48(-3) & 1.55(3)\phantom{-} & 1.22(2)\phantom{-} & 2.33\phantom{(-1)}   & 1.65(1)           & 2.25(1)           & 3.73(1)              & 49.8\phantom{3} & 43.1\phantom{3} & 40.6\phantom{3} & 15.4 & 20.8 & 22.1    & 1.38(-2)\\
       & 300  & \phantom{-}2.15(1)\phantom{-} &           -1.40(-1)           & 2.19(-3) & 1.31(2)\phantom{-} & 7.81\phantom{(-1)} & 1.01(-1)             & 4.56\phantom{(1)} & 5.67\phantom{(1)} & 1.21(1)              & 38.3\phantom{3} & 30.3\phantom{3} & 26.3\phantom{3} & 15.5 & 20.2 & 21.6    & 7.59(-4)\\
       & 3000 &           -2.36(-4)          &           -6.40(-1)           & 2.53(-9) & 1.05(-7)           & 1.69(-6)           & 4.40(-7)             & 1.02\phantom{(1)} & 1.02\phantom{(1)} & 4.07\phantom{(1)}    & \phantom{3}5.92 & \phantom{3}5.43 & \phantom{3}4.48 & 16.4 & 20.5 & 20.5    & 4.04(-5) \\ \\ \hline \\
       & 0    & \phantom{-}8.6{2}(1)\phantom{-} & \phantom{-}7.0{5}(1)\phantom{-} & 1.{06}(-2) & 5.04(1)\phantom{-} & 1.51(2)\phantom{-} & \nodata\phantom{(-)} & 1.54(1) & 9.{68}(1) & \nodata\phantom{(-)} & 24.8\phantom{3} & 26.0\phantom{3} & \nodata & 14.8 & 21.6 & \nodata & 5.41(-2) \\
       & 3    & \phantom{-}6.78(1)\phantom{-} & \phantom{-}2.26(1)\phantom{-} & 9.91(-3) & 2.23(2)\phantom{-} & 1.18(2)\phantom{-} & \nodata\phantom{(-)} & 2.30(1)           & 5.41(1)           & \nodata\phantom{(-)} & 30.9\phantom{3} & 29.8\phantom{3} & \nodata         & 15.3 & 20.7 & \nodata & 4.45(-2)\\
Simple & 30   & \phantom{-}1.34(2)\phantom{-} & \phantom{-}1.72\phantom{(-1)} & 5.56(-3) & 1.36(3)\phantom{-} & 1.47(2)\phantom{-} & \nodata\phantom{(-)} & 1.79(1)           & 2.56(1)           & \nodata\phantom{(-)} & 42.5\phantom{3} & 39.2\phantom{3} & \nodata         & 14.8 & 19.8 & \nodata & 1.33(-2)\\
       & 300  & \phantom{-}2.07(1)\phantom{-} & \phantom{-}5.87(-2)           & 1.27(-3) & 1.15(2)\phantom{-} & 8.71\phantom{(-1)} & \nodata\phantom{(-)} & 4.73\phantom{(1)} & 6.07\phantom{(1)} & \nodata\phantom{(-)} & 32.7\phantom{3} & 27.3\phantom{3} & \nodata         & 14.7 & 18.9 & \nodata & 6.29(-4)\\
       & 3000 &           -1.86(-4)           &           -6.40(-1)           & 2.04(-9) & 9.73(-7)           & 1.68(-6)           & \nodata\phantom{(-)} & 1.02\phantom{(1)} & 1.02\phantom{(1)} & \nodata\phantom{(-)} & \phantom{3}5.21 & \phantom{3}4.98 & \nodata         & 16.4 & 20.5 & \nodata & 2.74(-5)\\ \\ \hline \\
NoPair & 3    & \phantom{-}5.96(1)\phantom{-} & \phantom{-}1.78(1)\phantom{-} & 1.25(-2) & 2.50(2)\phantom{-} & 1.02(2)\phantom{-} & \nodata\phantom{(-)} & 2.12(1)           & 4.80(1)           & \nodata\phantom{(-)} & 36.0\phantom{3} & 32.4\phantom{3} & \nodata         & 15.6 & 21.4 & \nodata & 3.86(-2)\\
NoScat & 3    & \phantom{-}6.80(1)\phantom{-} & \phantom{-}2.02(1)\phantom{-} & 1.53(-2) & 2.50(2)\phantom{-} & 1.02(2)\phantom{-} & 2.11\phantom{(-1)}   & 2.23(1)           & 5.36(1)           & 7.08(1)              & 36.0\phantom{3} & 32.4\phantom{3} & 32.5\phantom{3} & 15.9 & 22.5 & 20.8    & 3.81(-2)\\
NoRel  & 3    & \phantom{-}6.17(1)\phantom{-} & \phantom{-}1.81(1)\phantom{-} & 5.76(-3) & 2.29(2)\phantom{-} & 9.49(1)\phantom{-} & 1.93\phantom{(-1)}   & 2.17(1)           & 4.70(1)           & 6.60(1)              & 31.0\phantom{3} & 28.5\phantom{3} & 28.1\phantom{3} & 15.0 & 20.0 & 20.2    & 5.26(-2)\\
NoWM   & 3    & \phantom{-}6.27(1)\phantom{-} & \phantom{-}1.89(1)\phantom{-} & 1.34(-2) & 2.43(2)\phantom{-} & 1.27(2)\phantom{-} & 2.11\phantom{(-1)}   & 2.13(1)           & 4.84(1)           & 6.89(1)              & 35.8\phantom{3} & 33.9\phantom{3} & 32.6\phantom{3} & 15.7 & 20.8 & 22.7    & 4.76(-2)\\
Shen   & 3    & \phantom{-}6.11(1)\phantom{-} & \phantom{-}1.78(1)\phantom{-} & 1.23(-2) & 2.45(2)\phantom{-} & 1.01(2)\phantom{-} & 2.10\phantom{(-1)}   & 2.12(1)           & 4.80(1)           & 6.93(1)              & 36.0\phantom{3} & 32.4\phantom{3} & 32.5\phantom{3} & 15.6 & 21.5 & 22.7    & 4.63(-2)\\
LS220  & 3    & \phantom{-}6.10(1)\phantom{-} & \phantom{-}1.77(1)\phantom{-} & 1.27(-2) & 2.45(2)\phantom{-} & 1.01(2)\phantom{-} & 2.11\phantom{(-1)}   & 2.11(1)           & 4.79(1)           & 6.93(1)              & 36.0\phantom{3} & 32.4\phantom{3} & 32.6\phantom{3} & 15.6 & 21.5 & 22.7    & 4.64(-2)\\
\tablenotetext{}{\textbf{Notes:} Volume-integrated quantities from the neutrino transport calculated by {\tt Sedonu}. Compare with Table~\ref{tab:flash}. The numbers in parentheses indicate the power of $10$ with which the data given must be scaled, e.g., $6.95(-1)$ is $6.95 \times 10^{-1}$. $\mathcal{L}_\mathrm{escape}$ and $\left<E_{\nu,\mathrm{escape}}\right>$ are the integrated luminosity and average energy of neutrinos that escape to infinity. $\mathcal{H_{\nu\bar{\nu}}}$ is the integrated annihilation rate within $45^\circ$ from the axis of symmetry. All other quantities are the same as in Table~\ref{tab:flash}: $\mathcal{C}_\nu-\mathcal{H}_\nu$ is the net rate of energy loss from the fluid by neutrinos. $\left<dY_e/dt\right>$ is the mass-weighted average of the rate of change of the electron fraction. $M_\nu$ is the mass in which neutrinos are a larger source of heat than viscosity, i.e. $\mathcal{H}_\nu-\mathcal{C}_\nu>\mathcal{H}_\mathrm{visc}$. $\mathcal{L}_\mathrm{emit}$ is the rate at which neutrino energy is emitted in the disk. $\left<E_{\nu,\mathrm{emit}}\right>$ is the energy density-weighted average energy of these emitted neutrinos.  $\nu_x$ represents the sum of all four heavy lepton neutrino species. $1\,\mathrm{B} = 10^{51}\,\mathrm{erg}$.}
\label{tab:results}
\end{deluxetable*}


\begin{deluxetable}{cccccccccccccccccc}
\tablecaption{Microphysics Resolution Study - Global Quantities}
\tablecolumns{18}
\tablewidth{9.5in}
\startdata
\tablehead{\colhead{Physics} & \colhead{$\mathcal{C}_\nu-\mathcal{H}_\nu$} & \colhead{$\left<\frac{dY_e}{dt}\right>$} & \colhead{$M_\nu$}            & \multicolumn{3}{c}{$\mathcal{L}_\mathrm{emit}\,(\mathrm{B\ s}^{-1})$} & \multicolumn{3}{c}{$\mathcal{L}_\mathrm{escape}\,(\mathrm{B\ s}^{-1})$} & \multicolumn{3}{c}{$\left<E_{\nu,\mathrm{emit}}\right>\,\mathrm{(MeV)}$} & \multicolumn{3}{c}{$\left<E_{\nu,\mathrm{escape}}\right>\,\mathrm{(MeV)}$} & \colhead{$\mathcal{H}_{\nu\bar{\nu}}$} \\
                             &  \colhead{$(\mathrm{B\ s}^{-1})$}            & \colhead{$(\mathrm{s}^{-1})$}            & \colhead{$(10^{-3}M_\odot)$} & \colhead{$\nu_e$} & \colhead{$\bar{\nu}_e$} & \colhead{$\nu_x$}         & \colhead{$\nu_e$} & \colhead{$\bar{\nu}_e$} & \colhead{$\nu_x$}           & \colhead{$\nu_e$} & \colhead{$\bar{\nu}_e$} & \colhead{$\nu_x$}            & \colhead{$\nu_e$} & \colhead{$\bar{\nu}_e$} & \colhead{$\nu_x$}      & \colhead{$(\mathrm{B\ s}^{-1})$}     }
\startdata
\cutinhead{BH Disk Global Quantities}\\
2xEnergy & 45.4 & 18.9 & 3.16 & 111 & 71.5 & 0.874 & 7.63 & 31.3 & 0.776 & 29.1 & 29.0 & 26.1 & 14.2 & 21.9 & 25.9 & 3.86(-3) \\
2x$\rho$ & 45.5 & 19.0 & 3.12 & 110 & 71.7 & 0.870 & 7.62 & 31.3 & 0.772 & 29.1 & 29.0 & 26.1 & 14.2 & 21.9 & 25.9 & 3.86(-3) \\
2xT      & 45.7 & 19.2 & 3.22 & 111 & 72.3 & 0.879 & 7.63 & 31.5 & 0.779 & 29.1 & 29.1 & 26.2 & 14.2 & 21.9 & 26.0 & 3.90(-3) \\
2x$Y_e$  & 45.4 & 18.9 & 3.16 & 110 & 71.5 & 0.867 & 7.63 & 31.2 & 0.769 & 29.1 & 29.0 & 26.2 & 14.2 & 21.9 & 26.0 & 3.85(-3) \\
2x$\phi$ & 45.4 & 18.9 & 3.12 & 110 & 71.5 & 0.872 & 7.63 & 31.2 & 0.775 & 29.1 & 29.0 & 26.2 & 14.2 & 21.9 & 26.0 & 3.79(-3) \\
2x$\mu$  & 45.4 & 18.9 & 3.14 & 110 & 71.5 & 0.873 & 7.62 & 31.2 & 0.773 & 29.1 & 29.0 & 26.1 & 14.2 & 21.9 & 25.9 & 2.55(-3) \\
2xSteps  & 45.5 & 18.9 & 3.13 & 110 & 71.5 & 0.871 & 7.66 & 31.3 & 0.773 & 29.1 & 29.0 & 26.2 & 14.2 & 21.9 & 25.9 & 3.86(-3) \\
\cutinhead{HMNS Disk Global Quantities}\\
2xEnergy & 61.5 & 17.8 & 12.5 & 250 & 102 & 2.11 & 21.2 & 48.0 & 69.3 & 36.0 & 32.4 & 32.5 & 15.6 & 21.4 & 22.7 & 4.64(-2) \\
2x$\rho$ & 61.5 & 17.9 & 12.5 & 249 & 102 & 2.11 & 21.2 & 48.1 & 69.3 & 36.0 & 32.4 & 32.6 & 15.6 & 21.4 & 22.7 & 4.65(-2) \\
2xT      & 61.8 & 18.0 & 12.7 & 250 & 103 & 2.12 & 21.2 & 48.3 & 69.3 & 36.1 & 32.5 & 32.7 & 15.6 & 21.5 & 22.7 & 4.67(-2) \\
2x$Y_e$  & 61.4 & 17.8 & 12.5 & 250 & 102 & 2.11 & 21.2 & 48.0 & 69.3 & 36.0 & 32.4 & 32.6 & 15.6 & 21.4 & 22.7 & 4.64(-2) \\
2x$\phi$ & 61.4 & 17.8 & 12.5 & 250 & 102 & 2.11 & 21.2 & 48.0 & 69.3 & 36.0 & 32.4 & 32.6 & 15.6 & 21.4 & 22.7 & 4.61(-2) \\
2x$\mu$  & 61.4 & 17.8 & 12.5 & 250 & 102 & 2.10 & 21.2 & 48.0 & 69.3 & 36.0 & 32.4 & 32.6 & 15.6 & 21.4 & 22.7 & 3.17(-2) \\
2xSteps  & 61.6 & 17.8 & 12.5 & 250 & 102 & 2.11 & 21.3 & 48.1 & 69.3 & 36.0 & 32.4 & 32.6 & 15.6 & 21.5 & 22.7 & 4.64(-2) \\
\enddata
\tablenotetext{}{\textbf{Notes:} Volume-integrated quantities from the neutrino transport calculated by {\tt Sedonu}. Each listed run doubles the number of points used in the indicated direction. $E_\nu$, $\rho$, $T$, $Y_e$ resolutions are properties of the NuLib opacity tables. $\phi$ and $\mu$ are the dimensions of the distribution function in each grid cell. In the 2xSteps run, the distance a neutrino moves in a single step is changed from 0.4 to 0.2 times the smallest dimension of the occupied grid cell. The numbers in parentheses indicate the power of $10$ with which the data given must be scaled, e.g., $6.95(-1)$ is $6.95 \times 10^{-1}$. The table quantities are the same as in Table~\ref{tab:results}: $\mathcal{L}_\mathrm{escape}$ and $\left<E_{\nu,\mathrm{escape}}\right>$ are the integrated luminosity and average energy of neutrinos that escape to infinity. $\mathcal{H_{\nu\bar{\nu}}}$ is the integrated annihilation rate within $45^\circ$ from the axis of symmetry. $\mathcal{C}_\nu-\mathcal{H}_\nu$ is the net rate of energy loss from the fluid by neutrinos. $\left<dY_e/dt\right>$ is the mass-weighted average of the rate of change of the electron fraction. $M_\nu$ is the mass in which neutrinos are a larger source of heat than viscosity, i.e. $\mathcal{H}_\nu-\mathcal{C}_\nu>\mathcal{H}_\mathrm{visc}$. $\mathcal{L}_\mathrm{emit}$ is the rate at which neutrino energy is emitted in the disk. $\left<E_{\nu,\mathrm{emit}}\right>$ is the energy density-weighted average energy of these emitted neutrinos. $\nu_x$ represents the sum of all four heavy lepton neutrino species. $1\,\mathrm{B} = 10^{51}\,\mathrm{erg}$.}
\label{tab:res_study}
\end{deluxetable}
\clearpage
\end{turnpage}

\end{document}